\documentclass[twocolumn]{aastex63}

\usepackage{amssymb,amsmath,amsfonts,amsbsy}
\usepackage{color}
\usepackage{enumerate}
\usepackage{graphicx}
\usepackage{pifont}

\usepackage{hyperref}

\newcommand{\orcid}[1]{\href{https://orcid.org/#1}}

\usepackage{textcase}
\usepackage{ulem}

\usepackage{CJKutf8}

\def\LKCs{Lipschitz-Killing curvatures}

\newcommand{\Mspace}{{\mathbb M}}

\newcommand{\Rspace}{{\mathbb R}}

\newcommand{\beq}{\begin{eqnarray}}
\newcommand{\eeq}{\end{eqnarray}}
\newcommand{\beqq}{\begin{eqnarray*}}
	\newcommand{\eeqq}{\end{eqnarray*}}

\newcommand{\beqn}{\begin{equation}}
\newcommand{\eeqn}{\end{equation}}
\newcommand{\beqqn}{\begin{equation*}}
\newcommand{\eeqqn}{\end{equation*}}

\newcommand{\Min}{{\mathcal M}}

\def\Min{\mathcal M}

\newcommand{\sqbinom}[2]{\begin{bmatrix}#1 \\ #2 \end{bmatrix}}
\newcommand{\ssqbinom}[2]{{\mbox{\scriptsize $\sqbinom{#1}{#2}$ }}}

\shorttitle{Minkowski Functionals of the SDSS-III BOSS catalogs}
\shortauthors{Appleby et al.}

\begin{document}

\title{Minkowski Functionals of SDSS-III BOSS : Hints of Possible Anisotropy in the Density Field?}

\correspondingauthor{Stephen Appleby}
\email{stephen.appleby@apctp.org}

\author[0000-0001-8227-9516]{Stephen Appleby} 
\affiliation{Asia Pacific Center for Theoretical Physics, Pohang, 37673, Korea}
\affiliation{Department of Physics, POSTECH, Pohang 37673, Korea}
\author[0000-0001-9521-6397]{Changbom Park}
\affiliation{School of Physics, Korea Institute for Advanced Study, 85
Hoegiro, Dongdaemun-gu, Seoul, 02455, Korea}
\author[0000-0003-1494-0856]{Pratyush Pranav}
\affiliation{Univ Lyon, ENS de Lyon, Univ Lyon1, CNRS, Centre de Recherche Astrophysique de Lyon UMR5574, F–69007, Lyon, France}
\author[0000-0003-4923-8485]{Sungwook E. Hong
\begin{CJK*}{UTF8}{mj}(홍성욱)\end{CJK*}}
\affiliation{Korea Astronomy and Space Science Institute, 776 Daedeokdae-ro, Yuseong-gu, Daejeon 34055, Korea}
\author[0000-0003-3428-7612]{Ho Seong Hwang}
\affiliation{Astronomy Program, Department of Physics and Astronomy,
Seoul National University, 1 Gwanak-ro, Gwanak-gu, Seoul 08826, Republic
of Korea}
\affiliation{SNU Astronomy Research Center,
Seoul National University, 1 Gwanak-ro, Gwanak-gu, Seoul 08826, Republic
of Korea}
\author[0000-0002-4391-2275]{Juhan Kim}
\affiliation{Center for Advanced Computation, Korea Institute for Advanced
Study, 85 Hoegiro, Dongdaemun-gu, Seoul, 02455, Korea}
\author[0000-0002-0828-3901]{Thomas Buchert}
\affiliation{Univ Lyon, ENS de Lyon, Univ Lyon1, CNRS, Centre de Recherche Astrophysique de Lyon UMR5574, F–69007, Lyon, France}

\begin{abstract}
We present measurements of the Minkowski functionals extracted from the SDSS-III BOSS catalogs. After defining the Minkowski functionals, we describe how an unbiased reconstruction of these statistics can be obtained from a field with masked regions and survey boundaries, validating our methodology with Gaussian random fields and mock galaxy snapshot data. From the BOSS galaxy data we generate a set of four density fields in three dimensions corresponding to the northern and southern skies of LOWZ and CMASS catalogs, smoothing over large scales such that the field is perturbatively non-Gaussian. We extract the Minkowski functionals from each data set separately, and measure their shapes and amplitudes by fitting a Hermite polynomial expansion. For the shape parameter of the Minkowski functional curves $a_0$, that is related to the bispectrum of the field, we find that the LOWZ-South data presents a systematically lower value of $a_0 = -0.080 \pm 0.040$ than its northern sky counterpart $a_0 = 0.032 \pm 0.024$. Although the significance of this discrepancy is low, it potentially indicates some systematics in the data or that the matter density field exhibits anisotropy at low redshift. By assuming a standard isotropic flat $\Lambda$CDM cosmology, the amplitudes of Minkowski functionals from the combination of northern and southern sky data give the constraints $\Omega_{\rm c} h^2 n_{\rm s} = 0.110 \pm 0.006$ and $0.111 \pm 0.008$ for CMASS and LOWZ, respectively, which is in agreement with the Planck $\Lambda$CDM best-fit $\Omega_{\rm c}h^{2} n_{\rm s} = 0.116 \pm 0.001$.\\
\end{abstract}


\section{Introduction} 

The Minkowski functionals (MFs) describe the morphological, i.e. geometrical and topological characteristics of a field. They have a long and venerable history within cosmology, both theoretical and computational, as well as observational, starting with the genus of isodensity excursion sets \citep{Gott:1989yj,1991ApJ...378..457P,Mecke:1994ax,Schmalzing:1997aj,Schmalzing:1997uc,1989ApJ...345..618M,1992ApJ...387....1P,2001ApJ...553...33P,Park:2009ja,doi:10.1111/j.1365-2966.2010.18015.x}. Often referred to merely as summary statistics in cosmological literature, they emerge from topo-geometrical description of space and manifolds, and together with the homology characteristics encoded in the Betti numbers \citep{Park:2013dga,Feldbrugge:2019tal,Pranav:2018lox,Pranav:2018pnu,Pranav:2016gwr,Shivshankar:2015aza,vandeWeygaert:2011ip}, and its hierarchical extension persistent homology \citep{edelsbrunnerharer10,Pranav:2016gwr}, as well as the Minkowski tensors \citep{Beisbart:2001vb,Beisbart:2001gk, Ganesan:2016jdk, Chingangbam:2017sap, Chingangbam:2017uqv,  Kapahtia:2019ksk, Appleby:2017uvb, Appleby:2018tzk, Kapahtia:2017qrg, K.:2018wpn,Wilding:2020oza,Goyal:2021nun,Chingangbam:2021kov}, present a high-level description of the topo-geometrical properties of the fields of interest in cosmology. Extracting information from these statistics remains an open and challenging program \citep{2001ApJ...553...33P,Hikage:2002ki,Hikage:2003fc,Park:2005fk,10.1111/j.1365-2966.2008.14358.x,Sheth:2002rf,Sheth:2005ys,Gott:2008kk,Choi:2010sx,Zhang:2010tha,Petri:2013ffb,Blake:2013noa,Wiegand:2013xfa,2014ApJ...796...86P,Wang:2015eua,Wiegand:2016ezl,Buchert:2017uup,Sullivan:2017mhr,Hikage_2001,Gott:2006yy,Appleby:2021lfq,Pranav:2021ted,Matsubara:2020fet,Lippich:2020vpy,Shim:2020qav}. 

For $D$-dimensional sets $\delta(x_{1},...x_{D})$, there are $(D+1)$ MFs, which correspond to topo-geometrical properties of the set. Considering a manifold $\Mspace$, in three dimensions, the MFs correspond to the enclosed volume ($W_0$) of $\Mspace$, as well as the surface area ($W_1$), mean ($W_2$) and Gaussian curvatures ($W_3$) of the boundary $\partial \Mspace$. In our particular setting, the manifolds of interest are the excursion sets of 3D cosmological density fields.

In a recent series of works, we have measured the genus of two-dimensional slices of the SDSS-III BOSS data, extracting cosmological information from the genus amplitude, and placing constraints on the scalar spectral index $n_{\rm s}$ and dark matter fraction $\Omega_{\rm c}h^{2}$, assuming a flat $\Lambda$CDM cosmology \citep{Appleby:2020pem}. Given that the BOSS data is spectroscopic, we have precise redshift information. Hence it is possible to extract the MFs from the full three-dimensional data, which contains more information than two-dimensional shells\footnote{Shells were analysed in \citet{Appleby:2017ahh,Appleby:2018jew,Appleby:2020pem,Appleby:2021lfq} with the long-term goal of comparing the results with photometric galaxy catalogs.}.

With this motivation in mind, in this work we extract the MFs from the full, three-dimensional galaxy distribution. Specifically, we use the SDSS-III BOSS DR12 catalog to reconstruct four distinct, three-dimensional density fields corresponding to the LOWZ/CMASS data in the northern and southern Galactic Planes. After generating smoothed number density fields from the point galaxy distribution, we measure the statistics, ensuring that the complex survey geometry does not impact our results. We then extract information from the amplitudes of these functions, using similar methodology to \cite{Appleby:2020pem}. We also measure the bispectrum components from the MFs, although we do not convert this information into cosmological parameter constraints in this work. 

The paper will proceed as follows. We define the MFs, and review their theoretical expectation values for a compact and boundary-less field in Section~\ref{sec:def}. In Section~\ref{sec:method}, we review our methodology, including the generation of a density field from a point distribution, and the mock data used to estimate statistical uncertainties. Time-poor readers can proceed directly to Section~\ref{sec:results}, which contains our principle results --- MF measurements extracted from the BOSS galaxy samples and cosmological parameter estimation using their amplitudes. We discuss our results in Section~\ref{sec:disc}.  

Some of the details of our methodology and analysis is contained in appendices. In Appendix \ref{app:MF} we define the Minkowski functionals for a generic manifold. In Appendix~\ref{sec:unbias} the numerical algorithm for MF extraction is elucidated, and we confirm that our method is unbiased by survey boundaries, using masked Gaussian random fields and mock galaxy snapshot data. The effect of redshift space distortion is briefly reviewed in Appendix~\ref{sec:rsd}. Potential systematics that could impact our analysis are described in Appendix~\ref{sec:systematics}.


\section{Integral-Geometry of Manifolds: Minkowski Functionals} 
\label{sec:def}

In a cosmological setting, we are typically interested in the properties of a scalar random field $u$, defined on a manifold $\Mspace$, which will be Euclidean and three-dimensional in this work. We take $u$ to be mean subtracted and root mean square normalised $\langle u^{2} \rangle = 1$, and Gaussian in this subsection. The Minkowski functionals $W_0, \ldots, W_3$ are topo-geometric quantifiers that encode the geometry and topology induced by the fluctuations of the field, in combination with the geometric characteristics of the manifold itself. The usual practice is to examine the properties of the \emph{excursion set} of the manifold, defined by 

\begin{equation} 
\label{intro:Au:equn}
E_\nu =\  \{x\in \Mspace: u(x)\geq \nu\},
\end{equation}

\noindent where $\nu$ is an iso-field threshold value. When dealing with compact boundary-less manifolds, such as the $2$-sphere or the periodically tiled Euclidean grid (found for example in cosmological simulation snapshot boxes), the expressions for the MFs simplify substantially, and the volume-normalised expressions may be written in terms of curvature integrals 

\begin{eqnarray} \label{eq:ci1}
W_{0} &=& {1 \over V} \int_{E_{\nu}} dV , \\ 
\label{eq:ci2} W_{1} &=& {1 \over 6V} \int_{\partial E_{\nu}} dA , \\
\label{eq:ci3} W_{2} &=& {1 \over 3\pi V} \int_{\partial E_{\nu}} G_{2}\, dA , \\ 
\label{eq:ci4} W_{3} &=& {1 \over 4\pi^{2} V} \int_{\partial E_{\nu}} G_{3}\, dA ,
\end{eqnarray} 
\noindent where $G_{2} = (\kappa_{1} + \kappa_{2})/2$ and $G_{3} = \kappa_{1} . \kappa_{2}$ are the mean and Gaussian curvatures of the boundary of the excursion set $\partial E_{\nu}$. Here, $\kappa_{1},\kappa_{2}$ are the principle curvatures of the surface $\partial E_{\nu}$ at a point, and $V$ is the total volume of the space under consideration. The equations for the $2$-sphere were derived by \cite{1970Ap......6..320D}, while the generic case including boundary effects was developed by \cite{Adler}. The usage of the Euler characteristic was introduced to cosmology by Gott and collaborators \citep{Gott:1989yj,1991ApJ...378..457P}, while the full set of volume normalised expressions in three dimensions was introduced in cosmology by \cite{Mecke:1994ax,Schmalzing:1997aj}.

For a Gaussian random field, the ensemble average of these curvature integrals can be written as \citep{1970Ap......6..320D,Adler,Gott:1986uz,10.1143/PTP.76.952,Hamilton:1986,Ryden:1988rk,1987ApJ...319....1G,1987ApJ...321....2W}
\begin{equation}\label{eq:mfg} 
\langle W_{k} \rangle  = {1 \over (2\pi)^{(k+1)/2}} {\omega_{3} \over \omega_{3-k}\omega_{k}} \left({\sigma_{1}^{2} \over 3\sigma_{0}^{2}} \right)^{k/2} e^{-\nu^{2}/2} H_{k-1}(\nu) ,
\end{equation}
where $H_k (\nu)$ is $k$-th Hermite polynomial, and $\sigma_i$ are the two-point cumulants
\begin{equation}\label{eq:cml} 
\sigma_{i}^{2} = {1 \over (2\pi)^{3}}\int d^3 k k^{2i} P(k) W^{2}(kR_{\rm G}) \, . 
\end{equation} 

\noindent $P(k)$ is the power spectrum of the field. $W(kR_{\rm G})$ is a smoothing kernel with corresponding scale $R_{\rm G}$, which we take to be Gaussian throughout this work: $W(kR_{\rm G}) \propto e^{-k^{2}R_{\rm G}^{2}/2}$.
 The amplitude of $\langle W_{k} \rangle$ is defined as 
\begin{equation}\label{eq:ampg} 
A_{k, {\rm G}} \equiv  {1 \over (2\pi)^{(k+1)/2}} {\omega_{3} \over \omega_{3-k}\omega_{k}} \left({\sigma_{1}^{2} \over 3\sigma_{0}^{2}} \right)^{k/2} . \end{equation} 

\noindent The constants $\omega_m$ are the volume of the unit ball in $\Rspace^m$:
\begin{equation}
\omega_m\,=\,\frac{\pi^{m/2}}{\Gamma(\frac{n}{2}+1)}\,,
\end{equation}
ie. $\omega_{0} = 1$, $\omega_1=2$, $\omega_2=\pi$ and $\omega_3=4\pi/3$. 

\noindent The curvature integrals ($\ref{eq:ci1}-\ref{eq:ci4}$) are equal to the MFs only when the manifold possesses no boundary. For a field defined on a generic manifold, the Gaussian Kinematic Formula (GKF) describes the MFs instead. The GKF is defined in Appendix \ref{app:MF}.

Practically, cosmological data sets are always defined on domains with boundaries. For example the Cosmic Microwave Background temperature field is measurable over the entire two-sphere, but realistically foreground masks ensure that only part of the all-sky data is used. Similarly, galaxy catalogs possess both angular and radial survey boundaries, and the resulting matter density field is measured only over a finite volume with boundary. The MFs of these cosmological data sets are therefore {\it not} described by the curvature integrals defined in ($\ref{eq:ci1}-\ref{eq:ci4}$). However, it is possible to generate an unbiased estimate of the curvature integrals from a field with boundary, and directly compare the results to the ensemble expectation values ($\ref{eq:mfg}$). 

To see this, we take one of the curvature integrals ($\ref{eq:ci2}$) as an example. By using an integral transform we can write 

\begin{equation}
    W_1 = \frac{1}{6V} \int_{\partial E_{\nu}} \textrm{dA} = \frac{1}{6V}   \int_{\Mspace} \textrm{dV} \,  \left| \nabla u \right| \, \delta_{D} \left( u - \nu \right)  .
\end{equation}

\noindent We recognise the final term in this equation as the volume average of the scalar quantity $\lambda \equiv  \delta_{D}(\nu - u) \sqrt{u_{i}u^{i}}/6$, where $u$ is the field, $u_{i}$ are the gradients with respect to some arbitrary coordinate system ($i=1,2,3$) and $\delta_{D}$ is the delta function, which is defined in a distributional sense as we are taking a volume average. If we generate a pixelated Gaussian random field, $W_{1}$ can be estimated as 

\begin{equation}\label{eq:p1} W_{1} \simeq {1 \over 6N_{\rm pix}} \sum_{\ell=1}^{N_{\rm pix}} \tilde{\delta}_{D}(u_{\ell} - \nu) \sqrt{u_{\ell,i}u_{\ell}^{i}} , \end{equation}

\noindent where $N_{\rm pix}$ is the number of pixels in the discretized field and we discretize the delta function as 

\begin{equation} \tilde{\delta}_{D}(u_{\ell} - \nu) = 
  \begin{cases}
    {1 \over \Delta \nu}, & \text{for } \nu - {\Delta \nu \over 2} \leq u_{\ell} < \nu + {\Delta \nu \over 2} \\
    0 , & {\rm otherwise}
  \end{cases} 
\end{equation}

\noindent The important point is that the numerical approximation of $W_{1}$ defined in ($\ref{eq:p1}$) can be estimated from a finite subset of a field. That is, we do not require an entire field defined on a boundary-less manifold to estimate the curvature integral ($\ref{eq:ci2}$), any unbiased sampling of pixels can be used to generate an unbiased estimator of this quantity.

The property of ergodicity is then used to equate the volume and ensemble averages of $\lambda$. The ensemble average of $\lambda$ can be written as

\begin{equation}\label{eq:p2} \langle W_{1} \rangle = {1 \over 6} \int P(u,u_{k}) du du_{k} \delta_{D}(u - \nu) \sqrt{u_{i}u^{i}} , \end{equation} 

\noindent where $P(u,u_k)$ is the joint pdf of the field and its derivatives. If we take $u, u_{k}$ to be uncorrelated and Gaussian distributed, this ensemble average yields the standard result in equation ($\ref{eq:mfg}$).

Again, we stress that the volume average ($\ref{eq:p1}$) does not require the field to be complete and boundary-less, and can be approximated by a subset of pixels of masked data. Hence we can estimate the curvature integrals from cosmological data, but we should understand that these quantities do not represent the MFs for a field defined on a manifold with boundary. This is intuitively obvious, as the curvature integrals are intrinsically `local' quantities in the sense that they can be estimated from the average properties of the field and its derivatives at points on the manifold. In contrast, the topology of the manifold is an intrinsically global quantity.

Throughout this work, we will use the terms `curvature integrals' and `Minkowski functionals' interchangeably, but the reader should understand the distinction made above, and this work is concerned solely with the curvature integrals.

\subsection{Weakly non-Gaussian Fields}

For a weakly non-Gaussian field, the amplitude and shape of the MFs is modified. To linear order in $\sigma_{0}$, the following {\it Edgeworth} expansion has been constructed \citep{Matsubara:1994wn,Matsubara:1994we,Matsubara:1995dv,1988ApJ...328...50M,Matsubara:1995ns,
2000astro.ph..6269M,10.1111/j.1365-2966.2008.12944.x,Pogosyan:2009rg,Gay:2011wz,Codis:2013exa}
\begin{multline} 
\langle W_{k} \rangle = A_{k, {\rm G}} e^{-\nu^{2}/2} \left\{\vphantom{\frac12} H_{k-1}(\nu)  \right. +   \\
\left[ {1 \over 6} S^{(0)}H_{k+2}(\nu) + {k \over 3} S^{(1)}H_{k}(\nu) \right. + \\
\label{eq:mat1} \left. \left.   {k(k-1) \over 6} S^{(2)} H_{k-2}(\nu) \right] \sigma_{0} + {\cal O}(\sigma_{0}^{2})  \right\} .
\end{multline}

 The quantities $S^{(0)}, S^{(1)}$, and $S^{(2)}$ are proportional to the three-point cumulants of the field:
\begin{eqnarray} \label{eq:sk0} 
S^{(0)} &=& {\langle \delta^{3} \rangle \over \sigma_{0}^{4}} , \\
\label{eq:sk1} S^{(1)} &=& - {3 \over 4} {\langle \delta^{2} (\nabla^{2} \delta) \rangle \over \sigma_{0}^{2} \sigma_{1}^{2}} , \\ 
\label{eq:sk2} S^{(2)} &=& -{9 \over 4}  {\langle (\nabla \delta \cdot  \nabla \delta)  (\nabla^{2} \delta) \rangle \over \sigma_{1}^{4}} . 
\end{eqnarray}  
The amplitude and shape of the MFs contain information pertaining to the $N$-point cumulants of the field, which in turn are sensitive to cosmological parameters.\footnote{See \citet{Matsubara:2020knr} for a recent expansion of $\langle W_k \rangle$ at ${\cal O}(\sigma_0^2)$ and \citet{Gay:2011wz} for a general expansion. For a model-independent approach applying Minkowski functionals to the CMB and using general Hermite expansions of the discrepancy functions with respect to the analytical Gaussian predictions, together with a generalization of Matsubara's $2$nd-order expansion, see \citet{Buchert:2017uup}.}

Throughout this work, we re-scale the iso-density threshold $\nu$ to $\nu_{\rm A}$, which is the threshold defined such that the excursion set has the same volume fraction as a corresponding Gaussian field:
\begin{equation}\label{eq:afrac} 
f_{\rm A} = {1 \over \sqrt{2\pi}} \int^{\infty}_{\nu_{A}} e^{-t^{2}/2} \, dt , 
\end{equation}
where $f_{\rm A}$ is the fractional volume of the field above $\nu_{\rm A}$. Expressing the MFs as a function of $\nu_{\rm A}$ as opposed to $\nu$ mitigates the non-Gaussianity in the MFs \citep{1987ApJ...319....1G,1987ApJ...321....2W,1988ApJ...328...50M}, although obviously does not completely remove it. Additional non-Gaussian information is retained in the mapping $\cal{F}_{\rm A}: \nu \rightarrow \nu_{\rm A}$ but is not used in this work. The non-Gaussian expansion of MFs as a function of $\nu_{\rm A}$ is \citep{2000astro.ph..6269M,10.1111/j.1365-2966.2008.12944.x}
\begin{multline} 
\langle W_{k} \rangle  = A_{k, {\rm G}} e^{-\nu_{\rm A}^{2}/2} \left\{\vphantom{\frac12} H_{k-1}(\nu_{\rm A}) \right.  +  \\
\left[ {k \over 3} \left( S^{(1)}-S^{(0)} \right) H_{k}(\nu_{\rm A}) \right.   + \\
\label{eq:nuaexp}  \left. \left.  {k(k-1) \over 6} \left(S^{(2)}- S^{(0)} \right) H_{k-2}(\nu_{\rm A}) \right] \sigma_{0} + {\cal O}(\sigma_{0}^{2})   \right\} .
\end{multline}

The amplitude of the MF $W_k$, which is the coefficient of Hermite polynomial $H_{k-1}(\nu_{\rm A})$ in the perturbative non-Gaussian expansion, predominantly contains the Gaussian information of the field (with second order corrections $\sim {\cal O}(\sigma_{0}^{2})$). All other Hermite polynomial coefficients contain only higher point cumulant information, which are induced by the non-Gaussianity of the field. 

As we have seen in equations~(\ref{eq:mfg}, \ref{eq:mat1}, \ref{eq:nuaexp}), the Hermite polynomial expansion is useful to extract information from the MF curves.
Especially, one can extract their coefficient by using the following orthogonality relation:
\begin{equation} \label{eq:orth} 
\int_{-\infty}^{\infty} H_{\rm m}(\nu) H_{\rm n}(\nu) e^{-\nu^{2}/2} \, d\nu = \sqrt{2\pi} n ! \, \delta_{mn} , 
\end{equation} 
where $\delta_{mn}$ is the Kronecker delta. Provided we are smoothing over scales $R_{\rm G}$ such that the $\sigma_{0}$ expansion of the MFs is applicable, we can multiply the measured MF by a Hermite polynomial and integrate over $\nu$ to obtain the coefficient of the polynomial. Performing the integral in equation (\ref{eq:orth}) over $\nu_{\rm A}$ rather than $\nu$ is recommended, as the MFs as a function of $\nu$  are more strongly asymmetric around $\nu = 0$. To reliably utilise the orthogonality property of the functions, one must integrate over large $|\nu|$ ranges. The extraction of polynomial coefficients using this method is only formally valid if we have measured the MF over sufficiently large threshold range. Any truncation in the integral will correlate the Hermite polynomial coefficients obtained using this method.

The intention of this paper is to provide a method of numerically reconstructing the curvature integrals ($\ref{eq:ci1}-\ref{eq:ci4}$) from a discretized and masked density field, and then extracting cosmological information from $A_{k, {\rm G}}$, the Gaussian amplitude of these functions. We are not measuring the `true' MFs of the bounded manifold on which the matter density field is defined, as explained in the previous section. We intend to pursue the difference between the curvature integrals and MFs in detail in future work, as there is additional information contained within the boundary. The presence of a mask can profoundly modify the global properties of the manifold.

\section{Methodology} 
\label{sec:method}

We now review and subsequently validate our methodology. First, we describe our reconstruction of a smoothed density field from the point distribution, and introduce the mock galaxy catalogs used to reconstruct the statistical error of our measurements. In appendix \ref{app:num} we provide a detailed explanation for our numerical algorithm for measuring the curvature integrals, and validate our analysis with mock data. 

\subsection{Density Field Reconstruction} 
\label{sec:data}

We measure the MFs of the SDSS-III Baryon Oscillation Spectroscopic Survey (BOSS) \citep{2000AJ....120.1579Y}. The $12^{\rm th}$ data release of the Sloan Digital Sky Survey (SDSS) \citep{2015ApJS..219...12A} imaged $9,376\,{\rm deg^{2}}$ in the $ugriz$ bands \citep{1996AJ....111.1748F}. The survey was executed with the 2.5m Sloan telescope \citep{2006AJ....131.2332G} at the Apache Point Observatory in New Mexico. The extra-galactic catalog contains 1,372,737 unique galaxies, with redshifts extracted using an automated pipeline described in  \citet{2012AJ....144..144B}. 

The SDSS-III BOSS data is decomposed into two catalogs. The LOWZ sample is composed of galaxies predominantly at redshift $z<0.4$, and are selected using numerous colour-magnitude cuts that are intended to match the evolution of a passively evolving stellar population. The purpose is to extend the bright and red ``low-redshift" galaxy population measured in the SDSS-II Luminous Red Galaxies (LRGs) to relatively higher redshift. 
The CMASS galaxies, on the other hand, are selected using a set of colour-magnitude cuts to identify ``high-redshift" galaxies at $0.4 < z < 0.7$. In contrast to the LOWZ data, the sample is not biased towards red galaxies as some of the colour limits imposed on the SDSS-II sample have been removed. The colour-magnitude cut is varied with redshift to ensure massive objects are sampled as uniformly as possible over the survey volume. We direct the reader to \cite{2016MNRAS.455.1553R} for further details of the galaxy samples. 

Throughout this work, we treat the LOWZ and CMASS catalogs separately, and also treat the north and south sky data as independent. Hence we have four practically independent data sets --- CMASS and LOWZ, north/south --- from which we extract the MF statistics. All steps below are repeated individually for each subset of the data. 

\begin{table}[tb]
\begin{center}
 \begin{tabular}{||c  c ||}
 \hline
 Parameter \, & Fiducial Value \\ [0.5ex] 
 \hline\hline
 $\Omega_{\rm m}$ & $0.307$   \\ 
 $h$ & $0.6777$   \\
 $w_{\rm de}$ & $-1$ \\
 $n_{\rm s}$ & $0.9611$   \\
 $\sigma_{8}$ & $0.8288$ \\ 
 $R_{\rm G}$ & $35 \, {\rm Mpc}$ \\
  \hline 
\end{tabular}
\caption{\label{tab:1}Fiducial cosmological parameters used to reconstruct distances in this work, and in the creation of the BOSS Patchy mocks. $R_{\rm G}$ is the smoothing scale used throughout this work, using a Gaussian kernel.}
\end{center} 
\end{table}

We begin by converting the galaxy data into a three-dimensional number density field. To do so, we bin the galaxies into redshift shells of thickness $\Delta z = 0.02$, over the range $0.10 < z < 0.48$ and $0.40 < z < 0.68$ for the LOWZ/CMASS data, respectively\footnote{We repeated our analysis using $\Delta z = 0.015$ and $\Delta z = 0.03$, and found no significant change to our results}. We then select galaxies in each shell to match the number density as $\bar{n} = 6.25 \times 10^{-5} \, {\rm Mpc}^{-3}$. The selection is made using a lower mass cut based on the predicted stellar mass of the galaxies from the {\it Portsmouth} model found in \citet{Tinker:2016zpi}. If the total number density in a given redshift shell is below $\bar{n}$, it is not used in our analysis. To generate a number density, which is a dimension-full quantity, we use the fiducial cosmology presented in Table~\ref{tab:1} to define the volume of the shells. For practical purposes, variation of this cosmology will not affect our results as the shells are used only to generate mass cuts. The number density cut restricts our analysis to the redshift ranges $0.20 < z < 0.40$ and $0.45 < z < 0.60$ for LOWZ/CMASS, respectively. We could use lower redshift data, but the volume over the range $0.1 < z < 0.2$ is insufficient to affect our results. In Appendix~\ref{sec:systematics}, we repeat our analysis for a randomly selected subset of the galaxies and find no significant change in our results. This is because the overall galaxy catalog is sparse, and we are using the majority of the sample in the given redshift range. In other words, the random and mass-selected subsamples are sufficiently similar so that the MF reconstruction is unaffected by this choice. In general, care must be taken with sampling, because the MFs can be sensitive to selection effects. The details of sampling becomes irrelevant only when the smoothing scale is much larger than the mean separation of the galaxies \citep{Kim:2014axe}. 

\begin{table}[tb]
 \begin{tabular}{|| c  c  c ||}
 \hline
 Data Set \, & $L_{\rm box}$ (Mpc) \, & $\Delta_g$ (Mpc) \\ [0.5ex] 
 \hline\hline
 CMASS N & $4200$ & $8.2$   \\ 
 CMASS S & $3330$  & $6.5$ \\
 LOWZ N  & $3050$ & $6.0$ \\
 LOWZ S & $2400$ & $4.7$  \\
  \hline 
\end{tabular}
\caption{\label{tab:2} Size of box that we enclose the four data sets in, and the resolution of each box. We use Gaussian smoothing with scale $R_{\rm G} = 35 \, {\rm Mpc}$ to mitigate any discrete pixel effects. }
\end{table}

The selected galaxies are aggregated into a uniform cubic grid of size $L_{\rm box}$ and resolution $\Delta_g$ along each dimension, where these values are provided in Table~\ref{tab:2} for the four different data sets. The three-dimensional positions of the galaxies are generated from their angular positions and redshifts using the fiducial cosmology in Table~\ref{tab:1}. At this stage, each galaxy is weighted to account for observational systematics. Specifically, the following combined weight was applied to each galaxy in the LOWZ and CMASS sample: 
\begin{equation}
w_{\rm tot} = w_{\rm systot}\left( w_{\rm cp} + w_{\rm noz} -1 \right) , \end{equation} 
where $w_{\rm cp}$ is the correction factor to account for the subsample of galaxies that are not assigned a spectroscopic fibre, $w_{\rm noz}$ is for the failure in the pipeline to assign redshifts for certain galaxies, and $w_{\rm systot}$ represents non-cosmological fluctuations in the CMASS target density due to stellar density and seeing. Each galaxy contributes $w_{\rm tot}$ to its nearest pixel, and it generates a discrete number field $n_{ijk}$, where $i,j,k$ subscripts run over the lattice in $x_{1,2,3}$ directions, respectively. The weights are also accounted for when generating the number density of the galaxy samples. We also generate a mask lattice $M_{ijk}$ by projecting the angular selection function $\omega_{\ell}$ into the same cubic grid and applying the radial boundaries. The angular selection function $\omega_{\ell}$ is a Healpix\footnote{http://healpix.sourceforge.net} \citep{Gorski:2004by} $N_{\rm side} =512$ map that takes value $\omega_{\ell} = 0$ if the $\ell$ pixel lies outside the survey boundary, and $0 \leq \omega_{\ell} \leq 1$ if $\ell$ is within the mask ($1 \leq \ell \leq 12\times N_{\rm side}^{2}$ is the pixel identifier). The angular mask is projected into a three-dimensional cube such that $M_{ijk} = \omega_{\ell}$ if the $(i,j,k)$ pixel lies inside the survey geometry, and $M_{ijk} = 0$ otherwise. We then apply a binary mask to $n_{ijk}$ by setting $n_{ijk} = 0$ if $M_{ijk} < M_{\rm cut}$, where we set $M_{\rm cut} = 0.9$. Finally, we define $\bar{n}$ as the average number of galaxies within the unmasked pixels, and define a mean subtracted number density as $n_{ijk} \to (n_{ijk}/\bar{n}-1)$. The remaining masked pixels have an arbitrary `bad pixel' value $\delta_{b}$ and are not used.

Next, we smooth both the field $n_{ijk}$ and mask $M_{ijk}$ with a Gaussian kernel $W(kR_{\rm G}) \propto \exp\left[-k^{2}R_{\rm G}^{2}/2\right]$, where $R_{\rm G} = 35 \, {\rm Mpc}$ is selected such that the field is expected to be in the weakly non-Gaussian regime \citep{Matsubara:2020knr}. We denote the smoothed field and mask as $\tilde{\delta}_{ijk}$ and $\tilde{M}_{ijk}$, respectively. We make a second mask cut and set $\tilde{\delta}_{ijk} = \delta_{\rm b}$ if $\tilde{M}_{ijk} \le M_{\rm cut}$, where $M_{\rm cut} = 0.9$. This second masking procedure cuts regions of the density field close to the survey boundary. Finally, we calculate the average $\mu$ and root mean square (R.M.S.) $\sigma$ of all unmasked pixels, and define a mean subtracted, unit variance field $\delta_{ijk} = (\tilde{\delta}_{ijk} - \mu)/\sigma$.

Having constructed a smoothed, discretized density field $\delta_{ijk}$, we next extract the MFs from the unmasked pixels in the following section using the method described in \citet{Appleby:2018tzk} but accounting for the presence of a mask. A discussion of how we adjust our algorithm to account for the mask can be found in appendix \ref{app:num}.

In this work we smooth on a relatively large scale $R_{\rm G} = 35 \, {\rm Mpc}$. Such a large smoothing is chosen because we intend to compare our measurements to the perturbative Edgeworth expansion derived in \citet{Matsubara:1994wn,Matsubara:1994we,Matsubara:1995dv}. If we smooth on smaller scales, the amplitude of the Minkowski functionals becomes increasingly contaminated by higher point cumulant contributions, and also by nonlinear redshift space distortion effects. An alternative approach is to smooth on small scales and correct for non-linear gravitational interactions using simulations \citep{Li:2016wbl,Appleby:2021lfq}. This method has the advantage of yielding much stronger constraining power, but requires more careful analyses to remove nonlinear systematics and to take into account their dependence on cosmological models. Throughout this work, we follow \citet{Appleby:2020pem} and avoid (as far as possible) correcting the measured MFs using simulations. We do correct for redshift space distortions using simulations, but this is a $\sim {\cal O} (1\%)$ effect.

In what follows we extract the $W_{0,1,2,3}$ MFs from the masked, smoothed galaxy density fields at $N=41$, $\nu_{\rm A}$ threshold values, equi-spaced over the range $-3 < \nu_{\rm A} < 3$. To determine the magnitude of the statistical fluctuations on these measurements, we repeat our analysis on a set of mock galaxies with similar properties to the data. The mock analysis is described next.

\subsection{Mock Galaxy Catalogs}

To estimate the statistical uncertainty of the $W_{k}(\nu_{A})$ measurements, we use $N_{\rm r}=250$ Multidark patchy mocks \citep{2016MNRAS.456.4156K,2016MNRAS.460.1173R}. A detailed explanation of their construction can be found in \citet{2016MNRAS.456.4156K}. Briefly, the mocks were generated using an iterative procedure to mimic a reference galaxy catalog using gravity solvers and statistical biasing models \citep{2014MNRAS.439L..21K}. The reference catalog is the Big-MultiDark $N$-body simulation, which used Gadget-2 \citep{Springel:2005mi} to gravitationally evolve $3840^{3}$ particles in a $(2.5 h^{-1} {\rm Gpc})^3$ volume. Halo abundance matching was utilized to reproduce the clustering of galaxy data. The Patchy code \citep{2014MNRAS.439L..21K,10.1093/mnras/stv645} matches the two- and three-point clustering statistics with the reference simulation in multiple redshift bins. Stellar masses are estimated and mock lightcones are generated, including masks and other selection effects. The mock catalogs accurately reproduce the number density, two-point correlation function, selection function and survey geometry of the SDSS-III BOSS DR12 observational data. The simulations adopted a Planck standard $\Lambda$CDM cosmology with $\Omega_{\rm m} = 0.307$, $\Omega_{\rm b} = 0.048$, $n_{\rm s} = 0.961$, $H_{0} = 67.77 ~{\rm km \,s^{-1} Mpc^{-1}}$, the same as the fiducial cosmology adopted in this study. 

\begin{figure}[htb]
  \centering 
  \includegraphics[width=0.47\textwidth]{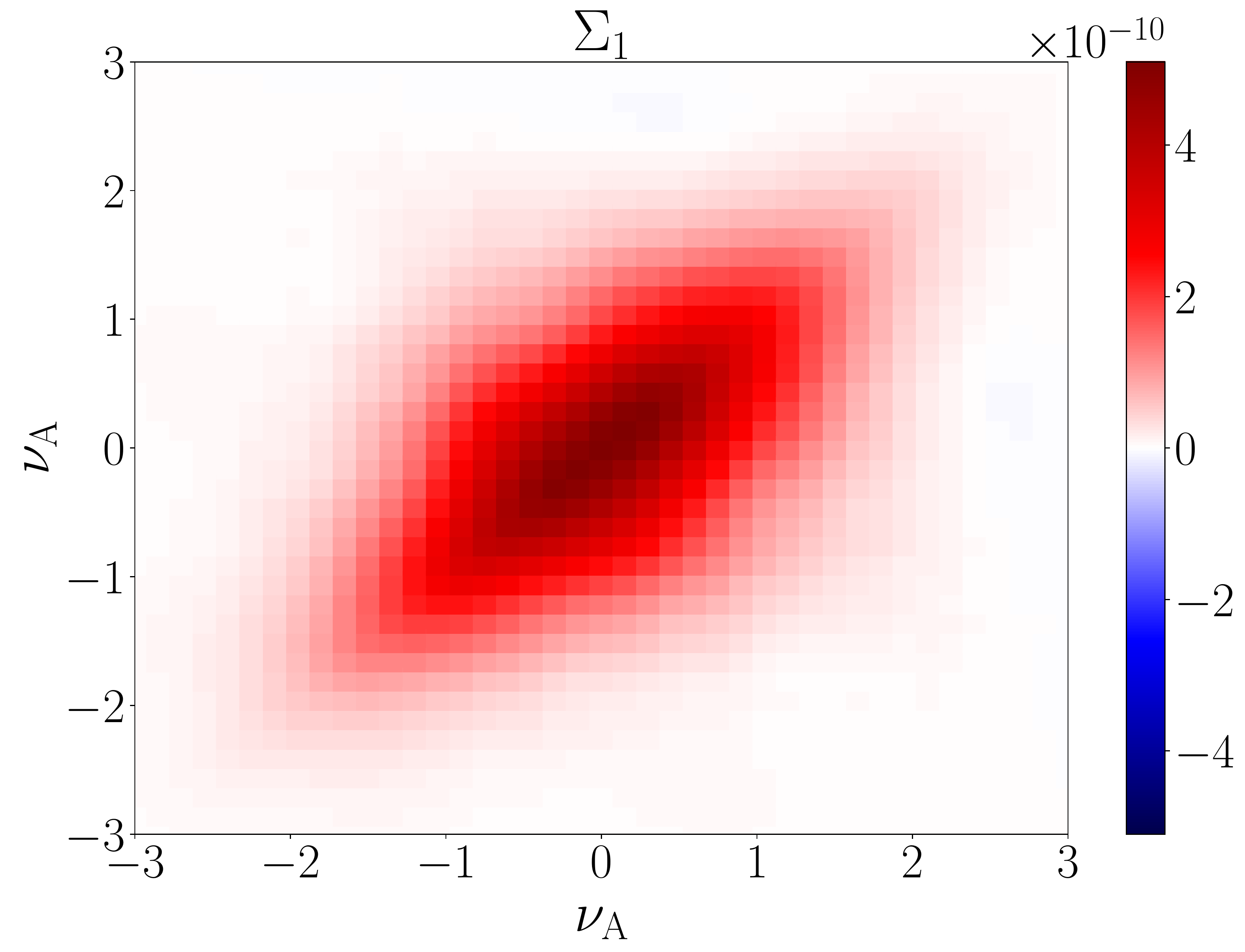} \\
  \includegraphics[width=0.47\textwidth]{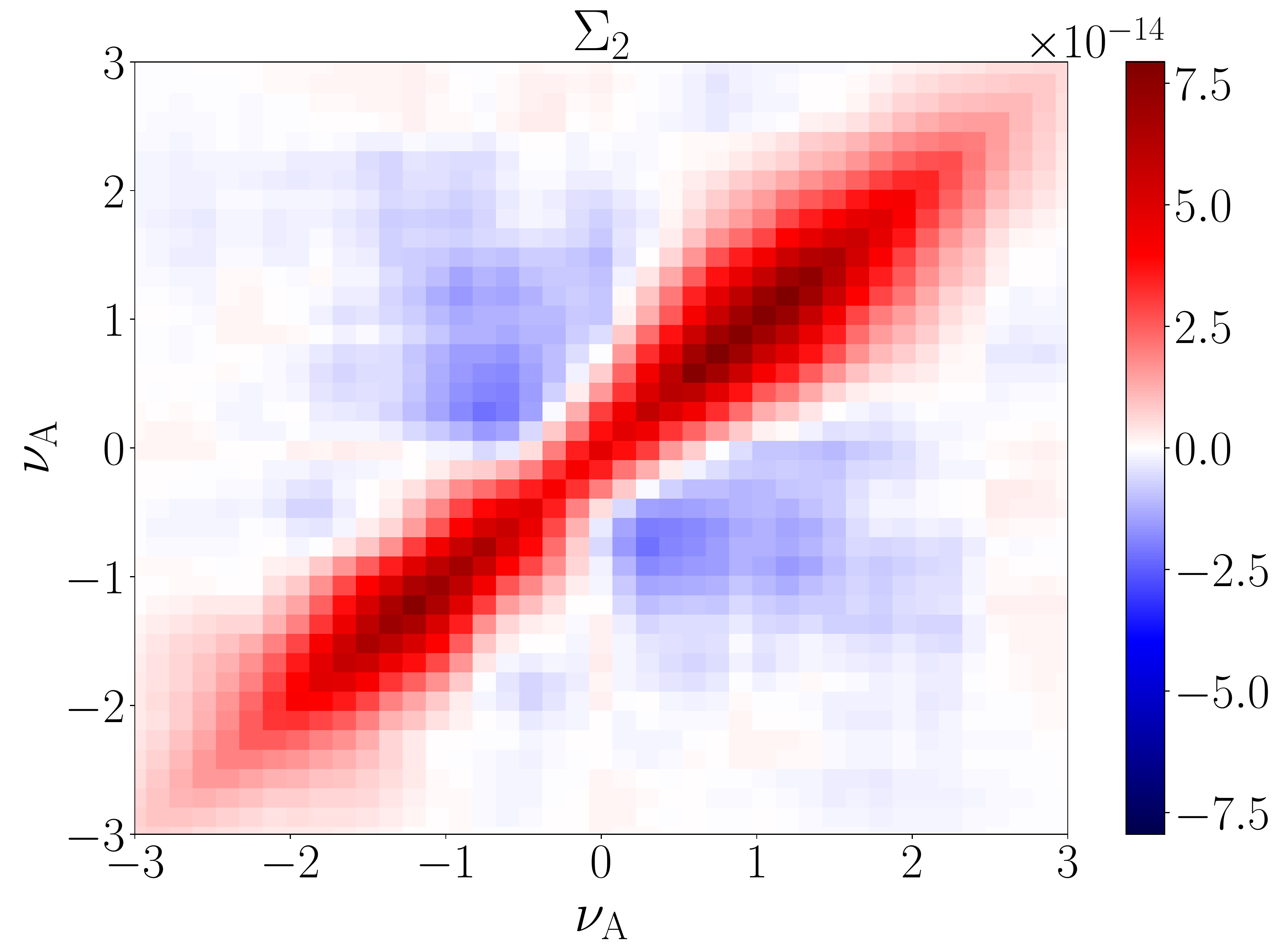} \\
  \includegraphics[width=0.47\textwidth]{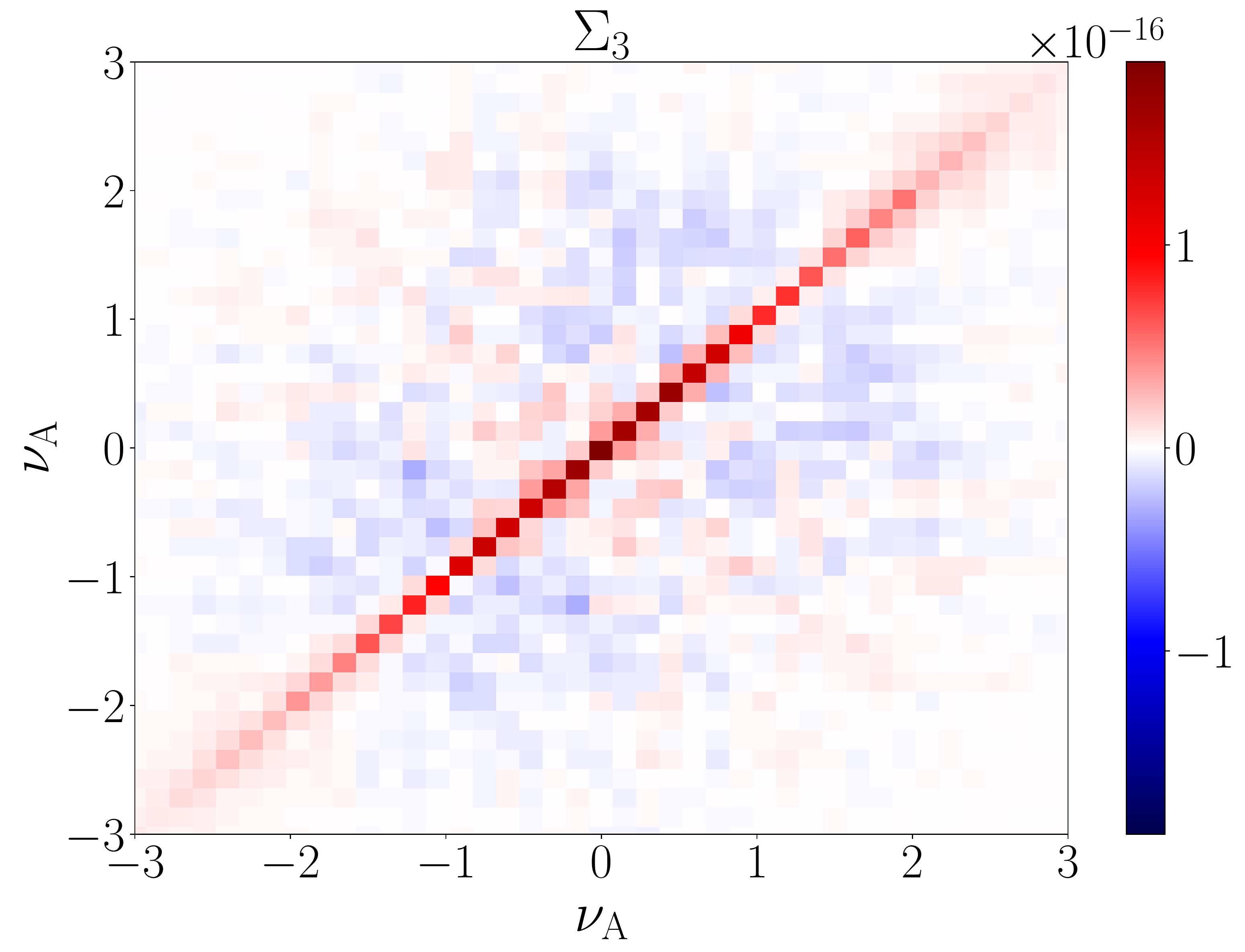} \\
  \caption{The covariance of the $W_{1}, W_{2}, W_{3}$ MFs (top - bottom) as a function of $\nu_{\rm A}$ threshold. There are  large correlations between threshold values for $W_{1}$ (red), and some anti-correlation in $W_{2}$, $W_{3}$ (blue regions). $W_{3}$ (bottom panel) exhibits the lowest level of cross-correlation between threshold values. }
  \label{fig:cov}
\end{figure}

For each mock catalog, we repeat our methodology. We begin by sorting the galaxies into redshift shells, apply a mass cut then project the surviving galaxies into a uniform lattice. We then apply our masking and smoothing procedure and use marching tetrahedra to extract the MFs from the density field $W_{k}^{p}(\nu_{\rm A})$, where $p$ represents the $p^{\rm th}$ mock realisation. As for the actual data, we measure the MFs at $41$ values of $\nu_{\rm A}$ over the range $-3 < \nu_{\rm A} < 3$. The method is repeated separately for mock CMASS/LOWZ north (`N') and south (`S') data. From these measurements, a set of covariance matrices $\Sigma^{m,n}_{k}$ can be generated as 
\begin{equation}\label{eq:cov2D} 
\Sigma^{m,n}_{k} = {1 \over N_{\rm r}-1} \sum_{p=1}^{N_{\rm r}} \left( W_{k, p}^{n} - \langle W_{k}^{n} \rangle \right) \left( W_{k, p}^{m} - \langle W_{k}^{m} \rangle  \right) ,
\end{equation}

\noindent where $W_{k,p}^{n}$ is the value of the $k^{\rm th}$ MF at the $n^{\rm th}$ threshold value $\nu_{{\rm A}, n}$ in the $p^{\rm th}$ mock realisation, and $\langle W_{k}^{n} \rangle$ is its average. Since we use a transformation to write the MFs as a function of $\nu_{\rm A}$, there is no information in the $W_{0}$ volume fraction\footnote{The information in $W_{0}$ has been transferred to the $\mathcal{F}_A:\nu\to\nu_{\rm A}$ mapping, which we do not use here.}, we restrict our analysis to $k=1,2,3$. Hence there are a total of three covariance matrices $\Sigma^{m,n}_{1,2,3}$, for each of our four data sets: LOWZ/CMASS N/S. We present these covariance matrices in Figure~\ref{fig:cov}; these are $W^{k}_{m,n}$ for the mock CMASS N catalogs. The other three sets are very similar and not exhibited. It is clear that the values of the MFs are strongly correlated between different threshold bins, and can also be anti-correlated. The correlation is largest for $k=1$ and lowest for $k=3$. When we utilise the covariance matrices for parameter estimation, we follow \citet{Hartlap:2006kj} and correct their inverses with a factor of $(N_{\rm r} - N_{\rm b} - 2)/(N_{\rm r} - 1)$, where $N_{\rm b}=41$ is the number of $\nu_{\rm A}$ bins. 

\begin{figure*}[tb]
  \centering 
  \includegraphics[width=0.48\textwidth]{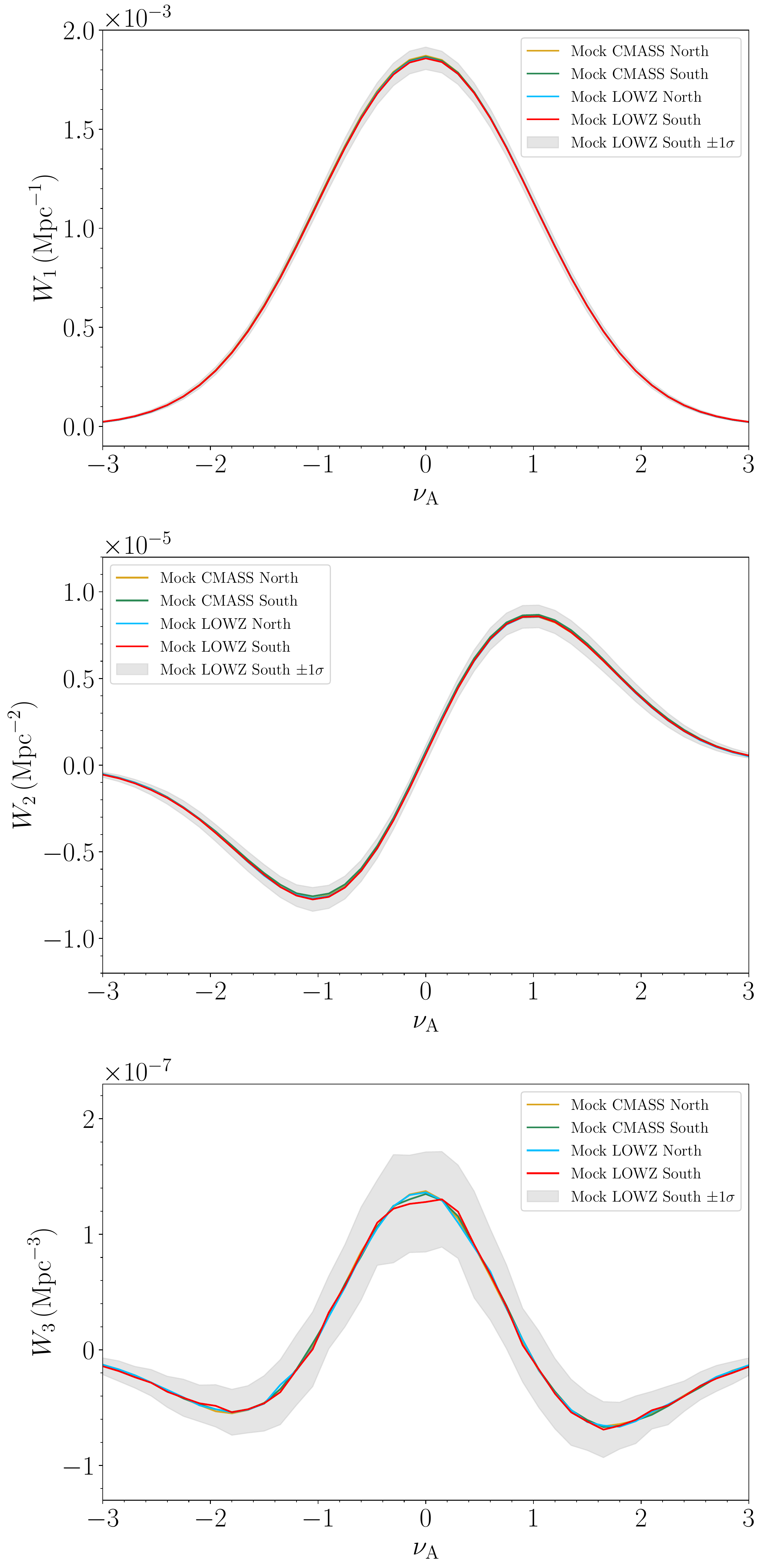}
  \includegraphics[width=0.48\textwidth]{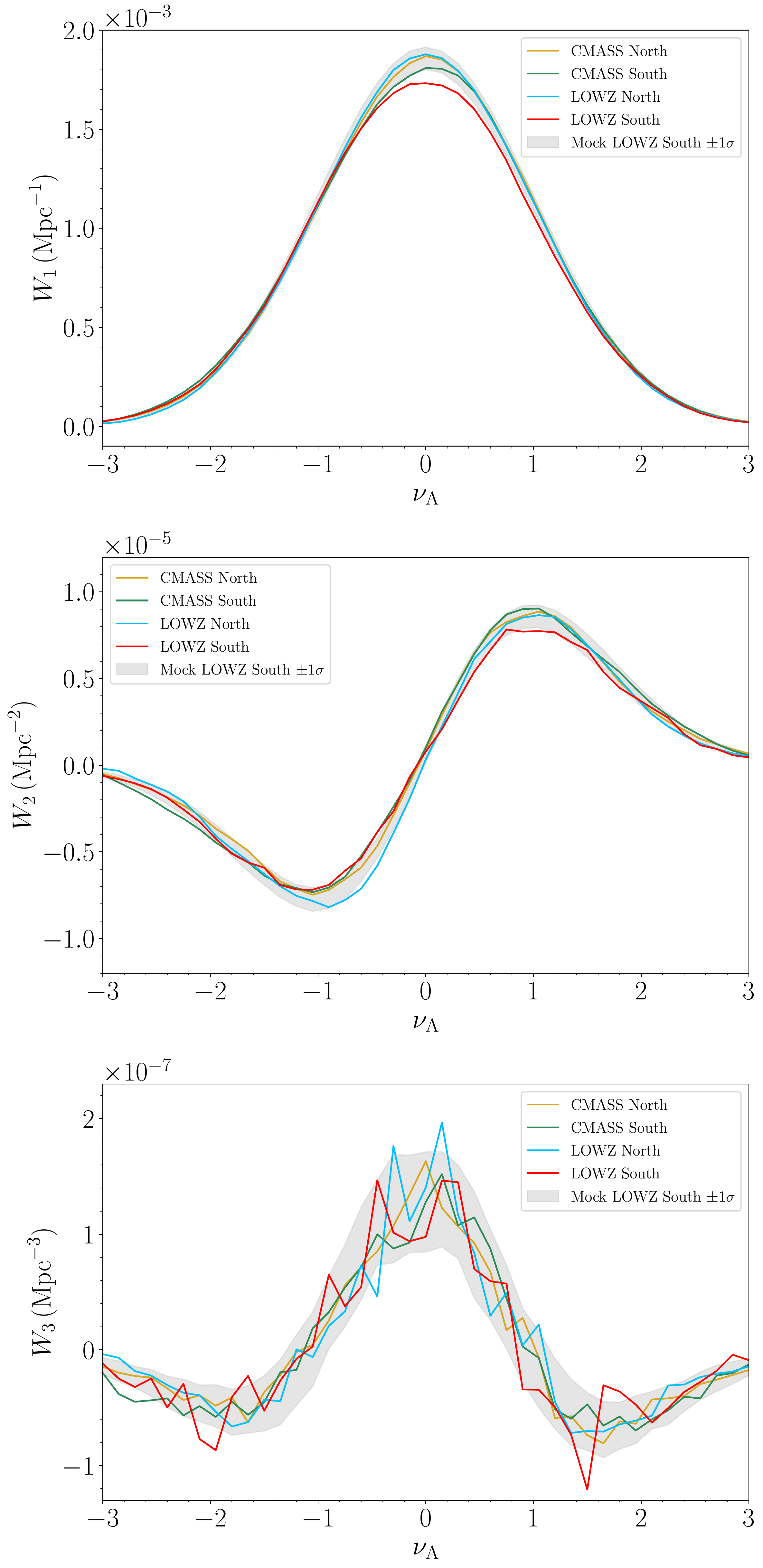}
  \caption{[Left panels] The MFs extracted from $N_{\rm rea} = 250$ Patchy mock catalogs as a function of volume fraction threshold $\nu_{\rm A}$. The grey filled region is the $\pm 1\sigma$ variation of $W_{1,2,3}$ (top/middle/bottom panels) from the LOWZ S realisations. The gold/green/blue/red solid lines are the mean values of the MFs extracted from the mock CMASS N/S and LOWZ N/S realisations, respectively. All datasets exhibit close agreement. [Right panels] The same statistics, extracted from the BOSS data. The colour schemes are the same as in the left panel. The LOWZ S data (red solid line) exhibits lower values than the other three data sets. All data sets have been smoothed with smoothing scale $R_{\rm G} = 35 \, {\rm Mpc}$.}
  \label{fig:mocks_data}
\end{figure*}

In Figure~\ref{fig:mocks_data} (left panels) we present the MFs obtained from the patchy mock data, treating the LOWZ/CMASS N/S data separately. We exhibit them as a function of $\nu_{\rm A}$, and present $W_{1,2,3}$. The solid gold/green/blue/red lines are the mean values extracted from $N_{\rm real} = 250$ mock realisations from the four different data sets, and the solid grey region is the standard deviation of the LOWZ S realisations. The mean values of all statistics are consistent within the statistical error of the measurements, which confirms the insensitivity of our analysis to the data mask. Each of the four subsets of  data have very different survey geometries and volumes.

\section{Results}
\label{sec:results}

We now present the $W_{1,2,3}$ statistics extracted from the BOSS galaxy data. We first present the results of our numerical analysis, and in Appendix~\ref{sec:systematics} test their robustness under variation of the assumptions implicit within our methodology.  

\subsection{Minkowski Functionals of BOSS Data}

\begin{figure}[htb]
    \centering 
    \includegraphics[width=0.45\textwidth]{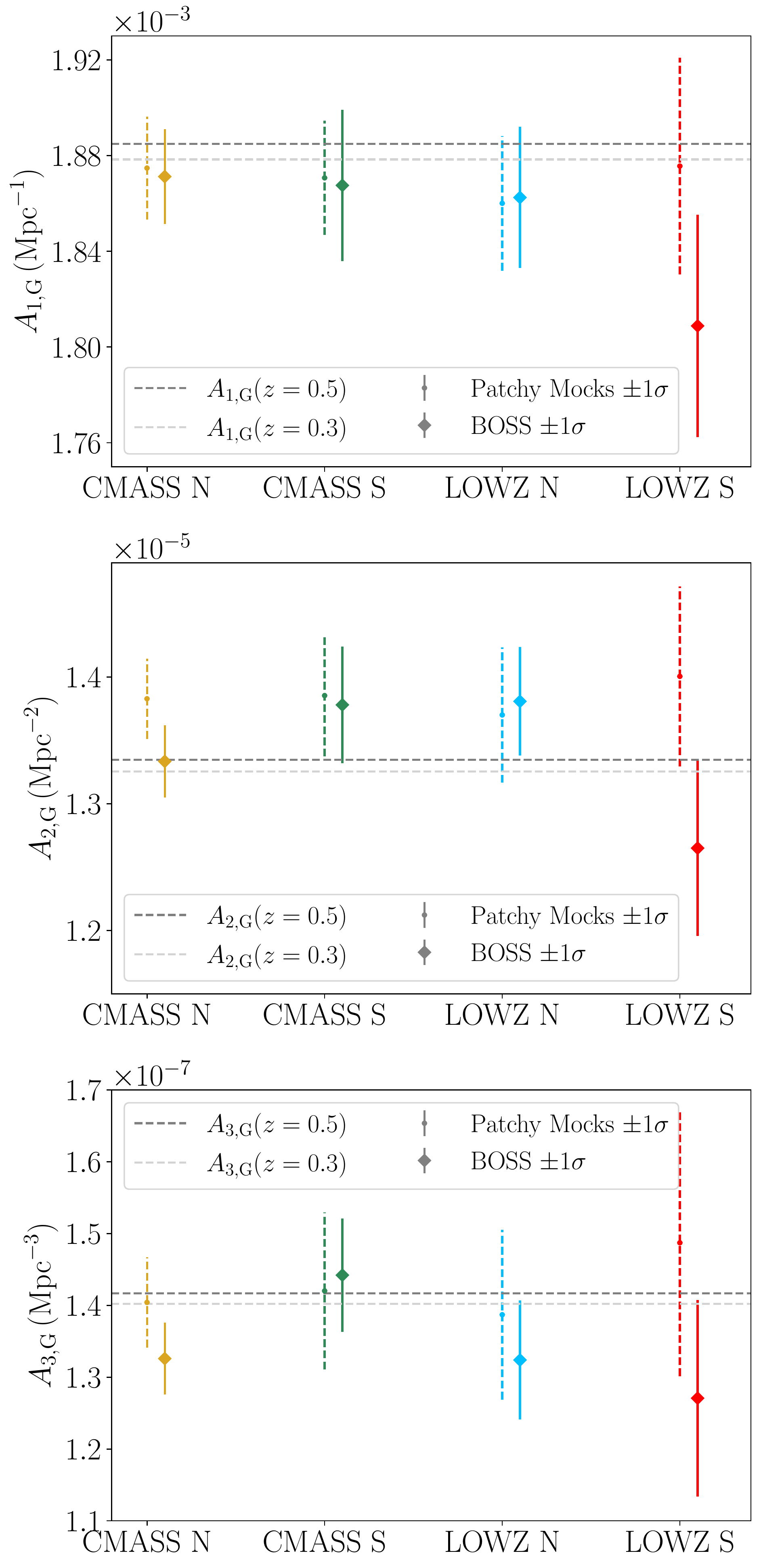}\\
  \caption{The best fit and 1-$\sigma$ uncertainties on the amplitudes $A_{k,{\rm G}}$ of the MFs $W_{1,2,3}$ (top/middle/bottom panels) for the four distinct BOSS data sets CMASS N/S and LOWZ N/S (gold/green/blue/red diamonds and solid error bars). The small points and dashed error bars are the mean and R.M.S. values of $A_{k,{\rm G}}$ extracted from the patchy mock catalogs. The dark/light grey dashed lines are the Gaussian expectation values (eq.~\ref{eq:ampg}) based on the patchy mock cosmological parameters and power spectrum at $z=0.5$ and $0.3$, respectively. }
  \label{fig:amp}
\end{figure}

In Figure~\ref{fig:mocks_data} (right panels), we present the MFs of the four BOSS data sets --- CMASS/LOWZ N/S --- as a function of $\nu_{\rm A}$. We again plot the 1-$\sigma$ standard deviation of the LOWZ south patchy mock realisations as grey filled regions, to provide a visual guide of the statistical errors on the measurements. The LOWZ S data have the largest uncertainties of the four, as it encompasses the smallest volume. The gold/green/blue/red solid lines represent the CMASS N/S and LOWZ N/S data, respectively. The right hand panels show much larger statistical fluctuations than the left because the left panels  present the mean of $N=250$ mock realisations whereas the right constitute a single data realisation. In contrast to the mock data, in the right panels one can observe some discrepancy between the MF curves in the northern and southern skies, with the LOWZ S (red lines) in particular presenting anomalously low values. The difference is most clearly observed in $W_{1}$ but is present in all three panels. The CMASS N and LOWZ N data occupy the largest volumes, and are consistent with the patchy mock realisations (blue/yellow lines). 

From the covariance matrices presented in Figure~\ref{fig:cov}, it is clear that $W_{1,2,3}$ are correlated between $\nu_{\rm A}$ threshold bins. For this reason, care should be taken not to perform statistical analysis `by eye', using Figure~\ref{fig:mocks_data}. To proceed, we fit the following functions to each curve,
\begin{multline} 
\tilde{W}_{k}  = A_{k, {\rm G}} e^{-\nu_{\rm A}^{2}/2} \left\{\vphantom{\frac12} H_{k-1}(\nu_{\rm A})  \right.   \\
\label{eq:fit} +   {k \over 3} a_{0} H_{k}(\nu_{\rm A})  \left.  + {k(k-1) \over 6} a_{2} H_{k-2}(\nu_{\rm A})   \right\} ,
\end{multline}
by minimizing the $\chi^{2}_{k}$ functions 
\begin{equation}\label{eq:chi2}
\chi^{2}_{k} = \sum_{n,m} (W_{k, n} - \tilde{W}_{k,n}) (\Sigma^{-1}){}_{k}^{n,m} (W_{k, m} - \tilde{W}_{k,m}) ,
\end{equation} 
assuming a Gaussian likelihood. In equation~(\ref{eq:chi2}), $W^{k}_{n}$ is the measured value of the $k^{\rm th}$ MF at the $n^{\rm th}$, $\nu_{\rm A}$ threshold, and $\Sigma_{k}^{m,n}$ is defined in equation~(\ref{eq:cov2D}). The parameters varied are $A_{k, {\rm G}}$, $a_{0}$, and $a_{2}$. The quantities $a_{0}$, $a_{2}$ contain information pertaining to the bispectrum, which can be explicitly written as \citep{2003ApJ...584....1M}
\begin{eqnarray} 
a_{0} &=& \left( S^{(1)}-S^{(0)} \right) \sigma_{0} , \\ 
a_{2} &=&  \left(S^{(2)}- S^{(0)} \right)  \sigma_{0} ,
\end{eqnarray} 
where $S^{(0)}$, $S^{(1)}$, $S^{(2)}$ are given in equations~(\ref{eq:sk0}-\ref{eq:sk2}). We do not use $a_{0}$ and $a_{2}$ for cosmological parameter estimation in this work, but each MF $W_{1,2,3}$ should measure consistent $a_{0}$, $a_{2}$ values when extracted from the same data set. This provides a consistency check of our methodology, assuming that equation~(\ref{eq:fit}) is a viable fitting function. Furthermore, if the galaxy distribution is isotropic then we should expect that the north and south data in each catalog will yield consistent $A_{k, {\rm G}}$ and $a_{0,2}$ values. However, CMASS and LOWZ will not necessarily yield consistent results, as they constitute two distinct galaxy samples with different selection criteria and redshifts. 

We minimize the function (eq.~\ref{eq:chi2}) for each of the four data sets and each $W_{k}$ function separately, to obtain a set of twelve measurements of $A_{k,G}$, $a_{0}$, and $a_{2}$. The prior ranges used are $-1 < a_{0,2} < 1$ and $-18.5 < \log[A_{k, G}] < -10$, and variation of these limits does not affect our results. When fitting equation~(\ref{eq:fit}) to $W_{3}$, we multiply the functions extracted from the data (c.f. Figure~\ref{fig:mocks_data}, bottom panels) by $-1$, as the convention in cosmology is to present $W_{3}$ in terms of $-H_{2}$.

\begin{figure}[htb]
    \centering 
    \includegraphics[width=0.45\textwidth]{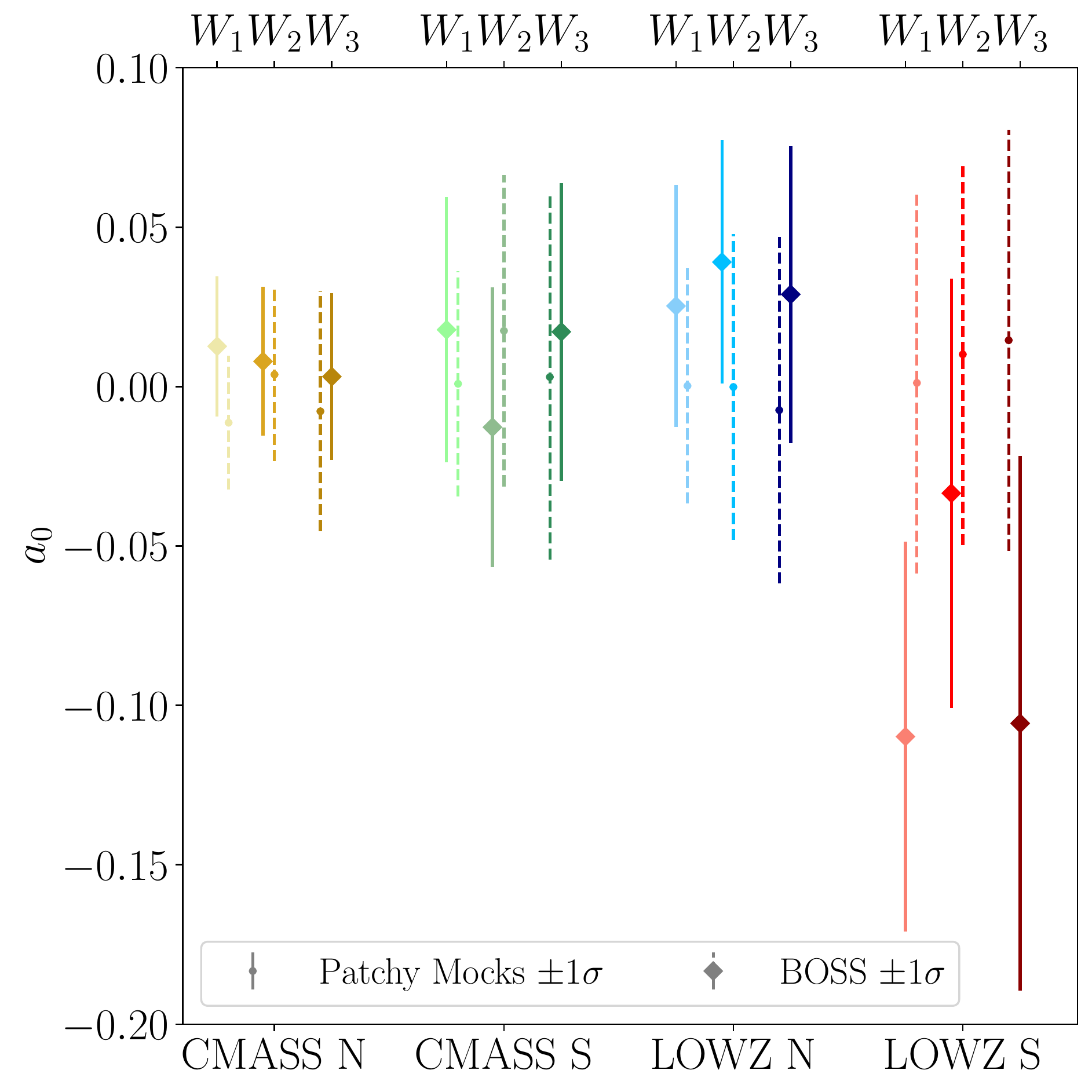}\\
    \includegraphics[width=0.45\textwidth]{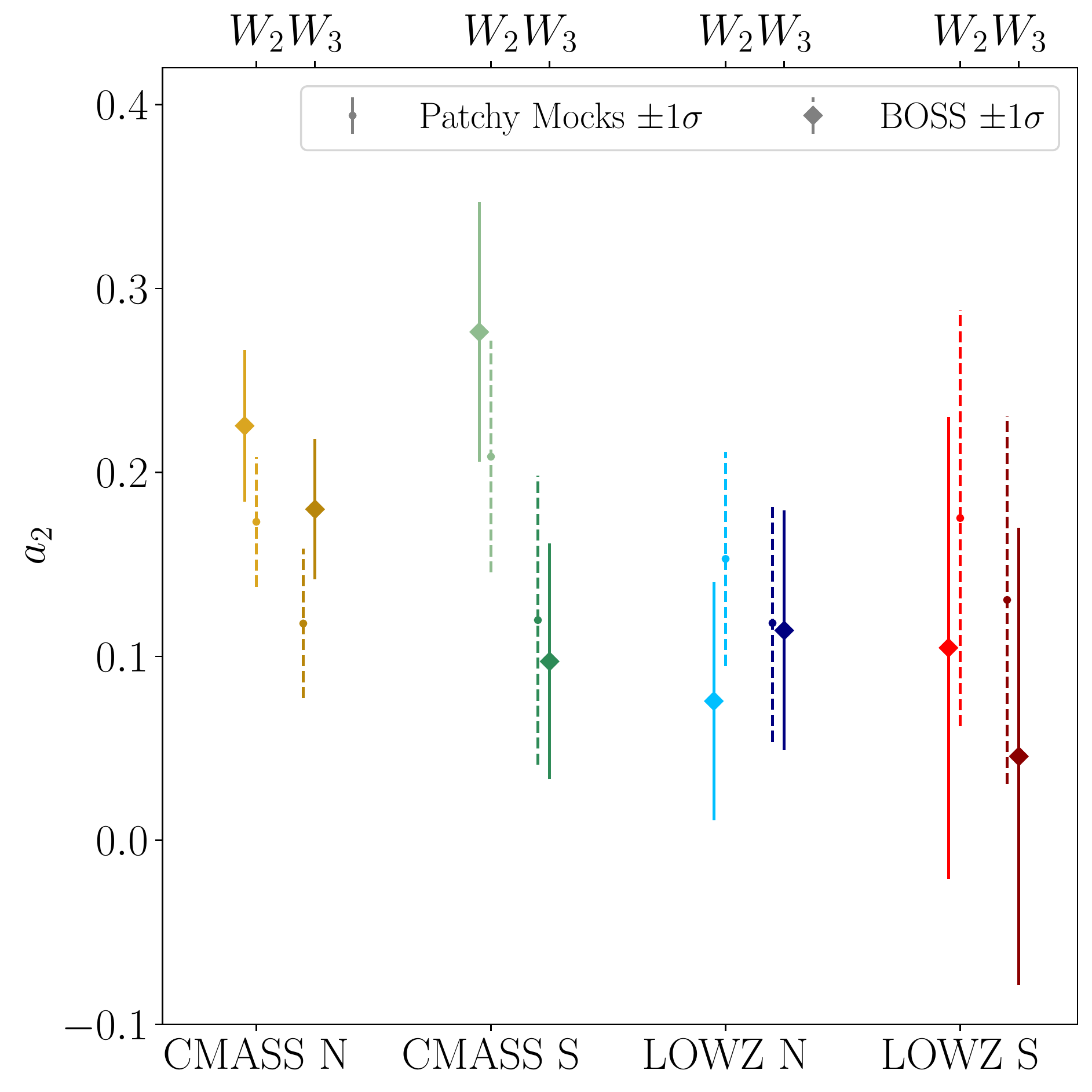}
  \caption{The best fit and 1-$\sigma$ uncertainties on $a_{0}$ (top panel) and $a_{2}$ (bottom panel) extracted from the BOSS data (diamonds and solid error bars), and the mean and R.M.S. value of these quantities from the patchy mocks. Gold/green/blue/red points correspond to CMASS N/S and LOWZ N/S, respectively. The light-to-dark range of colours represent the values of $a_{0}$, $a_{2}$ extracted from $W_{1}$, $W_{2}$, $W_{3}$, respectively, from the same data set.  }
  \label{fig:s0_s2}
\end{figure}

In Figures~\ref{fig:amp} and \ref{fig:s0_s2}, we present the best fit and marginalised 1-$\sigma$ limits of the parameters $A_{k,G}$ and  $a_{0,2}$, respectively, for each of the four data sets. The gold/green/blue/red diamonds and solid error bars are the best fit and $1$-$\sigma$ uncertainties obtained by minimizing the $\chi^{2}_{k}$ function (eq.~\ref{eq:chi2}) for CMASS N/S and LOWZ N/S data, respectively. In Figure~\ref{fig:s0_s2}, the light-to-dark points/error bars are the values of $a_{0}$ and $a_{2}$ obtained from each MF $W_{1,2,3}$ within the same subset of data. Note that $W_{1}$ is independent of $a_{2}$ due to the $(k-1)$ factor in the expansion (eq.~\ref{eq:fit}), so it is not included in the lower panel of Figure~\ref{fig:s0_s2}.

For comparison, the small points and dashed error bars in the figures are the mean and R.M.S. values of $A_{k, {\rm G}}$ and $a_{0,2}$ obtained from the patchy mock catalogs, obtained using the expressions 
\begin{eqnarray}
\label{eq:int1}  A_{k, {\rm G}} &\simeq& {1 \over \sqrt{2\pi} (k-1)!} \int_{-4}^{4}  W_{k}(\nu_{\rm A}) H_{k-1}(\nu_{\rm A}) \, d\nu_{\rm A} , \\ 
 \label{eq:int2} a_{0} &\simeq& {3 \over k \sqrt{2\pi} k! A_{k, {\rm G}}}\int_{-4}^{4}  W_{k}(\nu_{\rm A}) H_{k}(\nu_{\rm A}) \, d\nu_{\rm A} , \\
 \label{eq:int3} a_{2}|_{k>1} &\simeq& {6 \over \sqrt{2\pi} k! A_{k, {\rm G}}}\int_{-4}^{4}  W_{k}(\nu_{\rm A}) H_{k-2}(\nu_{\rm A}) \, d\nu_{\rm A} ,
\end{eqnarray}
where $W_{k}$ are the MFs extracted from the mocks. Finally, the dark/light grey horizontal dashed lines in Figure~\ref{fig:amp} are the Gaussian expectation values (eq.~\ref{eq:ampg}) for the MF amplitudes, assuming cosmological parameters in Table~\ref{tab:1} and $\bar{n} = 6.25 \times 10^{-5} \, {\rm Mpc}^{-3}$, and the linear galaxy bias $b=2$. The power spectrum adopted is $P(k,z) = b^{2} P_{\rm m}(k,z) + 1/\bar{n}$, where $P_{\rm m}(k,z)$ is the linear matter power spectrum at redshift $z=0.3$ and $0.5$ (light/dark dashed lines). We correct equation~(\ref{eq:ampg}) from real- to redshift-space by applying a constant factor $A^{\rm rsd}_{k,{\rm G}} = \alpha_{k} A_{k,{\rm G}}$, where $\alpha_{k} = 0.99, 0.98, 0.97$ for $k=1,2,3$, respectively. This correction factor is derived from mock catalogs and is discussed further in Appendix~\ref{sec:rsd}. 

The patchy mock results (small points and dashed error bars) are entirely self-consistent, in the sense that the four data sets yield values of $A_{k, {\rm G}}$ and $a_{0,2}$ that are in agreement within 1-$\sigma$. Furthermore, $W_{1,2,3}$ as measured within each data set yield consistent values of $a_{0,2}$, which serves as a check that the three-point cumulants are being correctly measured. The amplitudes $A_{k, {\rm G}}$ are in close agreement with the Gaussian expectation values (eq.~\ref{eq:ampg}), except the systematically high values of $A_{2, {\rm G}}$ from the patchy mocks (middle panel in Figure~\ref{fig:amp}). We can provide no compelling explanation for this discrepancy, other than our estimator for $W_{2}$ might be marginally biased by the presence of the mask. Although the mean values are practically consistent with the Gaussian prediction at 1-$\sigma$, the reconstructed values are systematically high. 

\begin{table*}[tb]
\begin{center}
 \begin{tabular}{|| c  c  c  c  c ||}
 \hline
 Data/MF \, & $\Omega_{\rm c}h^{2} \, n_{\rm s}$ \, & $a_{0}$ \, & $a_{2}$ & $\chi^{2}_{\rm r}$ \\ [0.5ex] 
 \hline\hline
 CMASS & $0.110\pm 0.006$ & $0.006\pm 0.012$ & $0.183 \pm 0.021$ &  $1.32$    \\
 LOWZ & $0.111\pm 0.008$ & $0.007\pm 0.020$ & $0.082 \pm 0.036$  &  $1.07$  \\
 \hline 
 CMASS N & $0.104\pm 0.008$ & $0.008\pm 0.014$ & $0.191\pm 0.023$  & $1.35$  \\
 CMASS S & $0.119\pm 0.010$ & $0.002\pm 0.025$ & $0.158 \pm 0.040$ & $1.27$   \\
 LOWZ N & $0.117\pm 0.009$ & $0.032\pm 0.024$ & $0.098 \pm 0.042$  & $1.10$  \\
 LOWZ S & $0.095\pm 0.012$ & $-0.080\pm 0.040$ & $0.066 \pm 0.073$ & $0.95$   \\
 \hline 
 CMASS N $W_{1}$ & $0.109\pm 0.010$ & $0.012\pm 0.022$ & $-$ &  $1.25$  \\
 CMASS S $W_{1}$ & $0.109\pm 0.015$ & $0.008\pm 0.040$ & $-$ &  $1.20$  \\
 LOWZ N $W_{1}$ & $0.109\pm 0.014$ & $0.025\pm 0.037$ & $-$ &  $1.35$ \\
 LOWZ S $W_{1}$ & $0.091\pm 0.017$ & $-0.116\pm 0.063$ & $-$ & $0.93$  \\
 \hline
 CMASS N $W_{2}$ & $0.116\pm 0.010$ & $0.008\pm 0.024$ & $0.224\pm 0.042$. &  $1.31$  \\
 CMASS S $W_{2}$ & $0.129\pm 0.017$ & $-0.007\pm 0.042$ & $0.284 \pm 0.066$ & $1.02$  \\
 LOWZ N $W_{2}$ & $0.133\pm 0.016$ & $0.038\pm 0.039$ & $0.070 \pm 0.065$ & $0.99$   \\
 LOWZ S $W_{2}$ & $0.100\pm 0.020$ & $-0.025\pm 0.067$ & $0.123 \pm 0.121$ & $1.25$  \\
 \hline
 CMASS N $W_{3}$ & $0.099\pm 0.010$ & $0.001\pm 0.026$ & $0.177 \pm 0.037$ & $1.42$  \\
 CMASS S $W_{3}$ & $0.122\pm 0.016$ & $0.028\pm 0.046$ & $0.113 \pm 0.062$ & $1.18$  \\
 LOWZ N $W_{3}$ & $0.103\pm 0.016$ & $0.030\pm 0.046$ & $0.114 \pm 0.064$ &  $0.86$ \\
 LOWZ S $W_{3}$ & $0.098\pm 0.023$ & $-0.102\pm 0.079$ & $0.043 \pm 0.116$ & $0.68$  \\
 \hline 
 Planck & $0.116 \pm 0.001$  & - & - & - \\
 \hline 
\end{tabular}
\caption{\label{tab:cos}Marginalised best fit and 1-$\sigma$ uncertainties on the parameters fit by minimizing the $\chi^{2}$ function (eq.~\ref{eq:chi2}) to the data sets. CMASS and LOWZ correspond to an overall fit of the functions (eq.~\ref{eq:fitcos}) to the north and south data combined $\chi^{2} = \chi_{1}^{2} + \chi_{2}^{2} + \chi_{3}^{2}$. CMASS N/S and LOWZ N/S are fits of equation~(\ref{eq:fitcos}) separately to each of the four data sets. The following twelve rows are fitting (eq.~\ref{eq:fitcos}) separately to each data set and each MF. The final row is the Planck best fit and 1-$\sigma$ uncertainty on the parameter combination $\Omega_{\rm c}h^{2}\, n_{\rm s}$, assuming $\Lambda$CDM. } 
\end{center} 
\end{table*}


The BOSS data results (diamond points, solid error bars) present some peculiarities. The amplitudes $A_{k, {\rm G}}$ of $W_{1,2,3}$ extracted from the LOWZ S data are systematically lower than the other three data sets. The statistical significance of this discrepancy is low, due to the large smoothing scales adopted in this work. The bispectrum term $a_{0}$ is also large and negative in LOWZ S, which suggests that the discrepancy in the data is not restricted to the two point cumulants. In addition, the data reconstruction of $a_{2}$ presents a mild discrepancy in the CMASS S data (c.f. Fig.~\ref{fig:s0_s2}, bottom panel). The reconstructed value of $a_{2}$ obtained from $W_{2}$ is high compared to the same quantity extracted from $W_{3}$. Some weak systematic offset is also observed in the mock reconstruction, which could again indicate some effect of the mask on the $W_{2}$ estimation. Extracting cosmological information from the bispectrum terms $a_{0}$ and $a_{2}$ will be considered in future work, and here we simply report the anomalous behaviour of the southern sky data. The marginalised best fit and 1-$\sigma$ uncertainties on the parameters $a_{0,2}$ for each of the measurements in Figure~\ref{fig:s0_s2} are presented in Table~\ref{tab:cos}.

So far, we have proceeded under the assumption that the perturbative expansion (eq.~\ref{eq:nuaexp}) can be applied to the data, and we truncated the expansion at order ${\cal O}(\sigma_{0})$. At order $\sigma_{0}^{2}$, multiple new terms are introduced that are related to the four-point cumulants $\sim \langle \delta^{4} \rangle/\sigma_{0}^{4}$, and the amplitude $A_{k,{\rm G}}$ also receives a correction \citep{Matsubara:2020knr}. Although we do not pursue higher order terms in this work, it is instructive to introduce a single additional Hermite polynomial coefficient to the fitting procedure, and check if it does not significantly alter our conclusions. We end this section by fitting the following functions to the data 
\begin{eqnarray} 
\label{eq:modfit1} \tilde{W}_{1} &=& A_{1, {\rm G}} e^{-\nu_{\rm A}^{2}/2}\Big(H_{0} + {a_{0} \over 3} H_{1} + h_{2} H_{2} \Big) ,\\
\label{eq:modfit2} \tilde{W}_{2} &=& A_{2, {\rm G}} e^{-\nu_{\rm A}^{2}/2}\Big(H_{1} + {2a_{0} \over 3} H_{2} + {a_{2} \over 3}H_{0}  + h_{3} H_{3} \Big) ,\\ 
\label{eq:modfit3} \tilde{W}_{3} &=& A_{3, {\rm G}} e^{-\nu_{\rm A}^{2}/2}\Big( H_{2} + a_{0} H_{3} + a_{2} H_{1}  + h_{0} H_{0} \Big) ,
\end{eqnarray}
where $h_{0,2,3}$ are additional free parameters. We select these terms as they correspond to coefficients of the lowest order Hermite polynomials introduced at order $\sigma_{0}^{2}$ for each MF. Additional, higher order polynomials should also be included, but they require increasing information from the large $|\nu_{\rm A}|$ tails to accurately measure. 

\begin{figure}[htb]
    \centering 
    \includegraphics[width=0.4\textwidth]{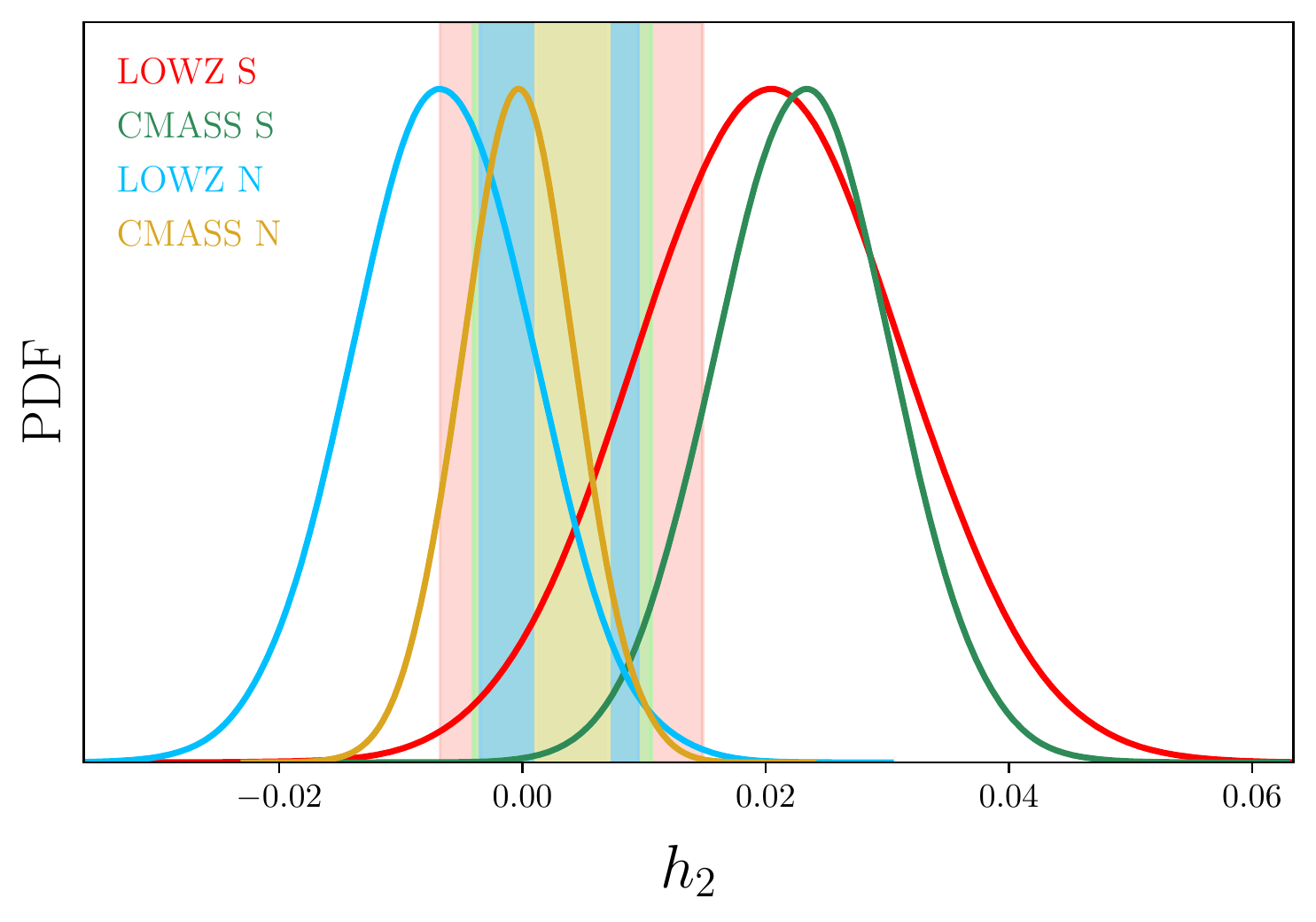}\\
    \includegraphics[width=0.4\textwidth]{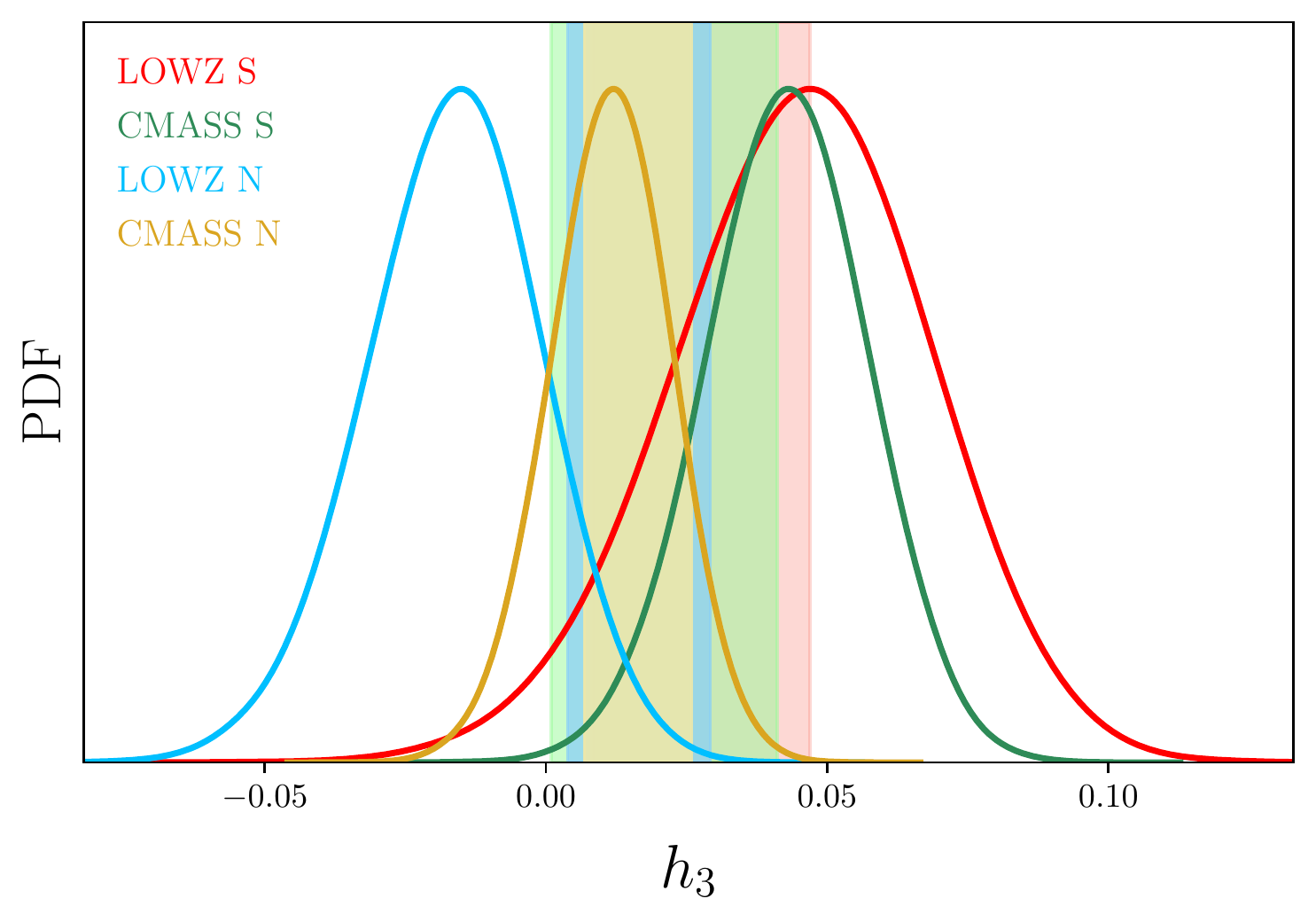}\\
    \includegraphics[width=0.4\textwidth]{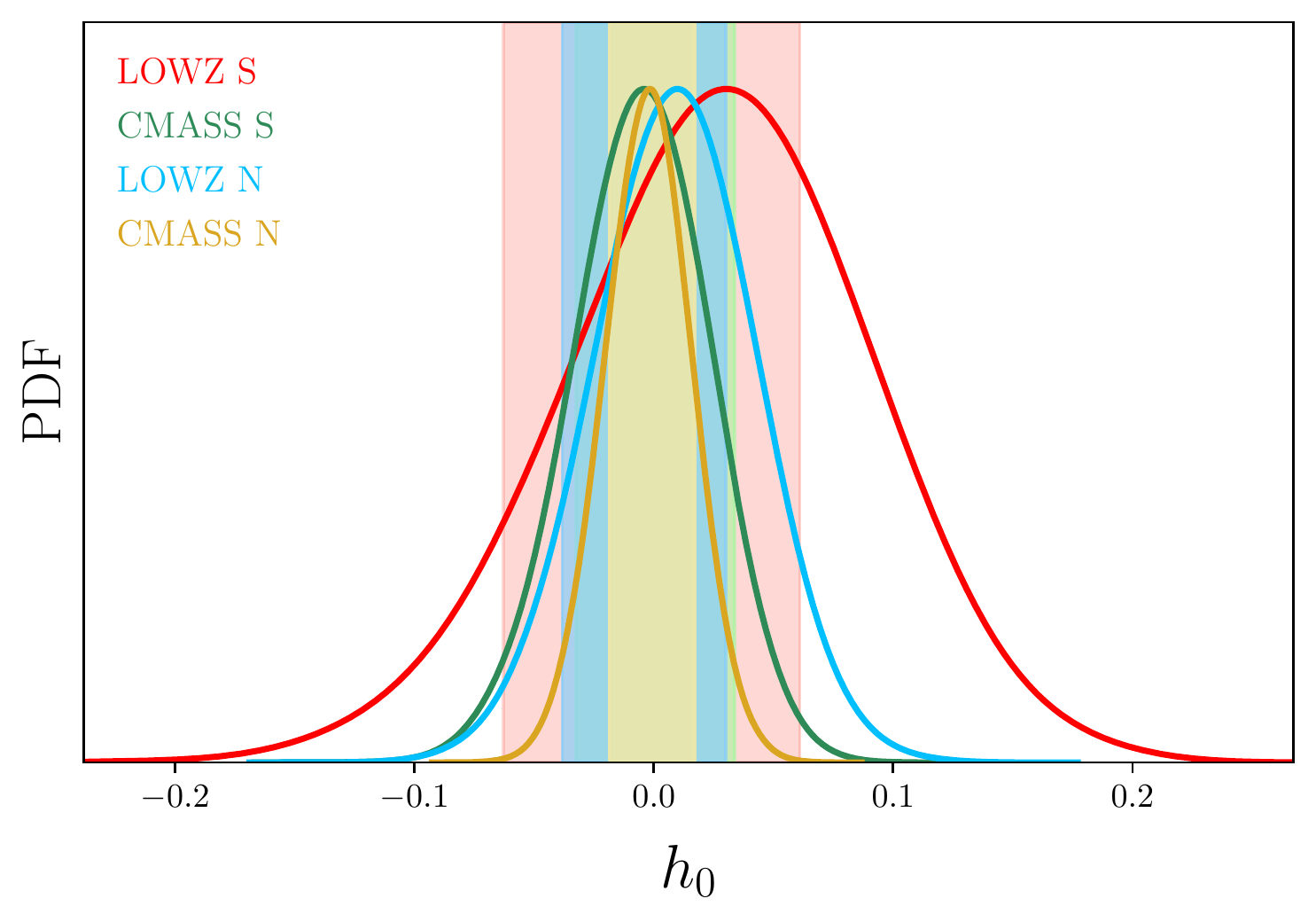}
  \caption{One-dimensional marginalised probability distribution functions for the parameters $h_{2}$, $h_{3}$ and $h_{0}$ (top - bottom panels) for the CMASS N/S and LOWZ N/S data (gold/green/blue/red solid lines, respectively). The vertical matching colour bars are the R.M.S. values of these parameters extracted from the patchy mock data. }
  \label{fig:h}
\end{figure}

We again minimize the $\chi^{2}_{k}$ functions (eq.~\ref{eq:chi2}), but with the additional parameters free to vary over the range $-1 < h_{0,2,3} < 1$. In Figure~\ref{fig:h}, we present the marginalised one-dimensional probability distribution functions for $h_{2}$, $h_{3}$, $h_{0}$ (top/middle/bottom panels). The vertical filled bars are the $1\sigma$ ranges of these parameters obtained directly from the patchy mocks, using 
\begin{equation} 
h_{m} \simeq {1 \over \sqrt{2\pi} m! A_{k, {\rm G}}} \int_{-4}^{4} H_{m}(\nu_{\rm A}) W_{k}(\nu_{\rm A}) \, d\nu_{\rm A} . 
\end{equation} 

There is slight evidence that the northern and southern sky data exhibits a dichotomy in $h_{2}$ [top panel], although again, the statistical significance is low. The introduction of $h_{0,2,3}$ does not move the best fit values of the other parameters $A_{k, G}$, $a_{0,2}$ outside their 1-$\sigma$ ranges, indicating that our results are stable under the addition of the higher point cumulants. The parameter values $A_{k,{\rm G}}$, $a_{0,2}$ with and without the $h_{0,2,3}$ terms are provided in Table~\ref{tab:4} of Appendix~\ref{sec:systematics}.

\subsection{Cosmological Parameter Estimation from the Minkowski Functional Amplitudes}
\label{sec:cos}

Finally, we repeat our $\chi^{2}_{k}$ minimization procedure of the previous section, but now fit the function 
\begin{multline} \label{eq:fitcos}
\tilde{W}_{k}  = {\alpha_{k} \over (2\pi)^{(k+1)/2}} {\omega_{3} \over \omega_{3-k}\omega_{k}} \left({\sigma_{1}^{2} \over 3\sigma_{0}^{2}} \right)^{k/2} e^{-\nu_{\rm A}^{2}/2} \left\{\vphantom{\frac12} H_{k-1}(\nu_{\rm A})  \right.   \\
+  {k \over 3} a_{0} H_{k}(\nu_{\rm A})  \left.  + {k(k-1) \over 6} a_{2} H_{k-2}(\nu_{\rm A})   \right\} ,
\end{multline}
to the MF curves extracted from the BOSS data. This is the same function as equation~(\ref{eq:fit}), but now we fit a cosmological model to the amplitudes rather than treating $A_{k,{\rm G}}$ as arbitrary constants. Cosmology enters via the ratio of two-point cumulants $\sigma_{1}$ and $\sigma_{0}$, which is given by  
\begin{equation}\label{eq:sigs} 
\frac{\sigma_{1}^{2}}{\sigma_{0}^{2}} = \frac{\int  k^{2} P_{g}(z,k) W^{2}(kR_{\rm G}) \, d^{3}k}{\int  P_{g}(z,k) W^{2}(kR_{\rm G}) \, d^{3}k } .
\end{equation} 
We approximate the galaxy power spectrum in real space as $P_{g}(z,k) = b^{2} P_{\rm m}(z,k)+ 1/\bar{n}$, where $b$ is the linear galaxy bias, $P_{\rm m}$ is the underlying linear matter power spectrum, and $\bar{n}$ is the number density of the galaxy sample being utilised. We fix $b=2$ based on the mock catalogs, but our results will be practically insensitive to variation of this parameter. This is a valid assumption provided we restrict our analysis to scales at which shot noise is negligible compared to the signal. We fix $z_{\rm LOWZ} = 0.3$ and $z_{\rm CMASS} = 0.5$ in the power spectrum $P_{g}(z,k)$, but these values will not affect our conclusions. We fix the baryon fraction and $\sigma_{8}$ to their Planck values $\Omega_{\rm b} h^{2} = 0.0224$, $\sigma_{8} = 0.8288$, as our statistics are only very weakly sensitive to these parameters. 

The quantities in equations~(\ref{eq:fit}, \ref{eq:sigs}) have been defined in real space, but the measured MFs are in redshift space. To account for this discrepancy, we correct the measured MF curves by a constant factor $\alpha_{k} = [0.99, 0.98, 0.97]$ for $W_{1}$, $W_{2}$, and $W_{3}$, respectively. These correction factors were obtained by measuring the MF statistics in a mock galaxy snapshot box in real and redshift space, and calculating the ratio of their amplitudes. Because the redshift space correction is so small, we do not expect any model dependence in this effect to be significant. This point is discussed further in Appendix~\ref{sec:rsd}. 

\begin{table}[tb]
\begin{center}
 \begin{tabular}{||c  c ||}
 \hline
 Parameter \, & Range \\ [0.5ex] 
 \hline\hline
 $\Omega_{\rm c}h^{2}$ & $[0.05, 0.30]$   \\
 $n_{\rm s}$ & $[0.6, 1.2]$ \\
 $a_{0}$  & $[-1, 1]$ \\ 
 $a_{2}$ & $[-1, 1]$ \\ 
 \hline 
\end{tabular}
\caption{\label{tab:3}Prior parameter ranges used in Section~\ref{sec:cos}.}
\end{center} 
\end{table}

In total, for each MF curve in each data set, we vary four parameters $\Omega_{\rm c}h^{2}$, $n_{\rm s}$, $a_{0}$, and $a_{2}$ over prior ranges given in Table~\ref{tab:3}. We again perform the $\chi^{2}_{k}$ minimization for each of the four data sets separately, and each MF separately (for a total of twelve sets of parameter constraints). We also combine the information from $W_{1}, W_{2}, W_{3}$ for each data set, by summing their $\chi_{k}^{2}$ values, to obtain four distinct measurements labelled CMASS N, CMASS S, LOWZ N, and LOWZ S. Finally, we combine north and south results, by simply summing their chi-squared values, to obtain overall CMASS and LOWZ results.

\begin{figure}[htb]
    \centering 
    \includegraphics[width=0.45\textwidth]{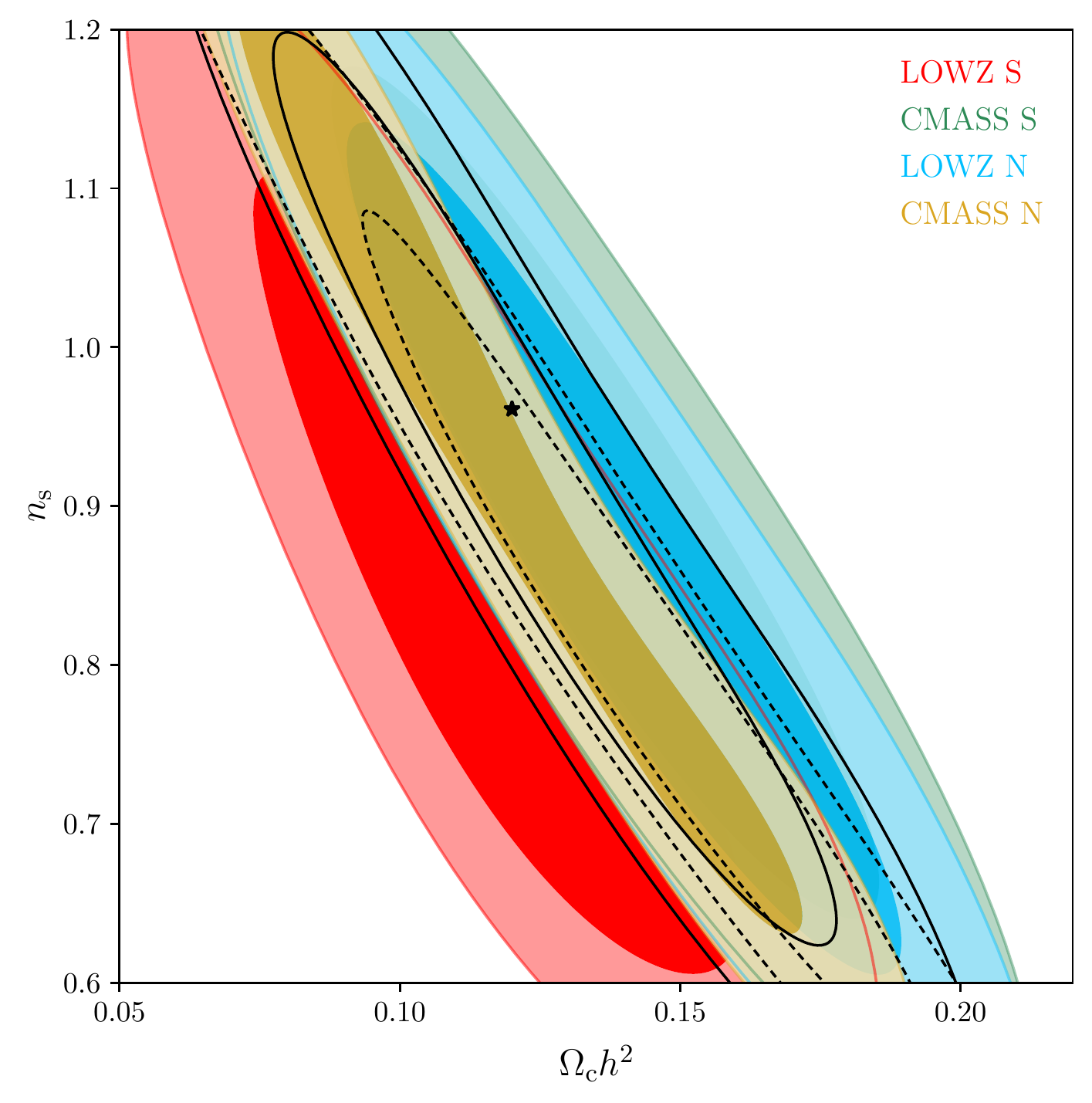}\\
    \includegraphics[width=0.45\textwidth]{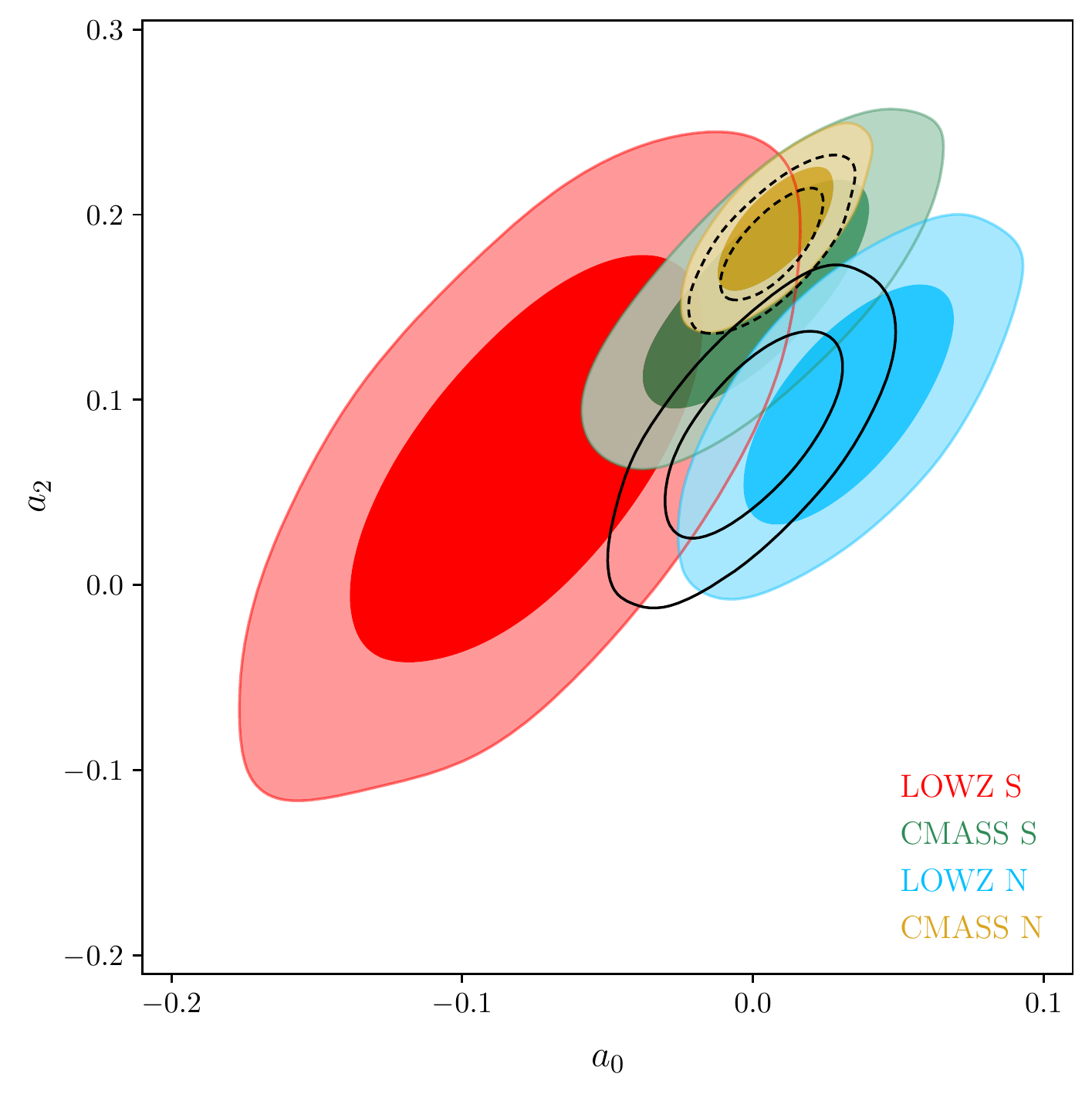}\\
  \caption{Marginalised two-dimensional contours in the $n_{s}$-$\Omega_{\rm c}h^{2}$ and $a_{2}$-$a_{0}$ planes. The gold/green/blue/red contours are the $68$-$95\%$ limits obtained from the CMASS N/S and LOWZ N/S data, respectively. The unfilled black dashed/solid lines are the contours obtained by combining north and south data into two catalogs CMASS and LOWZ, respectively. The black star in the top panel is the Planck best fit of these parameters, assuming $\Lambda$CDM.}
  \label{fig:cos}
\end{figure}

In Figure~\ref{fig:cos}, we present the two-dimensional contours for the parameters $\Omega_{\rm c}h^{2}$, $n_{\rm s}$, $a_{0}$, and $a_{2}$ for CMASS N/S and LOWZ N/S, (gold, green, blue, red filled contours), and the combined CMASS and LOWZ results (dashed/solid black empty contours). The $a_{0,2}$ parameters are orthogonal to $\Omega_{\rm c}h^{2}$ and $n_{\rm s}$, due to the orthogonal nature of the Hermite polynomials. Therefore, we do not present $a_{0}$,$a_{2}$-$\Omega_{c}h^{2}$,$n_{s}$ contours, as they are not informative.

There exists a strong degeneracy between $\Omega_{\rm c}h^{2}$ and $n_{\rm s}$; this was also observed in two dimensional slices of the BOSS data in \citet{Appleby:2020pem}. Since we cannot simultaneously constrain these two parameters, we rotate the parameter plane and obtain an effective one-dimensional constraint on the $\Omega_{\rm c}h^{2}\, n_{\rm s}$ combination\footnote{This is a different parameter combination to the two-dimensional results of \citet{Appleby:2020pem}; this is due to the different parameter sensitivity in the two- and three-dimensional statistics.}. The LOWZ S data presents a lower value of both $a_{0}$ and $\Omega_{\rm c}h^{2} \, n_{\rm s}$, but the uncertainties are large due to this data set occupying the smallest volume. In Table~\ref{tab:cos}, we present the one-dimensional marginalised best fit and 1-$\sigma$ uncertainties on $\Omega_{\rm c}h^{2} \, n_{s}$, $a_{0}$, and $a_{2}$ for each data set, and the corresponding reduced chi-squared values. We also present the Planck best fit of the combination of parameters $\Omega_{\rm c}h^{2} \, n_{\rm s}$ \citep{Aghanim:2018eyx}. Note that the combined CMASS N/S and LOWZ N/S data are consistent with the Planck cosmology, despite the LOWZ S data being systematically low. This is because the north data simply possesses more constraining power and is closer to the Planck cosmology. The difference in $a_{0}$ between LOWZ N and S ($a_{0} = 0.032 \pm 0.024$ and $-0.080\pm 0.040$) is the most significant discrepancy observed in this work, and brings into question the suitability of the expansion (eq.~\ref{eq:fit}) for the low-redshift galaxy data. The cosmological parameters inferred from LOWZ N and S are practically consistent ($\Omega_{\rm c}h^{2} \, n_{\rm s} = 0.117\pm 0.009$ and $0.095\pm 0.012$).

\section{Discussion} 
\label{sec:disc}

The Minkowski functionals provide a complementary approach to extracting information from cosmological data sets. In this work we have measured the MFs from the SDSS-III DR12 BOSS galaxy data. To do so, we binned the point distribution onto a uniform lattice and Gaussian smoothed the discrete field with comoving scale $R_{G} = 35 \, {\rm Mpc}$. At these large scales, the perturbative non-Gaussian expansion (eq.~$\ref{eq:mat1}$) can be used in principle. 

After validating our analysis with Gaussian random fields, and mock galaxy snapshot data, we measured the amplitude and shape of the MF curves obtained from the BOSS data. The resulting analysis yielded some quirks. Specifically, the LOWZ south data possesses systematically low Minkowski functional amplitudes compared to LOWZ north, and also low values of the shape parameter $a_{0}$, which contains information from the three point cumulants. The exact values can be found in Table~\ref{tab:cos}. The significance of these discrepancies is low, because we must smooth over relatively large scales to reconcile late Universe measurements of large-scale structure with the perturbative non-Gaussian expansion typically used in cosmology \citep{Matsubara:2020knr}. However, the presence of such anomalies could indicate either some unknown systematics in the data, or some physical anisotropy in the low redshift large-scale structure. The higher redshift CMASS data does not present any discrepancy between northern and southern sky data. Similarly, the patchy mock data is remarkably consistent between CMASS/LOWZ north/south subsamples, which suggests that the problem does not lie with our analysis pipeline. The LOWZ data considered in this work lies at cosmological distances $d_{\rm c} \sim 850 - 1600 \, {\rm Mpc}$ relative to the observer, scales at which we expect the data to be isotropic and homogeneous within the standard cosmological model. Measurements of the MFs from earlier large-scale structure catalogs support our findings \citep{Kerscher:1997xp,Kerscher:1998gs,Kerscher:2001gm}; the north and south sky data consistently present different morphological properties. 

If we overlook the north/south discrepancy and simply combine the data sets into two overall catalogs (CMASS and LOWZ), we find that the amplitude of the MFs are consistent with the Planck $\Lambda$CDM best fit. This is primarily due to the northern sky data simply occupying a larger volume, and the curious southern sky results are mitigated. Also, we are only using the two-point information contained within the amplitudes, and the two point function is not sensitive to the structure of the cosmic web.

The BOSS data has been exhaustively studied in the literature \citep{BOSS:2016apd,Ivanov:2019pdj,Zhai:2016gyu,Slepian:2015hca,Manera:2014cpa,Hamaus:2020cbu}. A direct comparison to our work and two-point correlation function and power spectrum analyses is difficult, because different parameters are varied, sampling choices are made and typically north and south data are not separately analysed. However, the literature consensus is that the BOSS data is consistent with the best fit Planck $\Lambda$CDM cosmology, when the two-point information is extracted\footnote{With the caveat that the Minkowski functional amplitudes are not sensitive to $\sigma_{8}$, only the shape of the power spectrum.}. We are in agreement with this conclusion. A recent, comprehensive analysis of the BOSS power spectrum in \citet{Ivanov:2019pdj} generated a constraint of $\Omega_{\rm c}h^{2} = 0.1127 \pm 0.0046$. This is consistent (both the best fit and approximately the statistical uncertainty) with our measurements if we fix $n_{s}=0.965 \pm 0.004$ using a Planck prior. Similarly, our results are consistent with a previous analysis of the two-dimensional genus extracted from shells of the BOSS data in \citet{Appleby:2020pem}. We provide a comparison of these results in Table \ref{tab:disc}.

\begin{table}[tb]
\begin{center}
 \begin{tabular}{||c  c ||}
 \hline
 Measurement \, & $\Omega_{\rm c}h^{2}$ \\ [0.5ex] 
 \hline\hline
 CMASS (This work) &  $0.114 \pm 0.005$  \\
 LOWZ (This work) &  $0.114 \pm 0.007$  \\
 \citet{Ivanov:2019pdj} & $0.113 \pm 0.005$   \\
 CMASS (\citet{Appleby:2020pem}) &  $0.121 \pm 0.006$ \\
 LOWZ (\citet{Appleby:2020pem}) & $0.116 \pm 0.008$ \\ 
 \hline 
\end{tabular}
\label{tab:disc}
\caption{A review of the constraints on $\Omega_{\rm c}h^{2}$ in this work (after applying a Planck prior on $n_{s}$), and corresponding measurements of the same parameter in other works in the literature.}
\end{center} 
\end{table}

Regarding the north/south comparison, \citet{Tojeiro:2014eea} noted some tension between north and south sky data in previous versions of the BOSS data, but the difference was not deemed significant. In this work, the discrepancy between the north and south sky presents predominantly in the non-Gaussian, higher point cumulants to which the Minkowski functionals are sensitive. The effect is modest, and future galaxy catalogs will provide more information on the non-Gaussian nature of the late-time gravitational field. In \cite{Sullivan:2017mhr}, the Minkowski functionals of the BOSS data were extracted using a germ-grain method, and the non-Gaussian properties were rigorously studied. The authors of \cite{Sullivan:2017mhr} concluded that there is no strong statistical evidence of any north/south discrepancy in the DR12 data. In Figure 6 and Table 3 of that work the LOWZ data presents a small offset between the north and south sky, which is most evident in the $W_{1}$ MF. The significance is low -- the quoted $p$-value associated with the hypothesis that the difference is consistent with random fluctuations is $p \simeq 0.02$, which is approximately in agreement with our $\gtrsim 2\sigma$ discrepancy in $a_{0}$ from the LOWZ data. In contrast, the CMASS data is fully consistent. The result is suggestive rather than conclusive -- given that other large scale structure data sets have presented north/south discrepancies \citep{Kerscher:1997xp,Kerscher:1998gs,Kerscher:2001gm}, it would be interesting to study the low redshift galaxy density field in more detail. 

There is increasing discussion in the literature on the potential existence of a dipole in various data sets \citep{Colin:2019opb,Mohayaee:2020wxf,Secrest:2020has,Luongo:2021nqh}, beyond the kinematic dipole observed in the CMB \citep{Planck:2013kqc}. A related observational framework to measure multipoles in low-redshift data can be found in \citet{Heinesen_2021}. The BOSS data is not a magnitude limited sample, and its complex selection criteria and incomplete sky coverage make it difficult to relate our findings to other claims in the literature. A study of the topology of all-sky density fields is an interesting direction of future study. 

At fixed comoving smoothing scales $R_{\rm G} = 35 \, {\rm Mpc}$, the majority of information contained in the late Universe density field is washed out. Also, the MFs themselves are `summary statistics' and do not contain all topological information. To proceed further, we should unmoor ourselves from the model-dependent non-Gaussian perturbative expansion in $\sigma_{0}$ cumulants (see footnote 2), and also consider the more complex class of topological statistics that can be applied to a point distribution. For point processes, a direct MF analysis using the decoration of galaxies with Boolean grains without constructing a density field,
hence without extra smoothing, provides an alternative strategy \cite{Mecke:1994ax, Kerscher:1997xp, Kerscher:1998gs, Kerscher:2001gm, Wiegand:2013xfa}.
This methodology naturally contains boundary corrections according to the Gaussian Kinematic Formula and is model-independent by construction. Such an analysis is currently being pursued by the authors to further determine the properties of the observed large-scale structure.  

\section*{Acknowledgements}
SAA is supported by an appointment to the JRG Program at the APCTP through the Science and Technology Promotion Fund and Lottery Fund of the Korean Government, and was also supported by the Korean Local Governments in Gyeongsangbuk-do Province and Pohang City. This work is also part of a project that has received funding from the European Research Council (ERC) under the European Union’s Horizon 2020 research and innovation programme (grant agreement ERC adG No. 740021–ARThUs, PI: TB).
SEH was supported by the project \begin{CJK}{UTF8}{mj}우주거대구조를 이용한 암흑우주 연구\end{CJK} 
(``Understanding Dark Universe Using Large Scale Structure of the Universe''), funded by the Ministry of Science.
HSH was supported by the New Faculty Startup Fund from Seoul National University.

The authors would like to acknowledge the support of the Korea Institute for Advanced Study (KIAS) grant funded by the government of Korea. Computing resources were supplied by the KIAS Center for Advanced
Computation Linux Cluster System. 

Funding for SDSS-III has been provided by the Alfred
P. Sloan Foundation, the Participating Institutions,
the National Science Foundation, and the U.S. Department
of Energy Office of Science. The SDSS-III web
site is http://www.sdss3.org/. SDSS-III is managed by
the Astrophysical Research Consortium for the Participating
Institutions of the SDSS-III Collaboration including
the University of Arizona, the Brazilian Participation
Group, Brookhaven National Laboratory, Carnegie Mellon
University, University of Florida, the French Participation
Group, the German Participation Group, Harvard
University, the Instituto de Astrofisica de Canarias, the
Michigan State/Notre Dame/JINA Participation Group,
Johns Hopkins University, Lawrence Berkeley National
Laboratory, Max Planck Institute for Astrophysics, Max
Planck Institute for Extraterrestrial Physics, New Mexico
State University, New York University, Ohio State
University, Pennsylvania State University, University of
Portsmouth, Princeton University, the Spanish Participation
Group, University of Tokyo, University of Utah,
Vanderbilt University, University of Virginia, University
of Washington, and Yale University.

The massive production of all MultiDark-Patchy mocks for the BOSS Final Data Release has been performed at the BSC Marenostrum supercomputer, the Hydra cluster at the Instituto de Fısica Teorica UAM/CSIC, and NERSC at the Lawrence Berkeley National Laboratory. We acknowledge support from the Spanish MICINNs Consolider-Ingenio 2010 Programme under grant MultiDark CSD2009-00064, MINECO Centro de Excelencia Severo Ochoa Programme under grant SEV- 2012-0249, and grant AYA2014-60641-C2-1-P. The MultiDark-Patchy mocks was an effort led from the IFT UAM-CSIC by F. Prada’s group (C.-H. Chuang, S. Rodriguez-Torres and C. Scoccola) in collaboration with C. Zhao (Tsinghua U.), F.-S. Kitaura (AIP), A. Klypin (NMSU), G. Yepes (UAM), and the BOSS galaxy clustering working group.

Some of the results in this paper have been derived using the healpy and HEALPix package

\bibliography{biblio}{}

\appendix

\section{Geometry of random fields on manifolds}
\label{app:MF}

Going under various names and orderings in different settings, such as Minkowski functionals, curvature integrals,  intrinsic volumes,  and   \LKCs, there are $(D+1)$ quantifiers associated with the geometry of a $D$-dimensional manifold $\Mspace$. Taking $V_D$ to be the $D$-dimensional Lebesgue measure, which quantifies the $D$-dimensional volume, and for convex $\Mspace$, there exist a set of numbers $W_0, \ldots, W_D$ known as \emph{Minkowski functionals}, which are associated with the volume of the \emph{tube} of radius $\epsilon$ around $\Mspace$, where $\epsilon$ is small, through the \emph{tube formula}:

\begin{equation}
\label{geometry:steiner:formula}
V_D (Tube (\Mspace, \epsilon)) = \sum_{j=0}^D  \frac{\epsilon^{j}}{j!}  W_j(\Mspace).
\end{equation}

Restricting to 3D, $W_0$ measures the volume, $W_1$ measures the surface area, and $W_2$ is associated with contour length and measures the caliper diameter of $\Mspace$. $W_3$, or equivalently $W_D$ of a $D$-dimensional manifold, is associated with a purely topological quantity called the \emph{Euler characteristic.}

The MFs of the excursion sets of {\it stochastic} fields on manifolds are defined in the usual sense of the tube formula in equation ($\ref{geometry:steiner:formula}$), with the exception that the Lebesgue measure is replaced by the probability measure, such that all measures of size are weighted with respect to probability content, giving the probabilistic version of the tube formula

\beqn
\label{rob:eq:GMdef}
\text{Pr}\left\{X\in \left\{x\in\mathbb R^d: \min_{y\in E_\nu} \|y-x\| \right\} \leq \epsilon \right\}
\ = \ \sum_{j=0}^{\infty} 
\frac{\epsilon^j}{j!} \Min^d_j(E_\nu).
\eeqn

The above equation is a Taylor series expansion, in which the coefficients $\Min^d_j$ are known as the \emph{Gaussian Minkowski functionals}; they play the role of the usual MFs and encode the geometric properties of the manifold induced by the random field $u$.

Restricting to cubical Euclidean grids, the MFs of the excursion sets are given via the \emph{Gaussian Kinematic formula}:

\begin{equation}
\label{rgeometry:EQ:equation} 
\langle W_i \left(E_\nu\right)\rangle  
=e^{-\nu^2/2}
\sum_{j=0}^{i} \sqbinom{D+j-i}{j}  \sqbinom{i}{j}
\frac{ \omega_j j! \lambda^{j}}{(2\pi)^{(j+1)/2}} 
 H_{j-1}\left(\nu\right)     W_{i-j}(\Mspace) .
\end{equation}

\noindent In the above equation, $\lambda$ is proportional to the second spectral moment of the power spectrum, or equivalently, proportional to the second order gradient of the correlation function. The  combinatorial `flag coefficients' are defined by
\beqn
\label{geometry:square-binom:def}
\ssqbinom{n}{j} = \binom{n}{j} \frac{\omega_n}{\omega_{n-j} \; \omega_j}\,,
\eeqn
where $H_n$ is the $n$-th Hermite polynomial, defined for $n\geq 0$, 
\beqqn
H_n(x) =  n!\, \sum_{j=0}^{\lfloor n/2\rfloor} 
\frac{(-1)^j x^{n-2j}}{j!\, (n-2j)!\, 2^j},
\eeqqn
\noindent while , for $n=-1$, we fix 
\beqn
H_{-1}(x)   \ = \ \sqrt{2\pi} e^{x^2/2}\Psi(x),
\eeqn
\noindent where
\beqn
\Psi(x) \ = \   \frac{1}{\sqrt{2\pi}} \int_u^\infty e^{-x^2/2}\,dx ,
\eeqn
\noindent is the Gaussian tail probability. 

The Gaussian Kinematic Formula describes the MFs for a field defined on a generic manifold, in the presence of boundaries or otherwise. It is equal to the curvature integrals ($\ref{eq:ci1} - \ref{eq:ci4}$) only when the field is boundary-less. Many works in the cosmological literature (including this one) actually extract the curvature integrals from data sets \cite{Appleby:2020pem,Schmalzing:1997uc}, rather than the Minkowski functionals \cite{Pranav:2021ted}.

\section{Unbiased Estimators of the Curvature Integrals}
\label{sec:unbias}

For large-scale structure catalogs, one must account for radial and angular selection functions, masks and complex survey geometries. Given that the ensemble averages quoted in Section~\ref{sec:def} apply only to unbounded fields, we must carefully construct unbiased estimators for these statistics when the data is masked. In this section, we review our numerical algorithm and then test our method by applying it to Gaussian random fields and mock galaxy snapshot data.

\subsection{Numerical Reconstruction of Minkowski Functionals} 
\label{app:num}

\begin{figure}[htb]
  \centering 
  \includegraphics[width=0.31\textwidth]{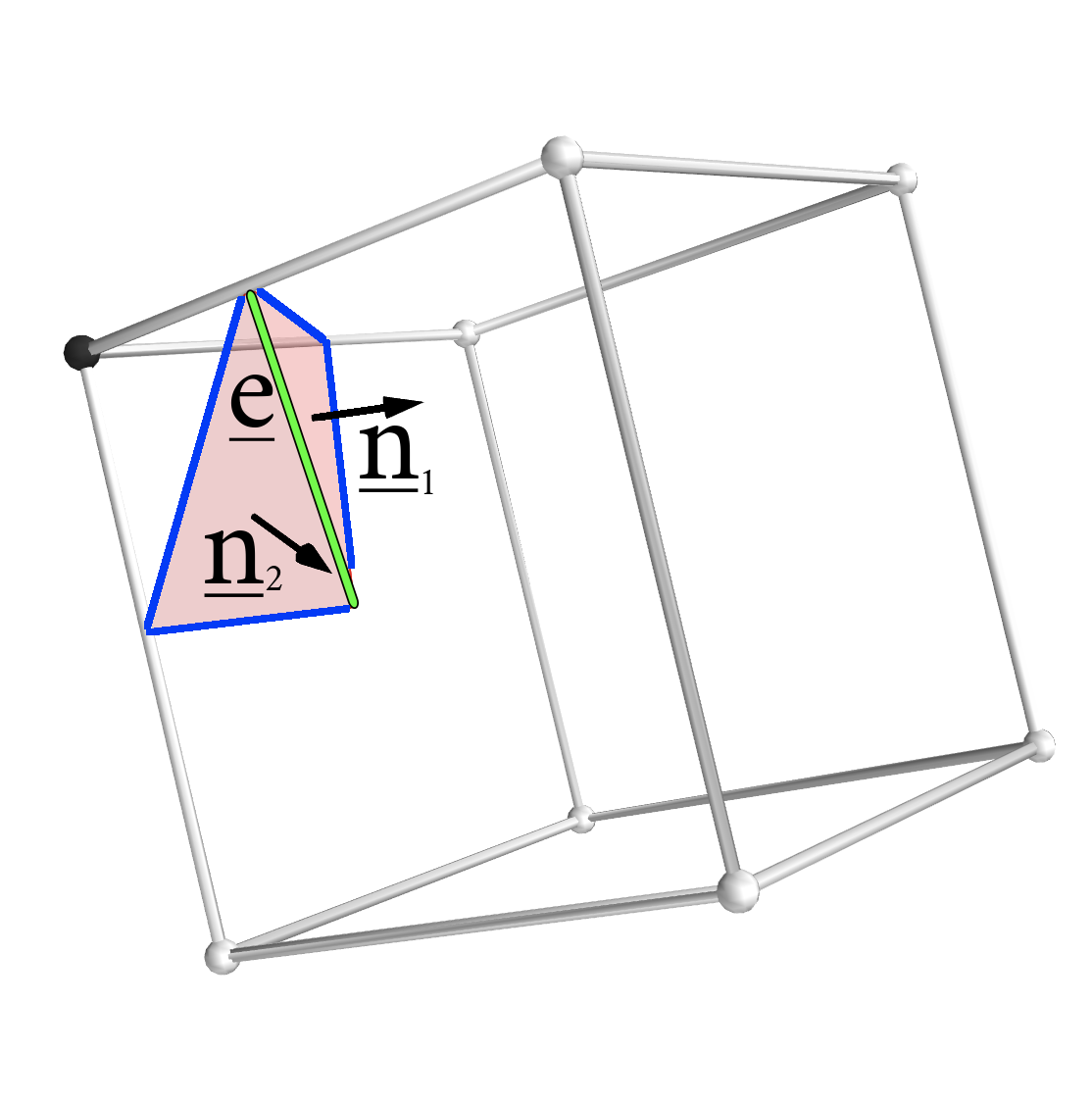} 
  \includegraphics[width=0.31\textwidth]{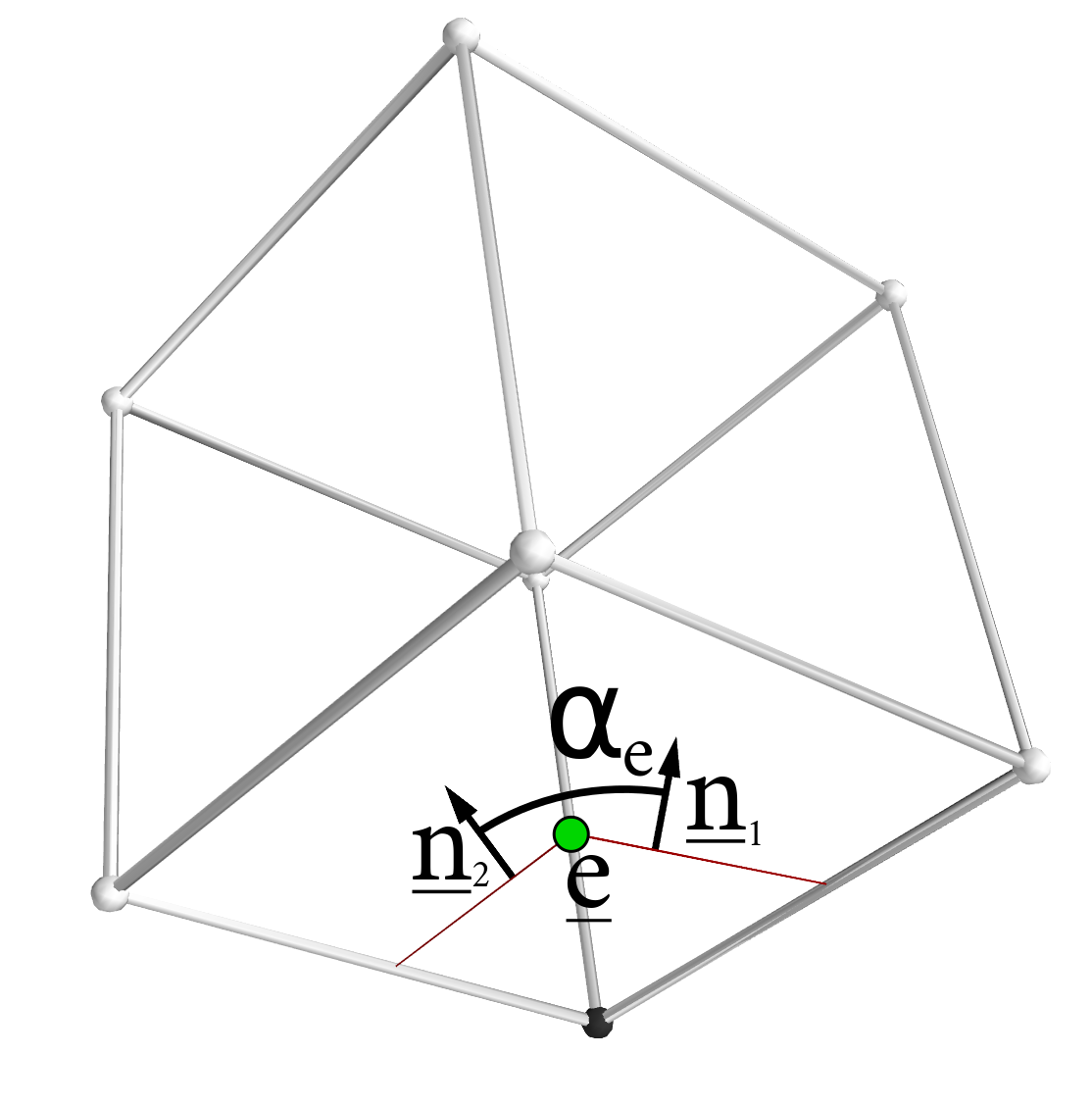} 
  \includegraphics[width=0.31\textwidth]{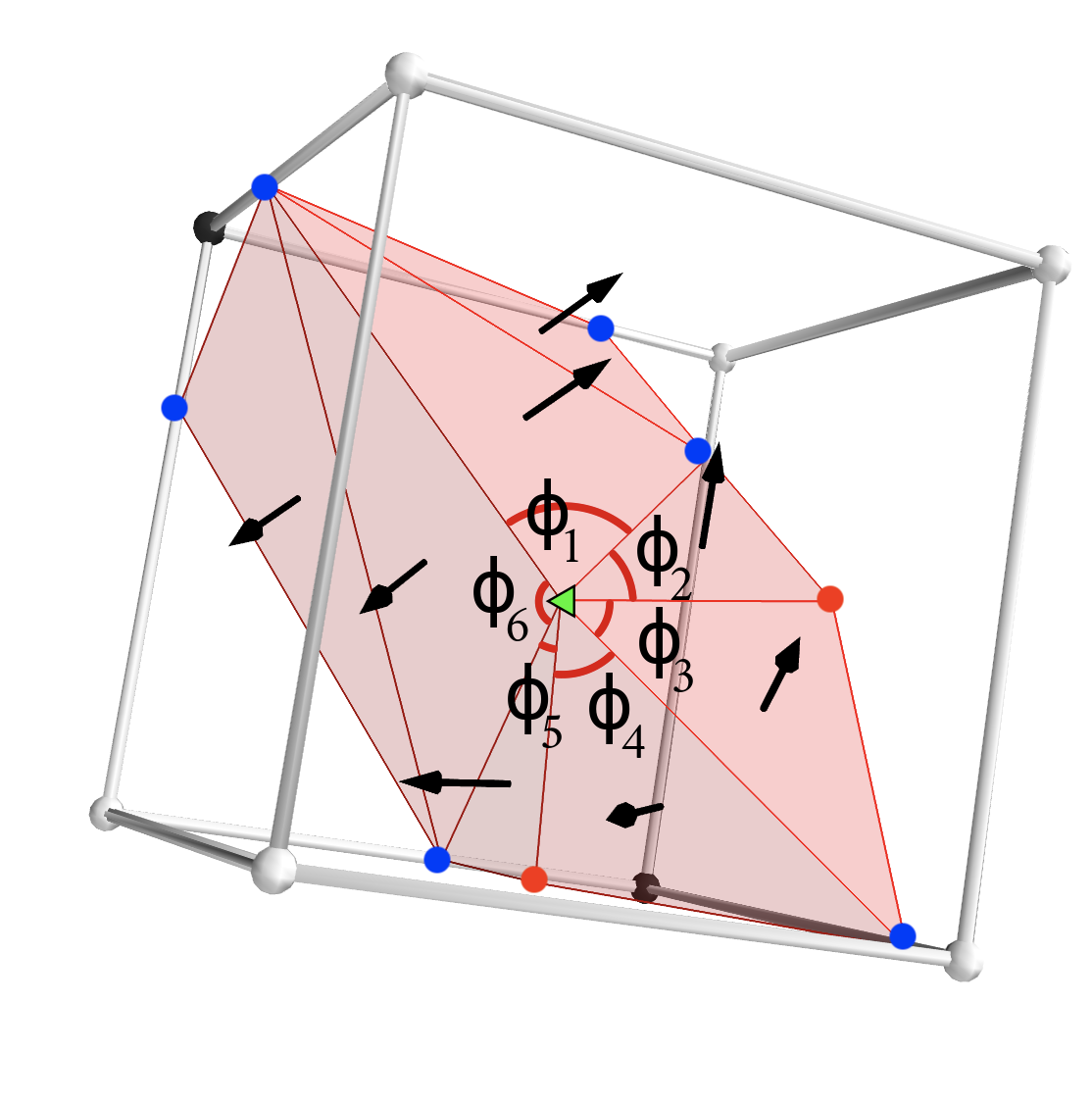}
  \caption{Two examples of pixel boxes that highlight our numerical algorithm. The black/white points are pixels in the discrete field lattice $\delta_{ijk}$ that are inside/outside the excursion set. The red triangles are the triangulated mesh of constant field value $\delta = \nu$ that our algorithm generates. The normal vectors, edges, and vertices of triangles are presented in the left and right panels. The quantity $\alpha_{e}$, used in the extraction of $W_{2}$, is presented in the middle panel. The middle panel is the same box as the left panel, rotated such that the line of sight is parallel to the green edge. The images presented here are modified versions of Figures 8,9 in \cite{Appleby:2018tzk}. \textcopyright AAS. Reproduced with permission. }
  \label{fig:pics}
\end{figure}

In \citet{Appleby:2018tzk}, we provided a detailed description on an algorithm to extract the MFs from a discretized field on a uniform lattice. To briefly review, the method requires a set of field values $u_{ijk}$ on a uniform lattice, where $i,j,k$ subscripts denote pixel identifiers in the $x_{1,2,3}$ directions, respectively. We then form `pixel boxes' from eight adjacent pixels ($u_{ijk}$, $u_{i,j,k+1}$, $u_{i,j+1,k}$, $u_{i,j+1,k+1}$, $u_{i+1,j,k}$, $u_{i+1,j,k+1}$, $u_{i+1,j+1,k}$, $u_{i+1,j+1,k+1}$). Decomposing each individual pixel box into six non-overlapping tetrahedra, we linearly interpolate along edges of the tetrahedra to find points at which $u = \nu$, where $\nu$ is some constant field value that we select. We then generate a triangulated surface mesh of constant $u = \nu$ from these points. This defines the excursion set boundary $\partial \Mspace$ as a triangulated mesh. Finally, we extract the MFs from the triangulated boundary according to 
\begin{eqnarray} \label{eq:w0dis}
W_{0} &=& {1 \over V} \sum_{i \in N} V_{t, i} , \\
\label{eq:w1dis} W_{1} &=& {1 \over 6V} \sum_{i \in N} A_{t, i} , \\
\label{eq:w2dis} W_{2} &=& {1 \over 6\pi V} \sum_{e}|{\bf e}| \alpha_{e} , \\
\label{eq:w3dis} W_{3} &=& {1 \over 4\pi^{2}V} \sum_{v} \left(1 - {1 \over 2\pi} \sum_{T \in v} \phi_{T}^{v}\right)  ,
\end{eqnarray}
where $V$ is the total volume occupied by the field, $i$, $e$, and $v$ in the summation denote pixel box, {\it unique} edge, and {\it unique} triangle vertex in the triangulated surface mesh, respectively. $V_{t, i}$ is the volume contained within the $i^{\rm th}$ pixel box that is enclosed by the triangulated mesh, and $A_{t, i}$ is the total area of the triangulated mesh within the $i^{\rm th}$ pixel box. $|{\bf e}|$ is the length of the edge ${\bf e}$ and $\alpha_{e}$ is the angle subtended by the normal's of the two triangles that share the edge ${\bf e}$. Finally, $\sum_{T \in v}\phi_{T}^{v}$ is the sum of internal angles of all triangles $T$ that share the common vertex $v$. The quantities $\alpha_{\rm e}$ and $\phi_{T}^{v}$ are presented pictorially in Figure \ref{fig:pics}, which is reproduced from \citet{Appleby:2018tzk}.

This methodology was applied to `complete' fields with no masked regions and periodic boundary conditions in \citet{Appleby:2018tzk}. We now highlight the modifications required to reconstruct the same statistics from a restricted field. As before, we define a field on some regular three-dimensional lattice $u_{ijk}$, but now some of the domain is masked. We assign all masked pixels a particular `bad' value $u_{\rm b}$ ; $u_{ijk} = u_{\rm b}$. We restrict our analysis only to pixel boxes for which all eight vertices ($u_{ijk}$, $u_{i,j,k+1}$, $u_{i,j+1,k}$, $u_{i,j+1,k+1}$, $u_{i+1,j,k}$, $u_{i+1,j,k+1}$, $u_{i+1,j+1,k}$, $u_{i+1,j+1,k+1}$) are not masked (that is, not assigned value $u_{\rm b}$). We call such pixels boxes as unmasked. We denote the total number of pixel boxes in the entire volume $V$ as $N$ and the total number of unmasked pixel boxes used in our analysis as $\ell \leq N$. The total volume of the domain is $V = N \Delta_{g}^{3}$ and the corresponding masked volume is $V_{\ell} = \ell \Delta_{g}^{3}$, where $\Delta_{g}^{3}$ is the volume of a single pixel box (the resolution along each $x_{1,2,3}$ dimension is $\Delta_{g}$).

For each unmasked pixel box, we perform the standard marching tetrahedron algorithm; generate six tetrahedra, interpolate along their edges to points at which $u = \nu$, and construct a triangulated mesh internal to this particular pixel box. From this, we can calculate the triangulated surface area of iso-field value $u = \nu$, and also the fractional volume enclosed by this triangulated surface. Hence the volume and surface area of the excursion set can be estimated locally within each pixel box. Our estimates of volume of the excursion set $W_{0}$ and surface area of its boundary $W_{1}$, {\it per unit volume}, are therefore given by 
\begin{eqnarray}
W_{0} &=& {1 \over V_{\ell}} \sum_{i \in \ell} V_{t, i} , \\
W_{1} &=& {1 \over 6V_{\ell}} \sum_{i \in \ell} A_{t, i} , 
\end{eqnarray} 
where the sums are over all unmasked pixel boxes $i \in \ell$, $A_{t, i}$ is the the total area of all triangles constructed within the $i^{\rm th}$ pixel box and $V_{t, i}$ is the volume enclosed by the triangulated mesh in the $i^{\rm th}$ box. $A_{t, i}$ and $V_{t, i}$ can be calculated using trigonometry from the tetrahedral decomposition.

The remaining MFs --- $W_{2}$, $W_{3}$ --- are also local quantities and can be estimated from a masked subset of data. However, unlike $W_{0}$ and $W_{1}$, they require information from adjacent boxes as they are determined by triangles in the surface mesh that share common edges and vertices. Each triangle edge can be shared by a maximum of two adjacent pixel boxes, and triangle vertices can be shared by a maximum of four adjacent pixel boxes. To estimate $W_{2}$, we only consider pixel boxes that are at least two pixels away from any mask or boundary, to ensure that all edges counted in the $W_{2}$ reconstruction have two matching triangles. This is necessary to construct $\alpha_{e}$ in equation~($\ref{eq:w2dis}$). The estimator is simply 
\begin{equation} \label{eq:w2mod} 
W_{2} = {1 \over 6\pi V_{\ell'}} \sum_{e'}|{\bf e'}| \alpha_{e'} ,
\end{equation}
where $\ell'$ identifies the set of all pixel boxes at least two pixels from the boundary ($\ell \neq \ell'$), $\sum_{e'}$ represents the sum of all triangle edges within this subset of pixel boxes and $V_{\ell'} = \ell' \Delta_{g}^{3}$. 

For $W_{3}$, we adopt the following modified estimator
\begin{equation}\label{eq:w3est} 
W_{3} = {1 \over 4\pi^{2}V_{\ell}} \sum_{\tilde{v}} \left(w_{\tilde{v}} - {1 \over 2\pi} \sum_{T \in \tilde{v}} \phi_{T}^{v}\right) ,
\end{equation} 
where now the sum is over $\tilde{v}$, which is all triangle vertices extracted using the marching tetrahedral algorithm, which counts vertices multiple times. For example, if a triangle vertex is generated on the surface/edge of a pixel box, then it will be counted two/four times in the $\tilde{v}$ sum, respectively, because the marching tetrahedron algorithm will extract it from two/four pixel boxes. If a triangle vertex is generated internally to a pixel box, then it will be counted once. To remove the multiple counting, we weight each triangle vertex in the reconstruction by $w_{\tilde{v}}$, where $w_{\tilde{v}} = 1, 1/2,$ and $1/4$ depending on the vertex being internal, on the surface, or edge of a pixel box, respectively. If we are considering a field without any boundary, then equation~(\ref{eq:w3est}) is equivalent to equation~(\ref{eq:w3dis}). In the presence of a mask, a triangle vertex may not contribute a total of unity to the first term in the sum (eq.~\ref{eq:w3est}), because the algorithm now skips masked boxes. However, we still obtain an unbiased reconstruction of $W_{3}$ because the sum over $\phi_{T}^{\tilde{v}}$ in equation~(\ref{eq:w3est}) only includes the triangle angles in the unmasked pixel boxes. 

We present an example of two pixel boxes used in our analysis in Figure~\ref{fig:pics}. The solid black/white points denote pixels in the lattice that are in/out of the excursion set (that is, they have values such that $\delta_{ijk} > \nu$ and $\delta_{ijk} < \nu$,  respectively). The pixel boxes are decomposed into six non-overlapping tetrahedra as described in \citet{Appleby:2018tzk}, and a triangulated mesh of $\delta = \nu$ is generated (red triangles in the figure). $\alpha_{e}$ is presented in the middle panel, which is the pixel box in the left panel rotated to align with the green triangle edge. In the left panel, the green triangle edge is completely internal to the pixel box, and hence both triangles incident to it are internal. This means that $\alpha_{e}$ for the green edge can be obtained within the box, from the normal vectors $n_{1}$ and $n_{2}$. On the contrary, the blue triangle edges lie on the surfaces of the pixel box, and each require a triangle in adjacent boxes to define their corresponding $\alpha_{e}$. For this reason, this pixel box will only be used to calculate $W_{2}$ if all adjacent pixel boxes also contain no bad pixels. 

In the left panel, the volume enclosed is the volume occupied between the red triangles and the black pixel and the surface area is the total area of the triangles. These are the contributions to $W_{0}$ and $W_{1}$ from this particular pixel box.

In the right panel of Figure~\ref{fig:pics}, we present a different pixel box. The triangulated surface $\delta = \nu$ is presented as a set of red triangles, and the triangle vertices are coloured green/blue/red. The green vertex in the center is completely internal to the pixel box, and hence will contribute $w = 1$ to equation~($\ref{eq:w3est}$), and all incident triangles are present. The red dots are triangle vertices on the surface of the pixel box, and will contribute $w=1/2$ when this particular box is encountered in the algorithm. The blue dots lie on the edges of the box and will contribute $w=1/4$, as they are potentially shared by four other boxes. All triangle internal angles in the Figure are counted in the $\sum \phi$ term in equation~($\ref{eq:w3est}$). 
 
Our methodology can be used to extract the local properties of a surface per unit volume. To perform this numerical calculation we do not need to sample the entire data domain, hence the presence of a mask is practically irrelevant. Due to the local nature of the curvature integrals, they can be extracted by sampling a subset of the surface, and hence our numerical algorithm can provide an unbiased estimate of the Edgeworth expansion of $W_{k}$. Only local statistics can be extracted in an unbiased manner using our methodology -- the average curvature per unit volume, for example. Global properties such as topology cannot be extracted using this approach. 

Next, we verify that our estimators provide an unbiased estimate of the curvature integrals on an unbounded domain, by applying them to masked Gaussian random fields and mock galaxy snapshot boxes. We also address the separate issue of smoothing masked fields.

\subsection{Gaussian Random Fields}
\label{sec:gass} 


To confirm that the estimators described above can be used to reconstruct the underlying curvature integrals, we generate mock data. Initially, we take realisations of a Gaussian random field (GRF) in a periodic box. We draw the random fields from a linear $\Lambda$CDM matter power spectrum with parameters given in Table~\ref{tab:1}, in a periodic box of volume $V = (3.15 \, {\rm Gpc})^3$. We adopt a resolution of $\Delta = 6 \, {\rm Mpc}$ and smooth the field with Gaussian kernel of scale $R_{\rm G} = 35 \, {\rm Mpc}$. We denote the unmasked, smoothed field $\delta_{ijk}$. We then mask the field. First, we set all vertices within distance $R_{\rm G}$ of the boundary of the (periodic) volume to $\delta_{ijk} = \delta_{\rm b}$, where $\delta_{\rm b}$ is some arbitrary `bad pixel' value. We then generate $200$ cylinders through the box in the $x_{3}$ direction, of radius $r_{n}$ and center $x_{1, n}$, $x_{2, n}$, where $1 \leq n \leq 200$. The values of $0 < r_{n} < 200 \, {\rm Mpc}$ and $0 < x_{1, n} < 3150 \, {\rm Mpc}$, $0 < x_{2, n} < 3150 \, {\rm Mpc}$ are randomly generated from a uniform distribution. Any $\delta_{ijk}$ vertex within the cylinders is also assigned a `bad pixel' value $\delta_{ijk} = \delta_{\rm b}$. This simple mask is representative of an angular mask on the sky, which generates cylinders through data in the distant observer limit (more precisely, cones but we do not pursue this distinction here). We measure the MFs of the unmasked GRFs and the masked equivalents.

\begin{figure}[htb]
  \centering 
  \includegraphics[width=0.95\textwidth]{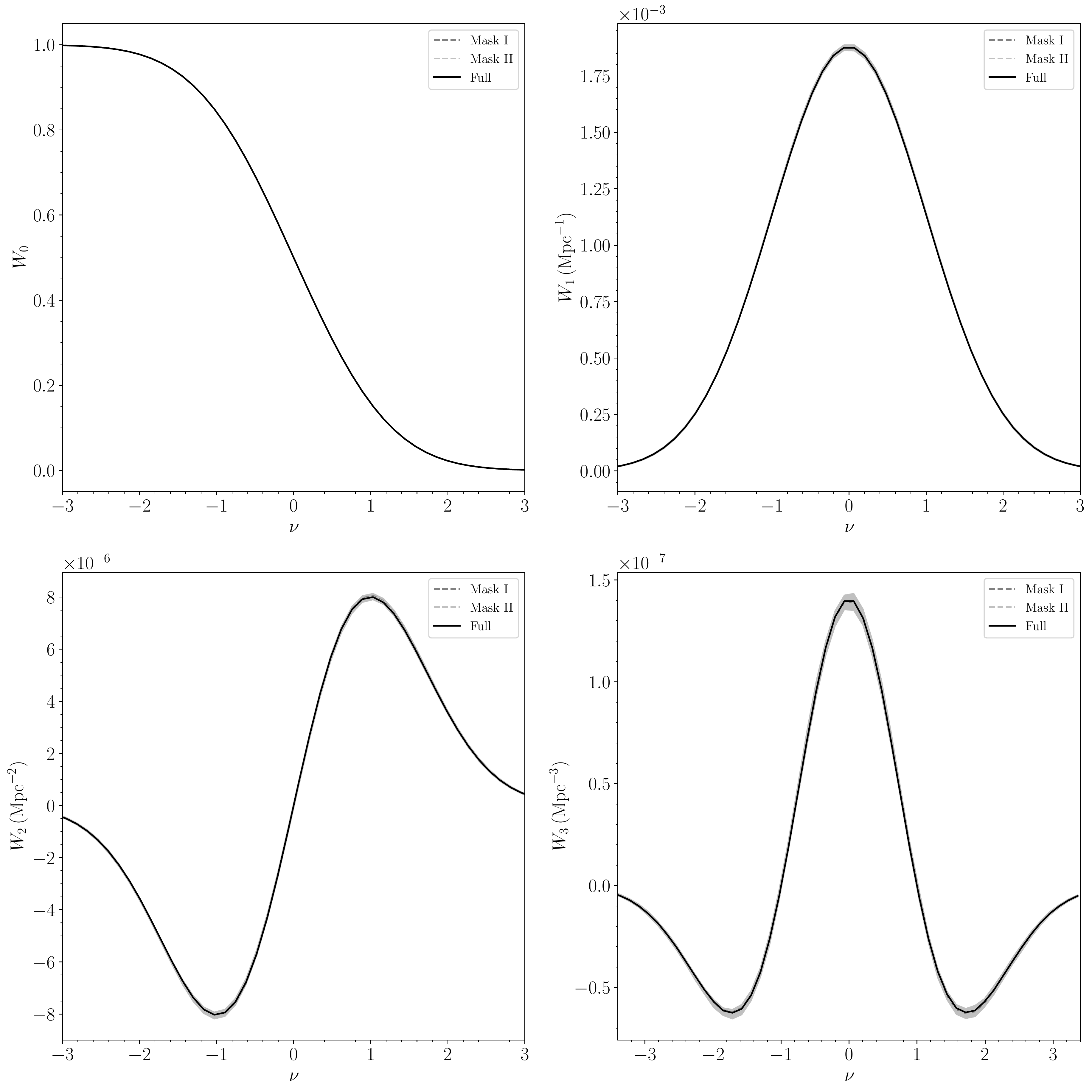} 
  \caption{The MFs obtained from $N_{\rm r} = 50$ realisations of a Gaussian random field (GRF) with $\Lambda$CDM linear matter power spectrum. The solid grey region is the $\pm 1$-$\sigma$ statistical uncertainty from the `Mask I' realisations. The black solid/dark grey dashed/light grey dashed lines are the mean values of the statistics obtained from the full unmasked field, and the result of two masking procedures explained in Sections~\ref{sec:gass} and \ref{sec:smooth}. }
  \label{fig:1}
\end{figure}

In Figure~\ref{fig:1}, we present the mean and standard deviation of the MFs for $N_{\rm real} = 50$ GRF realisations. The dark grey dashed lines (labelled `Mask I') represent the mean of the masked fields extracted using the algorithms described in Section~\ref{sec:unbias}, and the solid black lines are the corresponding MFs extracted from the full, unmasked data. The grey solid region is the R.M.S. fluctuations of the statistics from the masked realisations. We observe no systematic deviation between the bounded and unbounded domain, and the solid and dashed lines practically overlap.

\begin{figure}[htb]
  \centering 
  \includegraphics[width=0.94\textwidth]{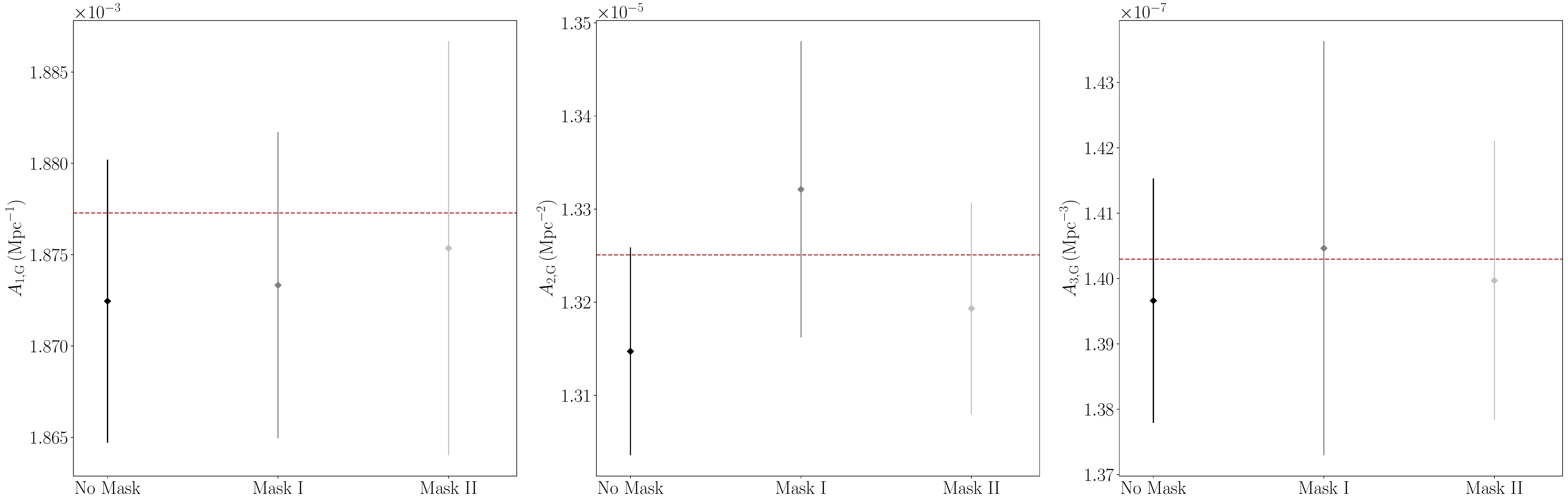} \\
  \includegraphics[width=0.32\textwidth]{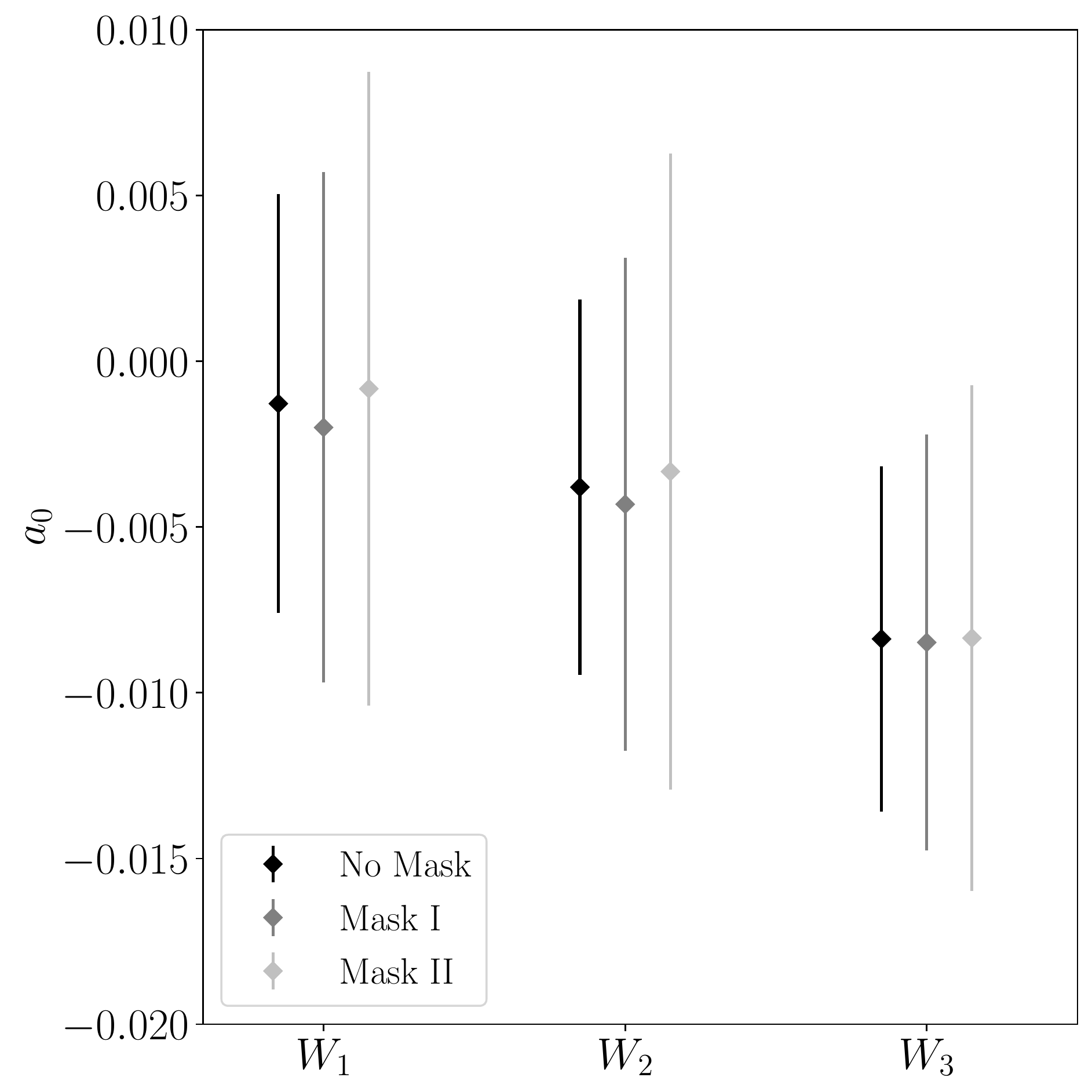} 
  \includegraphics[width=0.32\textwidth]{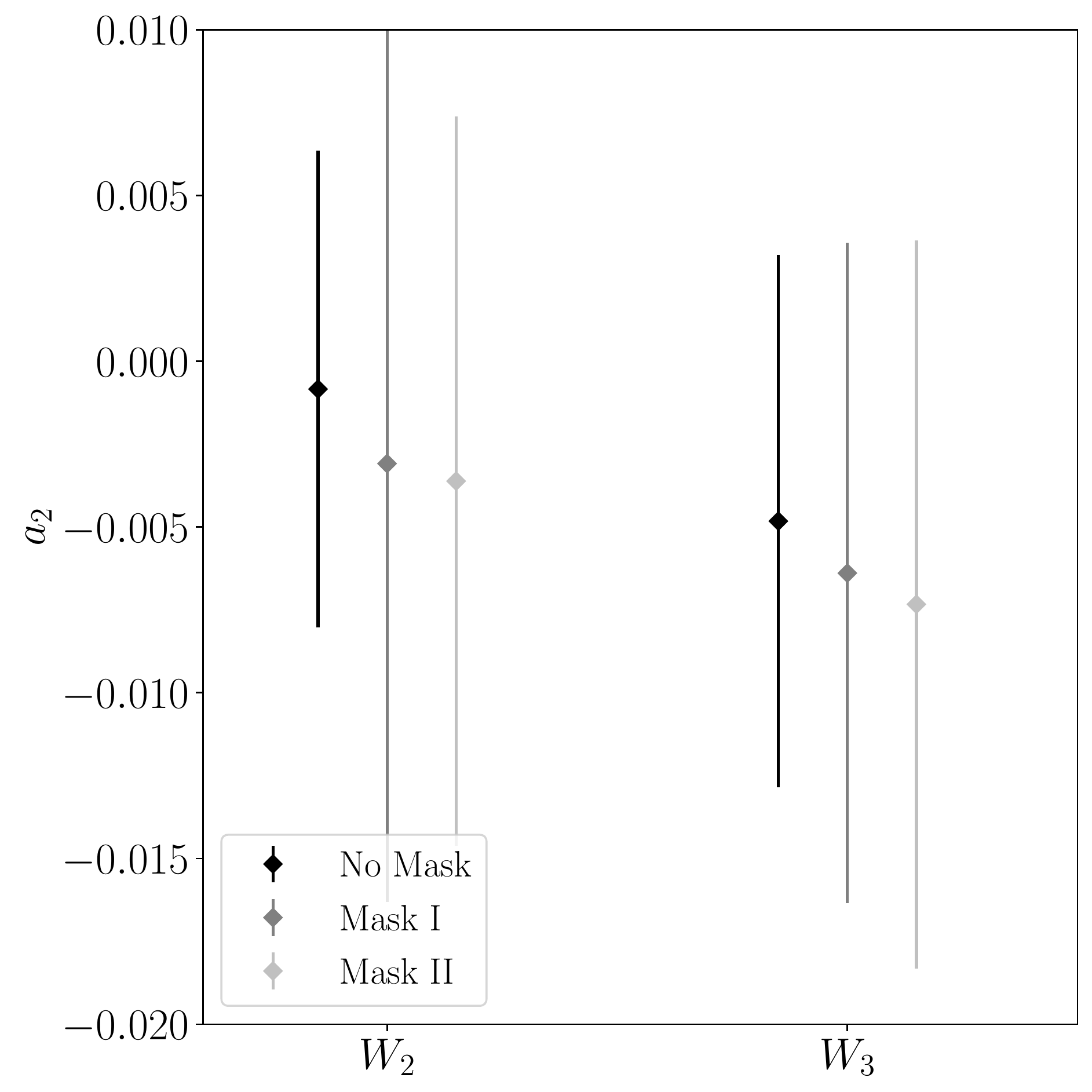} \\
  \caption{The amplitudes of the MFs (top three panels) and $a_{0}$, $a_{2}$ extracted from the $N_{\rm r} = 50$ realisations of a GRF. The brown dashed lines in the top panels are the Gaussian expectation value from the Edgeworth expansion, and the black/dark grey/light grey diamonds and error bars are the mean and R.M.S. fluctuations of the quantities numerically extracted from the unmasked and masked data respectively. For a GRF, $a_{0} = a_{2} = 0$.}
  \label{fig:gauss_c}
\end{figure}

Following the main body of the text, we then extract the amplitudes $A_{k,{\rm G}}$ and $a_{0,2}$ from each realisation using equations~(\ref{eq:int1}-\ref{eq:int3}). In Figure~\ref{fig:gauss_c}, we present the mean and R.M.S. values obtained from the unmasked/masked data (black/dark grey diamonds and error bars respectively, labelled `No Mask' and `Mask I'). In the top panel, we also present the Edgeworth expansion, Gaussian expectation values of $A_{k, {\rm G}}$ from the expression (eq.~\ref{eq:ampg}) as brown dashed horizontal lines. In all instances, the masked data is consistent with the unmasked equivalents, and consistent with the ensemble expectation value. The $a_{0,2}$ parameters should be consistent with zero for a Gaussian field, and this expectation is recovered in our analysis. There is a $\lesssim 1\%$ systematic discrepancy in $a_{0}$ extracted from the $W_{3}$ curve (lower left panel, right hand side); this is due to our method of extracting this parameter. For $W_{3}$, $a_{0}$ is the coefficient of the cubic Hermite polynomial $H_{3}$, which has a relatively large tail in the high $|\nu_{A}|$ regime, whereas we truncate the integrals in equations~(\ref{eq:int1}-\ref{eq:int3}) at $|\nu_{A}| = 4$.

\subsection{Mock Galaxy Catalogs}

A GRF is a special example in the sense that all information is contained in the underlying power spectrum. In terms of the MFs, all information in $W_{k}$ is contained within the $H_{k-1}$ Hermite polynomial coefficient. The matter density field in the late Universe is not well described by a GRF, even when smoothing on large scales $R_{\rm G} \sim 35 \, {\rm Mpc}$. We now check that our estimators are also unbiased for gravitationally evolved, non-linear matter fields. 

To this end, we repeat our test on a mock galaxy snapshot box, gravitationally evolved to $z=0$. Specifically, we use Horizon Run 4 (HR4) --- a cosmological $N$-body simulation containing $N = 6300^{3}$ particles in a volume of $V = (3150 {\rm Mpc}/h)^{3}$. The simulation uses a modified GOTPM code\footnote{For a description of the original GOTPM code, please see \cite{DUBINSKI2004111}. A description of the modifications introduced in the Horizon Run project at \href{https://astro.kias.re.kr/~kjhan/GOTPM/index.html}{https://astro.kias.re.kr/\textasciitilde kjhan/GOTPM/index.html}}. The cosmological parameters used are $h=0.72$, $n_{\rm s} = 0.96$, $\Omega_{\rm m} = 0.26$, $\Omega_{\rm b} = 0.048$. Details of the numerical implementation and the method by which mock galaxies are constructed can be found in \cite{Hong:2016hsd}. The mock galaxies are defined using the most bound halo particle galaxy correspondence scheme, and the survival time of satellite galaxies post merger is estimated via a modification of the merger timescale model described in \cite{Jiang:2007xd}.

\begin{figure}[htb]
  \centering 
  \includegraphics[width=0.95\textwidth]{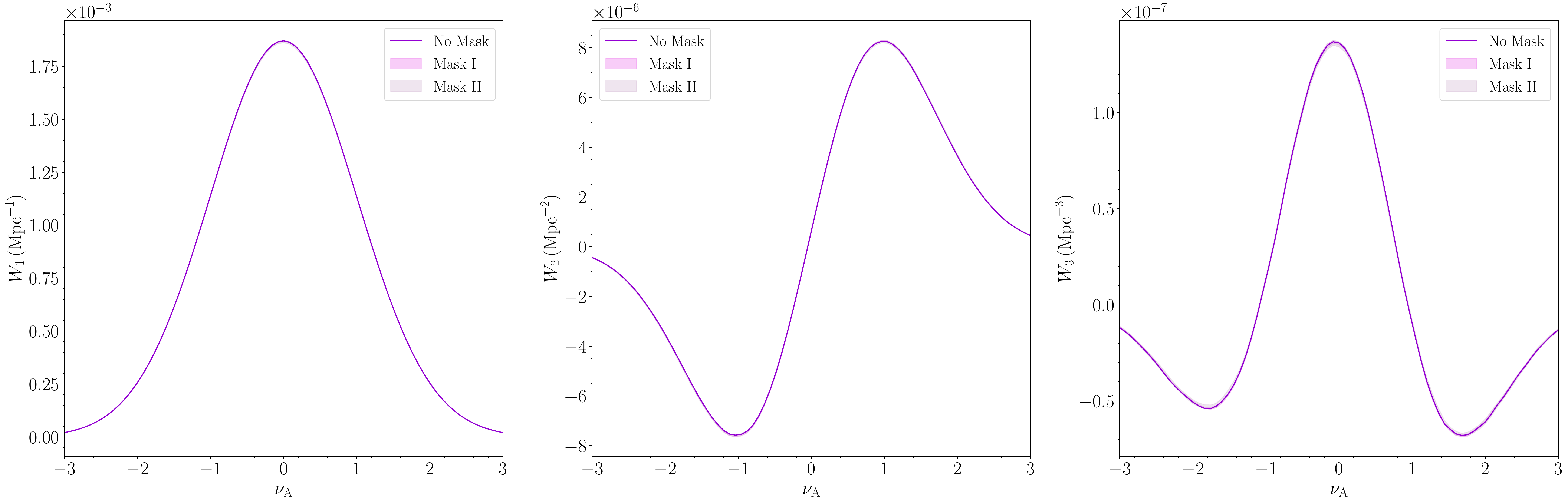} 
  \caption{The MFs $W_{1}$, $W_{2}$, $W_{3}$ extracted from the Horizon Run 4 (HR4) snapshot box at $z = 0$. The purple line is the value obtained from the unmasked box, and the lighter solid regions are the 1-$\sigma$ R.M.S. fluctuations obtained from $N_{\rm real} = 50$ random realisations of the masking procedures outlined in Sections~\ref{sec:gass} and \ref{sec:smooth}.}
  \label{fig:HR4_1}
\end{figure}

\begin{figure}[htb]
  \centering 
  \includegraphics[width=0.94\textwidth]{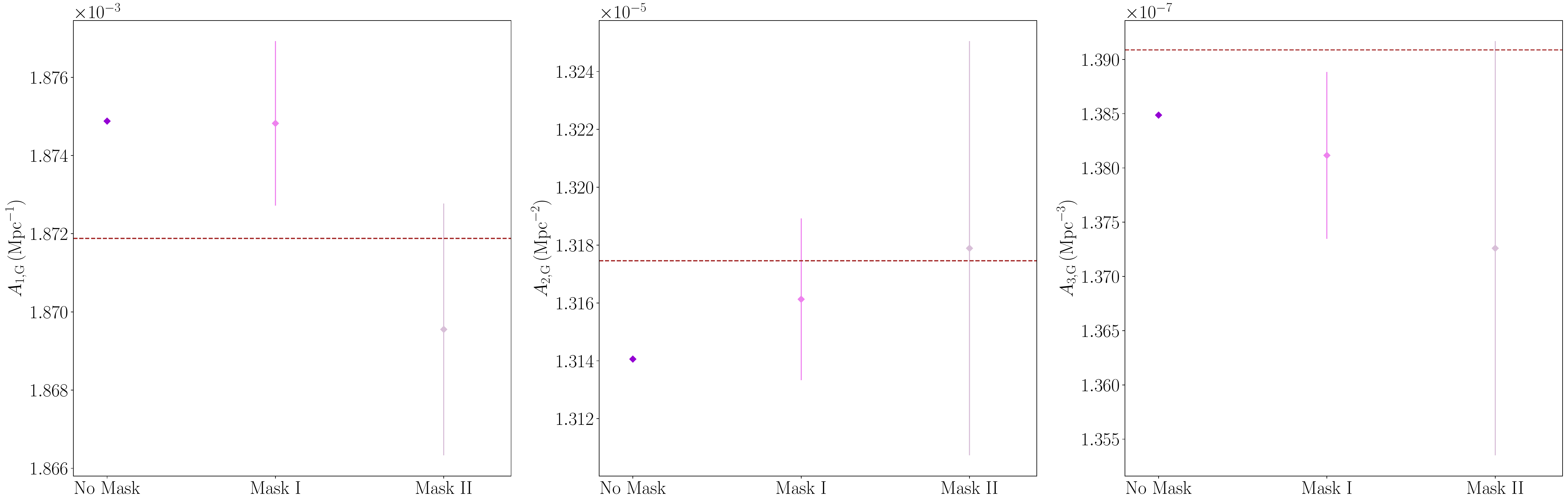} \\
  \includegraphics[width=0.32\textwidth]{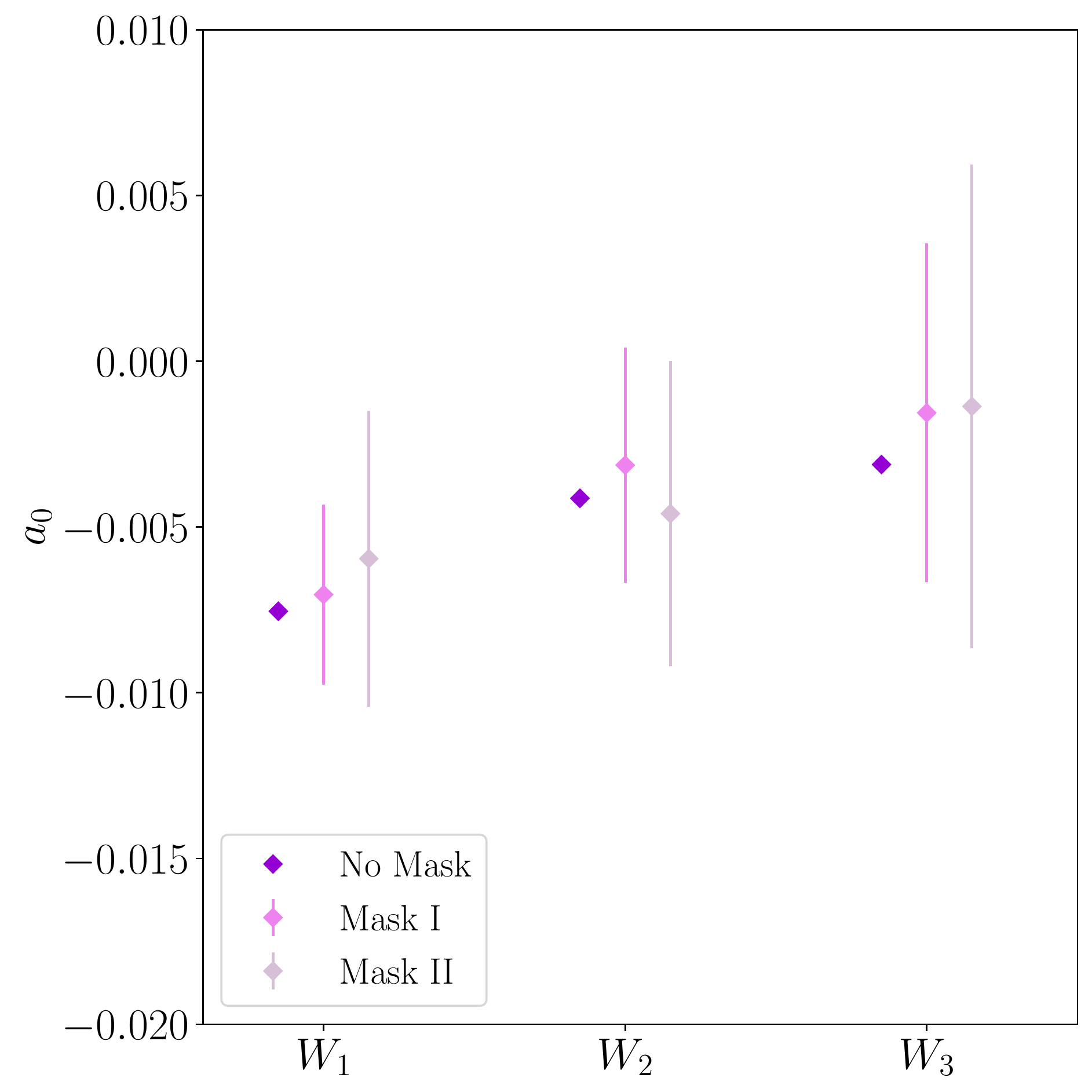} 
  \includegraphics[width=0.32\textwidth]{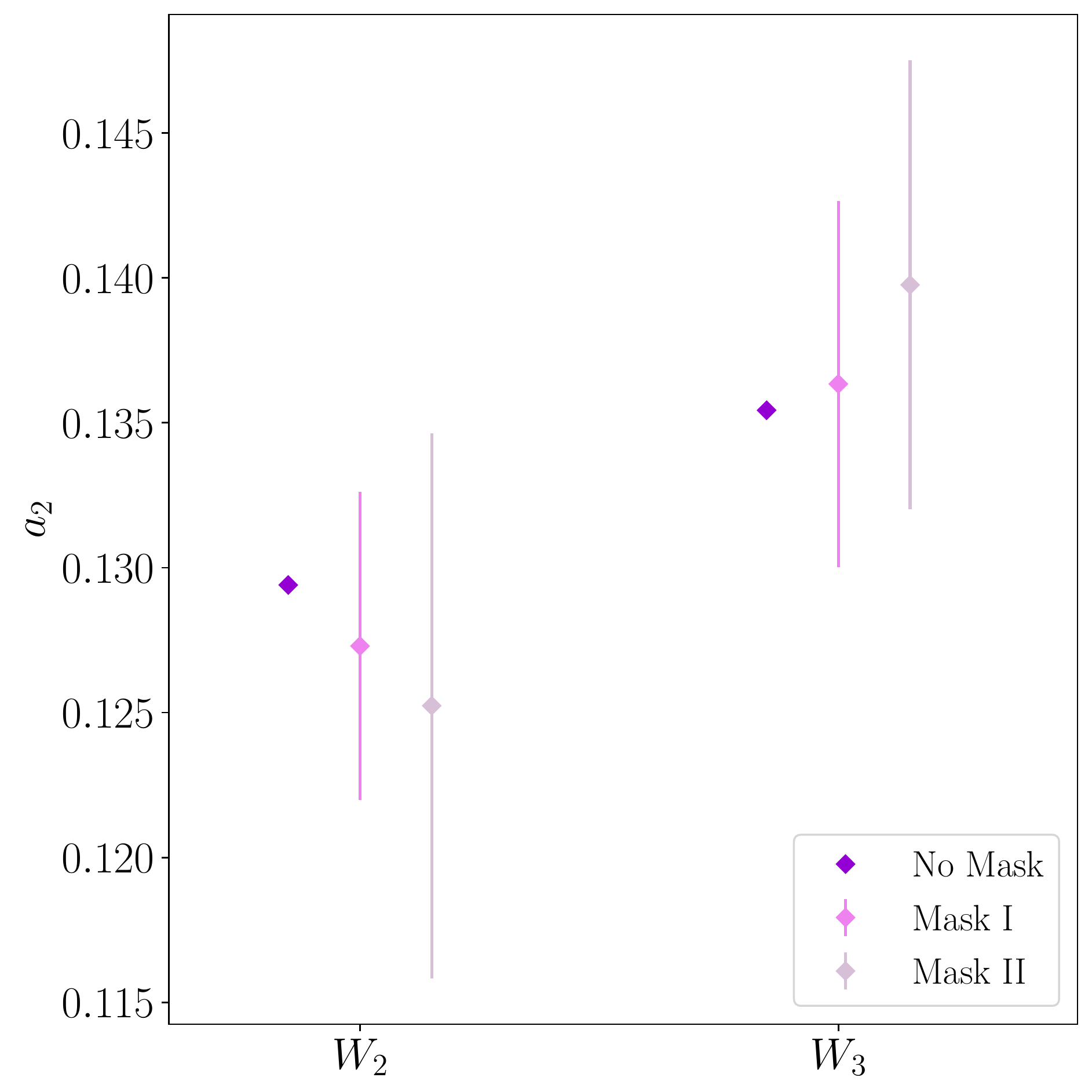} \\
  \caption{The amplitudes of the MFs and $a_{0}$, $a_{2}$ parameters extracted from the HR4 snapshot box. The deep purple diamond is the value obtained from the unmasked box (a single realisation; no error bar) and the lighter purple diamonds/error bars are the values of the statistics from the $N_{\rm real} = 50$ realisations of the mask. The error bars will not represent the true statistical uncertainty, as we are not accounting for cosmic variance by using a single data realisation. The horizontal dashed lines in the top panel are the Gaussian expectation values (eq.~\ref{eq:ampg}) assuming HR4 cosmological parameters.   }
  \label{fig:amp_HR4}
\end{figure}

We use the mock galaxy snapshot box at $z=0$, making no redshift space corrections to the galaxy positions. We bin the galaxies into a regular lattice and generate a mean subtracted number density field $(n_{ijk}-\bar{n})/\bar{n}$, where $n_{ijk}$ is the number of galaxies in ($i,j,k$)-pixel and $\bar{n}$ is the mean number of galaxies in all pixel boxes. We smooth the field with Gaussian kernel, and then mask the data using the same procedure as for the GRFs. Defining the smoothed field as $\tilde{\delta}_{ijk}$, we finally define a mean subtracted, unit variance field $\delta_{ijk} = (\tilde{\delta}_{ijk} - \mu)/\sigma$, where $\mu$, $\sigma$ are the mean/rms of $\tilde{\delta}_{ijk}$. Despite only having a single realisation of the data, we generate sub-samples by repeating the masking procedure over $N_{\rm real} = 50$ realisations, each time drawing a different set of masked cylinder positions and radii. Each time we measure the MF curves, and resulting Hermite polynomial coefficients. Although the resulting error bars will under-represent the `true' statistical uncertainty because the samples are not independent, we can use the mean values to confirm that the masked MFs are unbiased with respect to the MFs of the full unbounded field. We present the results in Figure~\ref{fig:HR4_1}. The dark solid curves are the MFs obtained from the full, unmasked box. The dark violet solid region is the 1-$\sigma$ uncertainty of the statistics over the $N_{\rm real} = 50$ re-samplings. We observe no statistically significant bias in our reconstruction of the full data from the masked subsets, indicating that our methodology can be used to extract the `Edgeworth expansion' theoretical prediction from masked data. We also present the parameters $A_{k, {\rm G}}$ and $a_{0,2}$ in Figure~\ref{fig:amp_HR4} (top panels, bottom left, and right, respectively). The darkest diamonds are the results from the unmasked, periodic field, and the dark pink diamonds/error bars (labelled `Mask I') are the mean and R.M.S. from the masked field. All parameters are consistent between masked and unmasked data. The dashed horizontal lines in the top panels are the Gaussian expectation values of the amplitudes, which are consistent with the galaxy data.

\subsection{Smoothing a Masked Field}
\label{sec:smooth}

In the previous section, we took a `complete' data set (that is, unmasked and with periodic boundary conditions), smoothed the field, then applied a mask. The goal was to confirm that the MF reconstruction used in this work is unbiased. However, in realistic situations, we cannot mask the data after smoothing, but rather must smooth the masked data. This introduces an additional complication, as the `true' field is not well reconstructed in regions close to the boundary after smoothing. We must therefore cut a region in the vicinity of the boundary from our analysis. We now quantify this statement. 

Initially returning to GRFs, we perform the following steps. We generate a discrete field $\delta_{ijk}$ within a periodic box, and then apply the same masking procedure as in the previous sub-section. We define the field as $\delta_{ijk}$. We also define a discrete mask field $M_{ijk}$, where $M_{ijk} = 0$ if $\delta_{ijk} = \delta_{\rm b}$ is masked and $M_{ijk} = 1$ otherwise.  

We then smooth both the masked field $\delta_{ijk}$ and the mask $M_{ijk}$, both with the same Gaussian smoothing kernel and scale $R_{\rm G} = 35 \, {\rm Mpc}$. We denote the smoothed fields as $\tilde{\delta}_{ijk}$ and $\tilde{M}_{ijk}$. We then apply the smoothed mask to the smoothed data, taking $\tilde{\delta}_{ijk} = \delta_{\rm b}$ if $\tilde{M}_{ijk} < M_{\rm cut}$, where $M_{\rm cut}$ is some arbitrary value selected to be $M_{\rm cut} = 0.9$. Increasing $M_{\rm cut}$ increases the volume of data in the vicinity of the boundary that is cut during this second masking procedure. We then measure the MFs of the double-masked field $\tilde{\delta}_{ijk}$. The results are presented in Figures~\ref{fig:1} and \ref{fig:HR4_1} for GRFs and the HR4 mock galaxy snapshot box at $z = 0$, respectively (light grey dashed line, pale pink filled region, labelled `Mask II'). We again observe an unbiased reconstruction of the MFs of the unmasked field. The parameters $A_{k, {\rm G}}$ and $a_{0,2}$ are also presented in Figures~\ref{fig:gauss_c} and \ref{fig:amp_HR4} labelled `Mask II'. The parameter reconstruction is unbiased within the uncertainties of the measurements.

We stress that the issues described in the previous subsection and here are not related. The former relates to the ability to reconstruct the MF statistics from a masked field, and the latter is the ability to faithfully reconstruct a field from a masked domain. We have shown that both issues can be ameliorated, with a judicious choice of statistical estimator and smoothing algorithm respectively.

\section{Redshift Space Distortion} 
\label{sec:rsd}

Redshift space distortion (RSD) presents some difficulty with MF analysis. In \citet{Matsubara:1995wj}, the effect was first theoretically predicted in the Kaiser limit and under the plane parallel approximation. The analysis was continued in \citet{Codis:2013exa}, in which the Kaiser approximation was applied to higher order cumulants. This represents the current state-of-the-art in MF analysis. However, two outstanding issues remain. One is that the Finger-of-God effect could also impact the statistics, even on relatively large scales, as the cumulants $\sigma_{1}$ and $\sigma_{0}$ are integrated quantities. Second, large area sky surveys generically violate the plane parallel, distant observer approximation, which breaks the homogeneity of the RSD signal. 

\begin{figure}[htb]
  \centering 
  \includegraphics[width=0.95\textwidth]{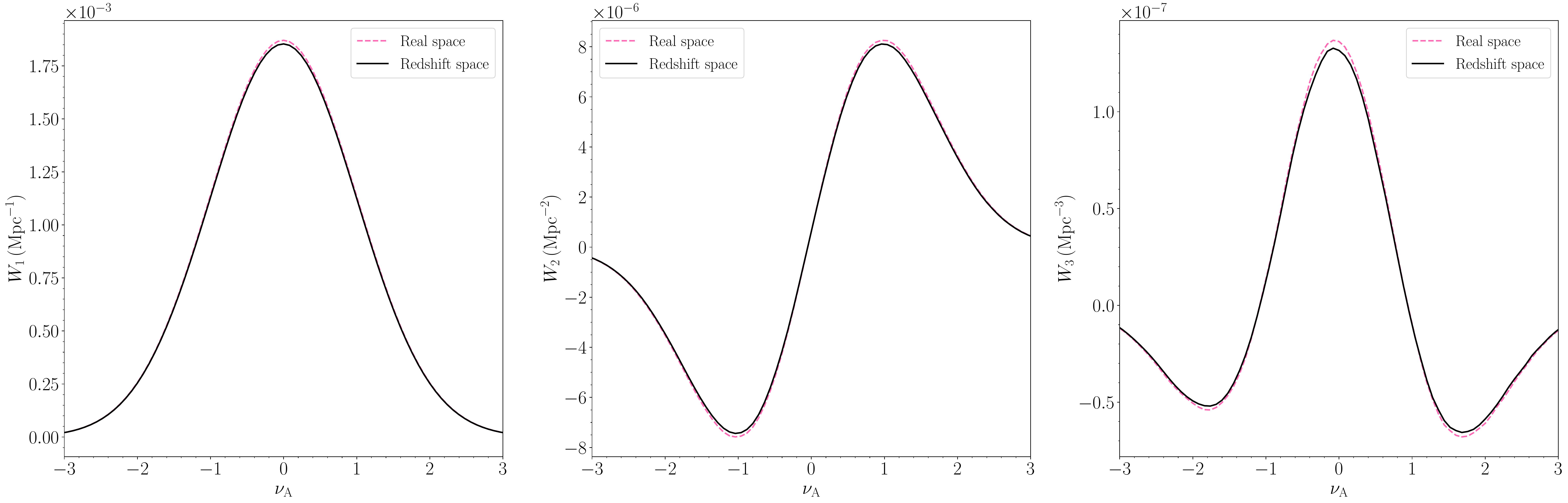} 
  \caption{The MFs obtained from the HR4 $z=0$ snapshot box in real (pink dashed line) and redshift (black solid line) space. The effect of redshift space distortion is to decrease the amplitude of the MFs. }
  \label{fig:HR4_rsd_MF}
\end{figure}

Fortunately, the effect of redshift space distortion on the three-dimensional MFs is small. In Figure~\ref{fig:HR4_rsd_MF}, we present the MFs of the HR4 mock galaxy snapshot box at $z = 0$ in real and plane-parallel redshift space. To generate the redshift space box, we perturb the galaxies along an arbitrary $x_{3}$ direction according to 
\begin{equation}
x_{3} \to x_{3} + {(1+z_{\rm b}) v_{3} \over H(z_{\rm b})} , 
\end{equation} 
where $z_{b}=0$, and $v_{3}$ is the galaxy velocity in the $x_{3}$ direction. In Figure~\ref{fig:HR4_rsd_MF}, the pink dashed/black solid lines are the MF statistics extracted from the real/redshift space boxes, respectively. We extract the amplitudes of these functions using the integral in equation~(\ref{eq:int1}). We define the ratio $\alpha_{k} = A^{(\rm s)}_{k, {\rm G}}/A^{(\rm r)}_{k, {\rm G}}$, where r/s superscripts denote the amplitudes in real/redshift space, respectively. We find $\alpha_{1} = 0.991$, $\alpha_{2} = 0.981$, and $\alpha_{3} = 0.971$. Our results indicate that the effect of RSD constitutes an approximately $1,2,3\%$ decrease in the amplitude of $W_{1,2,3}$ respectively. In the main body of the text, we correct the theoretical expectation $A_{k, {\rm G}}$ MF amplitudes by $\alpha_{k}$ to convert them into redshift space. 

Predicting the consequences of the breakdown of the plane-parallel approximation is beyond the scope of this work, and will be considered elsewhere. However, if the overall effect on the MFs due to RSD is $\sim {\cal O}({\rm few}\,\%)$ on quasi-linear scales, we can expect this subtlety to not significantly affect our conclusions. Similarly, we expect the cosmological parameter dependence of the RSD effect to be a small correction to $\alpha_{k}$, which is itself a small correction. We therefore treat $\alpha_{k}$ as constant correction factors in the main body of the paper.

\section{Testing for Systematics} 
\label{sec:systematics}

Finally, we consider the various systematics that could impact our measurements of the MFs, to test the robustness of our measurements. We perform a number of stress-tests on our results by significantly varying the assumptions made. Each time we vary a single assumption used in the main body of the paper, and run the new data through our analysis pipeline. We fit the functional form in equation~(\ref{eq:fit}) to the data sets, and present the best fit and 1-$\sigma$ marginalised uncertainties on $A_{k, {\rm G}}$. 

We identify the following potential issues:
\begin{enumerate}[I.]
    \item The galaxy weights significantly affect the MF reconstruction.
    \item Our estimator of the MFs is affected by the mask/boundaries of the data.
    \item Changing the sampling method will yield sub-samples of galaxies with different properties.
    \item Finite pixel resolution effects.
\end{enumerate}
We consider each point in turn.  

The galaxy weights are one possible issue. We have accounted for systematic variations in the galaxy number density by weighting each galaxy, as detailed in \citet{2016MNRAS.455.1553R}. We have also removed areas of the data in which the completeness is below $90\%$. Due to the high completeness of the sample, and the large smoothing scales adopted, we do not anticipate that our conclusions will be affected by the galaxy weighting scheme. However, to test this we repeat our analysis on each of the data sets without weighting the galaxies with {\it any} observational systematic correction, assuming that each galaxy contributes a total of unity to the number density. This is obviously an inappropriate procedure, but will inform us how significant the effect of the galaxy weights are on our conclusions. 

Given that the northern/southern sky data possess very different survey geometries, it is possible that our methodology is affected by the boundaries of the data. However, this seems unlikely given that our treatment of the boundary is repeated exactly for the patchy mock data, and all subsets of the mock catalogs present excellent consistency. To test the impact of the mask further, we decrease the mask cut $M_{\rm cut}$ applied to the data to be $M_{\rm cut} = 0.8$, to test the sensitivity of our results to this parameter. As we lower $M_{\rm cut}$, we are including regions closer to the survey boundary, where the field may not be well reconstructed. 

Regarding the sampling, in principle the MFs are insensitive to galaxy sampling and linear bias. This implies that on large scales any dependence on linear galaxy bias drops out. However, in practise this is not always precisely true because the presence of shot noise in a field reconstructed from a point distribution makes the measured power spectrum the sum of two distinct quantities -- the cosmological and noise power spectra. The ratio between these two contributions will impact the MF amplitudes, and is a function of the matter power spectrum amplitude and galaxy bias. To test the sensitivity of the statistics to sample selection, we randomly select galaxies in the CMASS and LOWZ data to match the fiducial number density $\bar{n} = 6.25\times 10^{-5} \, {\rm Mpc}^{-3}$, rather than mass selecting as in the main body of the paper. It was shown in \citet{Kim:2014axe} that the difference between randomly and mass selecting galaxies can significantly impact the genus statistic. However, for the BOSS data we do not expect such a pronounced difference because the entire data set constitutes mostly massive, highly biased galaxies, and with our number density cut we are using the majority of the sample over the range $0.2 < z < 0.4$ and $0.45 < z < 0.6$. Hence randomly and mass selecting galaxies will yield a similarly biased sample.

Finally, we consider finite resolution pixel effects. This is also an unlikely source of contamination because we are Gaussian smoothing over at least four pixels in each data set, and also the patchy mock catalogs are subject to the same pixel resolution and are self-consistent. However, we repeat our analysis, using the same box sizes as in Table~\ref{tab:2} but with resolutions $\Delta = 5.5, 4.3, 4.0, 3.1 \, {\rm Mpc}$ for the CMASS N/S and LOWZ N/S data, respectively. This corresponds to a change from $512^3$ to $768^3$ pixels within the uniform lattices onto which we aggregate the data.

\begin{figure}[htb]
  \centering 
  \includegraphics[width=0.94\textwidth]{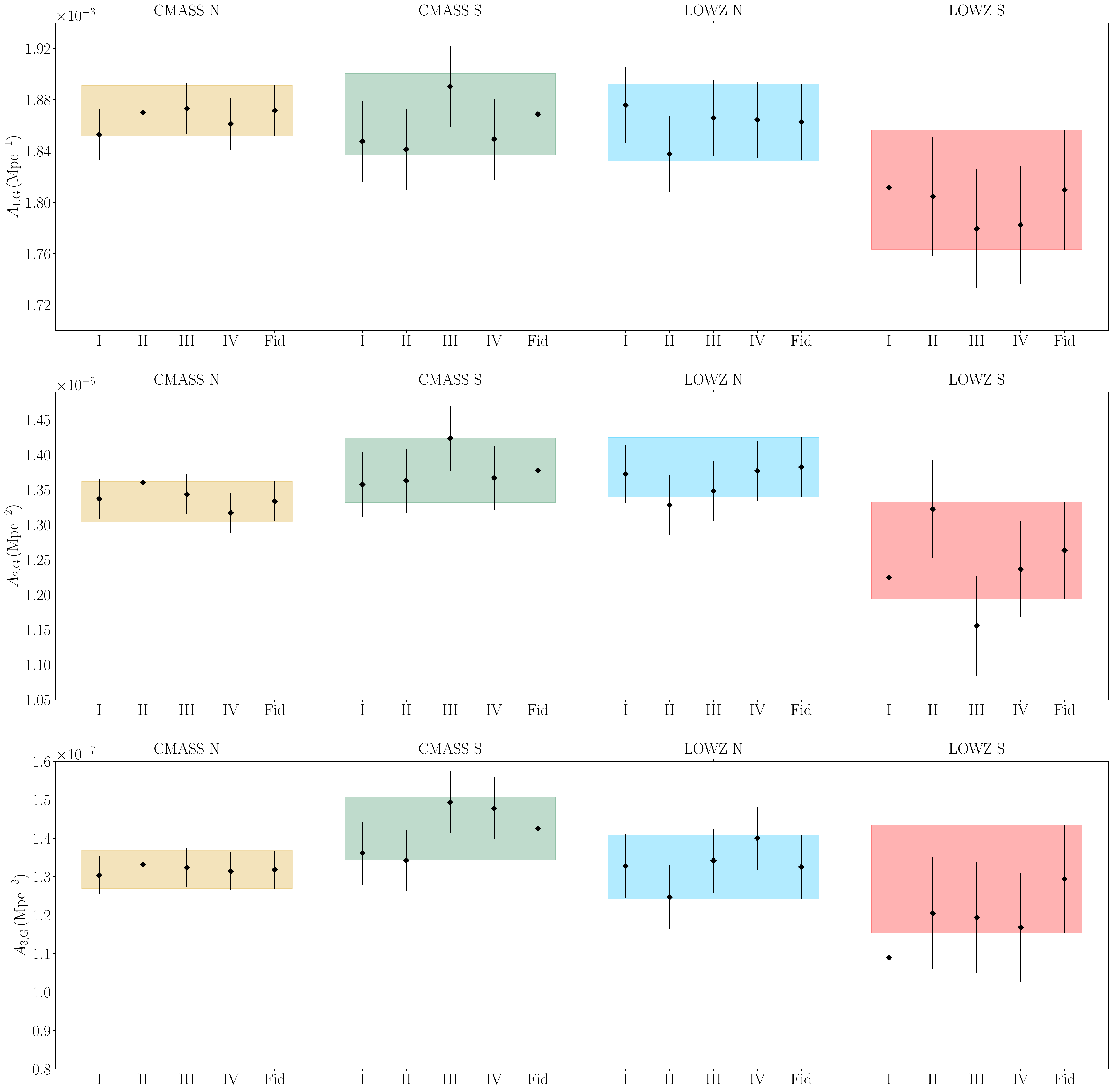}
  \caption{The best fit and 1-$\sigma$ uncertainties of the amplitudes of the MFs $W_{1}$, $W_{2}$, $W_{3}$ (top-bottom panels) obtained by varying the assumptions made in the main body of the paper and repeating our analysis. `Fid' represents the fiducial analysis in the paper, and I-IV are the four variations of our analysis discussed in Appendix~\ref{sec:systematics}. The solid gold/green/blue/red regions are the 1-$\sigma$ uncertainties from the fiducial analysis of the CMASS N/S and LOWZ N/S data, respectively.}
  \label{fig:sys_a}
\end{figure}

\begin{figure}[htb]
  \centering 
  \includegraphics[width=0.45\textwidth]{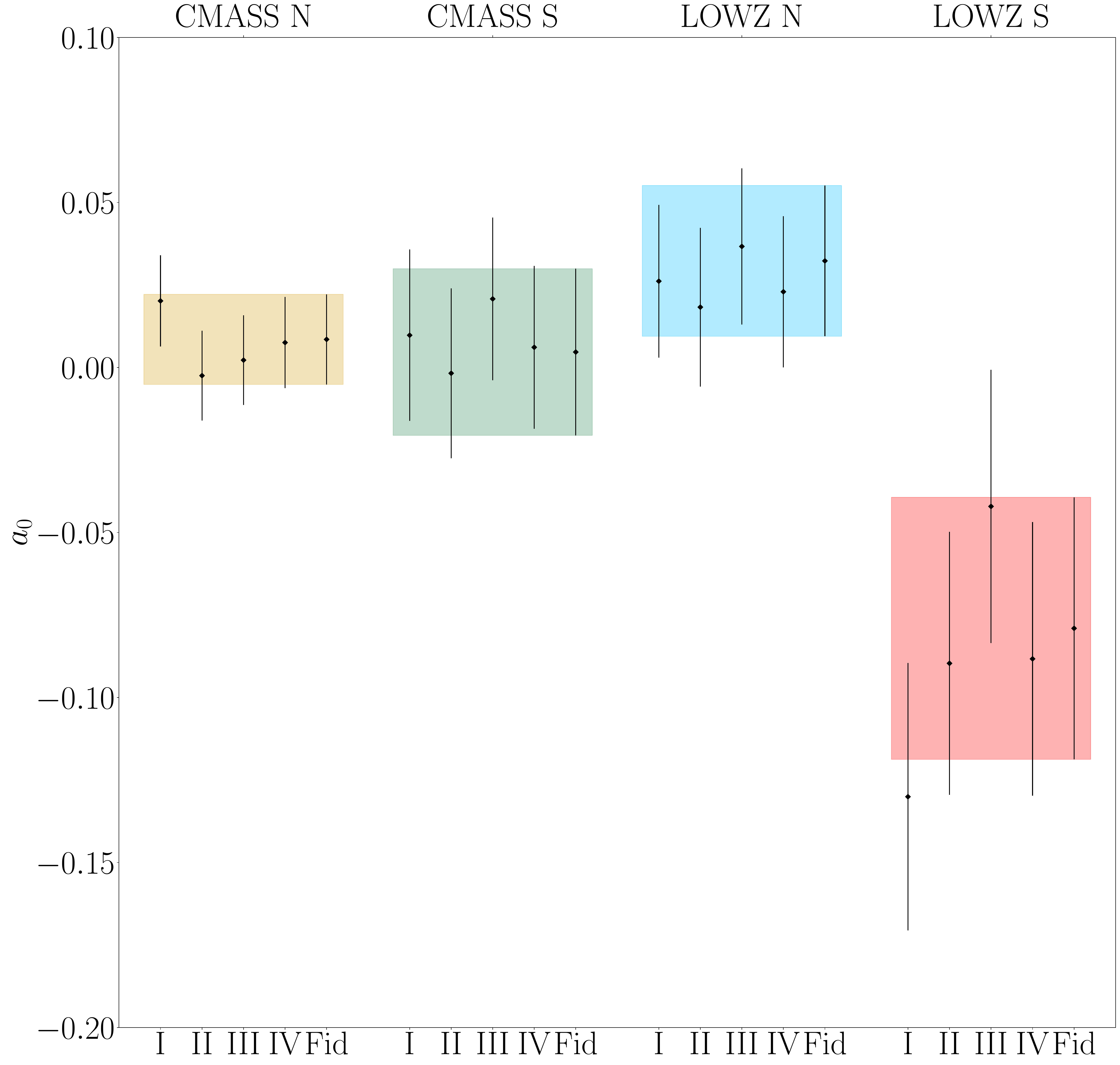} 
  \includegraphics[width=0.45\textwidth]{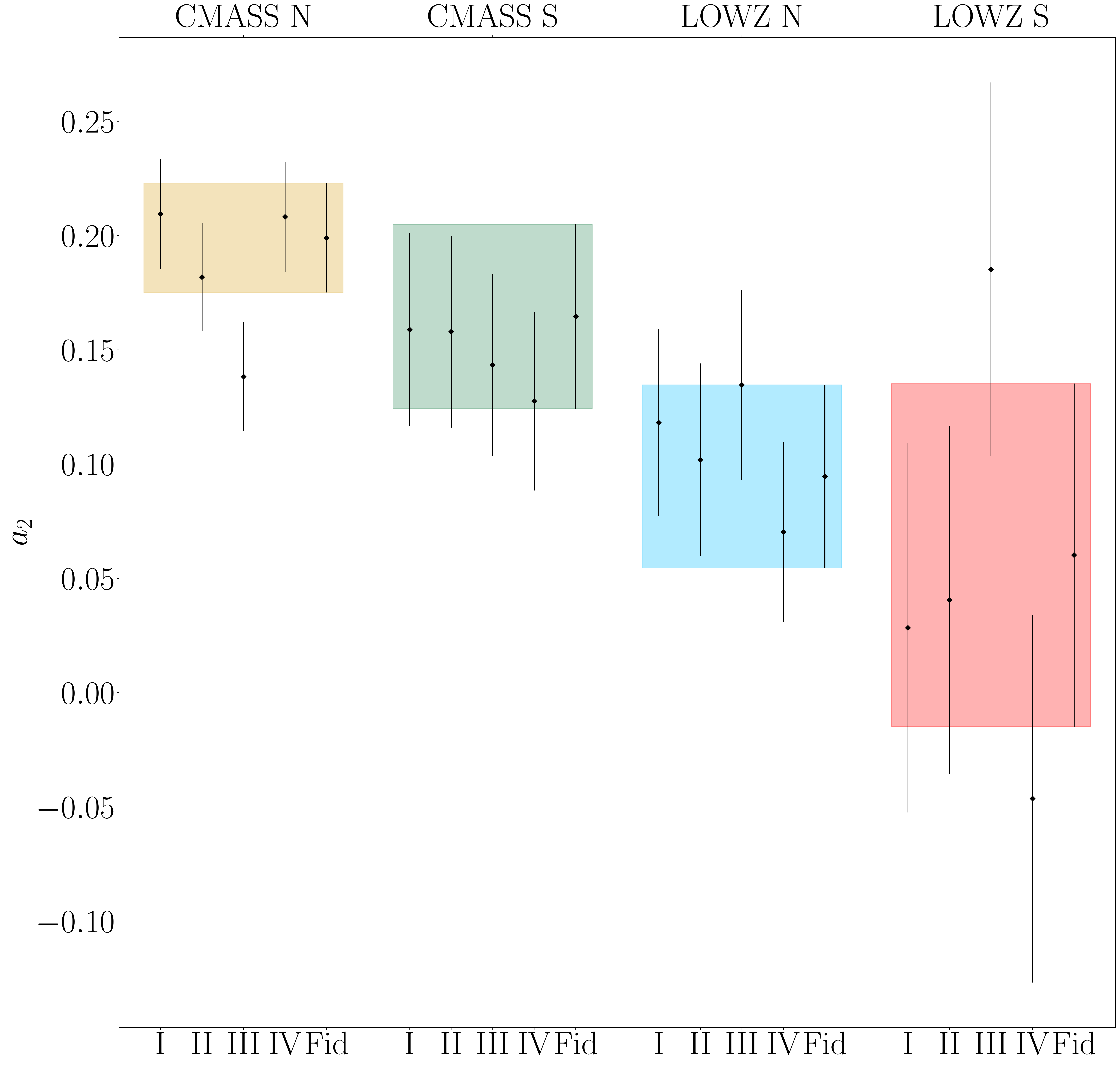}  
  \caption{Same as Figure~\ref{fig:sys_a}, but with parameters $a_0$ and $a_2$ (left/right panels).}
  \label{fig:sys_s0s2}
\end{figure}

The results of these tests are presented in Figure~\ref{fig:sys_a} for the amplitudes $A_{k,{\rm G}}$ and Figure~\ref{fig:sys_s0s2} for $s_{0}$ and $s_{2}$, respectively (left/right panels). We present a combined fit of five parameters $A_{k, {\rm G}}$, $s_{0}, s_{2}$ to $W_{1,2,3}$ extracted separately from the CMASS/LOWZ N/S data. That is, we fit the functional form in equation~(\ref{eq:fit}) to $W_{1,2,3}$, by minimizing $\chi^{2} = \chi_{1}^{2} + \chi_{2}^{2} + \chi_{3}^{2}$, with $\chi_{k}^{2}$ defined in equation~(\ref{eq:chi2}). The black diamonds and error bars are the result of repeating our analysis in the main body of the paper, relaxing each assumption made in points I-IV above. The `Fid' points in the figures are the fiducial values obtained in the main text body, and the solid gold/green/blue/red filled areas are the $\pm 1\sigma$ ranges of the parameters from the fiducial analysis for CMASS N/S and LOWZ N/S, respectively, included as a visual guide.

The CMASS N/S and LOWZ N/S data present self-consistent results for $A_{k, {\rm G}}$ and $a_{0,2}$ for practically all tests performed in this section. The only seemingly significant peculiarity is the low value of $a_{2}$ in the CMASS N data for test III (randomly sampling the galaxies; right panel in Figure~\ref{fig:sys_s0s2}). However, it is clear that $a_{2}$ is the least well measured quantity that we extract from the data, and it is not clear if we are obtaining a systematically high value of this quantity in the CMASS N data. This could also simply be a statistical fluctuation, as the discrepancy between this point and the fiducial measurement is not high (less than 2-$\sigma$). The amplitudes that we use for cosmological parameter estimation present excellent stability when we modify our analysis, which indicates that they are robust cosmological measurements. 

\begin{table}[tb]
\begin{center}
\resizebox{\columnwidth}{!}{%
 \begin{tabular}{|| c  c  c  c  c  c  c ||}
 \hline
 Data/MF \, & $\log[A_{k,{\rm G}}]$ \, & $a_{0}$ \, & $a_{2}$ & $h_{0}$ & $h_{2}$ & $h_{3}$  \\ [0.5ex] 
 \hline\hline
 CMASS N $W_{1}$ & $-6.281(-6.281) \pm 0.010$ & $0.013(0.013) \pm 0.022$ & $-$ &  $-$ & $0.000 \pm 0.005$  & $-$  \\
 CMASS S $W_{1}$ & $-6.284(-6.283)\pm 0.017$ & $0.010(0.018) \pm 0.041$ & $-$ &  $-$ & $0.023 \pm 0.007$ & $-$   \\
 LOWZ N $W_{1}$ & $-6.287(-6.286)\pm 0.016$ & $0.024(0.025)\pm 0.038$ & $-$ & $-$ & $-0.007 \pm 0.007$ &  $-$ \\
 LOWZ S $W_{1}$ & $-6.317(-6.315)\pm 0.026$ & $-0.110(-0.110)\pm 0.062$ & $-$ & $-$ & $0.020 \pm 0.011$ & $-$  \\
 \hline
 CMASS N $W_{2}$ & $-11.211(-11.225)\pm 0.025$ & $0.006(0.008)\pm 0.025$ & $0.216(0.225)\pm 0.042$ &  $-$ & $-$ & $0.012 \pm 0.011$  \\
 CMASS S $W_{2}$ & $-11.145(-11.192)\pm 0.036$ & $-0.021(-0.013)\pm 0.041$ & $0.244(0.276) \pm 0.066$ & $-$ & $-$ & $0.043 \pm 0.015$  \\
 LOWZ N $W_{2}$ & $-11.215(-11.190)\pm 0.038$ & $0.039(0.035)\pm 0.039$ & $0.075(0.076) \pm 0.066$ & $-$ & $-$ & $-0.017 \pm 0.016$   \\
 LOWZ S $W_{2}$ & $-11.227(-11.288)\pm 0.058$ & $-0.029(-0.033)\pm 0.064$ & $0.100(0.104) \pm 0.116$ & $-$ & $-$ & $0.050 \pm 0.023$  \\
 \hline
 CMASS N $W_{3}$ & $-15.838(-15.836)\pm 0.038$ & $0.003(0.003)\pm 0.026$ & $0.180(0.0175) \pm 0.038$ & $-0.002 \pm 0.019$ & $-$ & $-$   \\
 CMASS S $W_{3}$ & $-15.757(-15.752) \pm 0.058$ & $0.019(0.017)\pm 0.047$ & $0.099(0.097) \pm 0.065$ & $-0.006 \pm 0.029$ & $-$ & $-$  \\
 LOWZ N $W_{3}$ & $-15.837(-15.838)\pm 0.062$ & $0.030(0.029)\pm 0.046$ & $0.116(0.114) \pm 0.065$ &  $0.008 \pm 0.034$ & $-$ & $-$  \\
 LOWZ S $W_{3}$ & $-15.873(-15.879)\pm 0.108$ & $-0.097(-0.106)\pm 0.086$ & $0.062(0.046) \pm 0.129$ & $0.025 \pm 0.064$ & $-$ & $-$  \\
 \hline 
\end{tabular}
}
\end{center} 
\caption{\label{tab:4}Marginalised best fit and 1-$\sigma$ uncertainties on the parameters $\log[A_{k,{\rm G}}]$, $a_{0,2}$, and $h_{0,2,3}$ obtained by fitting the functions in equations~(\ref{eq:modfit1}-\ref{eq:modfit3}) to the data by minimizing the $\chi_{k}^{2}$ function (eq.~\ref{eq:chi2}). The values inside the brackets are the corresponding `fiducial' values with $h_{0,2,3}=0$. } 
\end{table}

Finally in this appendix, we include the results of including $h_{0,2,3}$ when fitting a Hermite polynomial expansion to the BOSS MF curves. In Table~\ref{tab:4}, we present the best fit and 1-$\sigma$ uncertainties in all parameters with $h_{0,2,3}$ included in the fits. Also included are the best fit values with $h_{0,2,3}=0$ (shown in brackets).

\end{document}